\documentclass[twocolumn,english,superscriptaddress,amsmath,amssymb,aps,prb,floatfix,nofootinbib]{revtex4-2}

\usepackage[english]{babel}
\usepackage{booktabs}
\usepackage{graphicx}% Include figure files
\usepackage{dcolumn}% Align table columns on decimal point
\usepackage{bm}% bold math
\usepackage{hyperref}% add hypertext capabilities
\usepackage[]{cleveref} %http://www.ctan.org/pkg/cleveref

\usepackage{multirow}
\usepackage{makecell}
\usepackage[caption=false]{subfig}
\usepackage{wasysym}

\renewcommand{\d}{\ensuremath{\mathrm{d}}}

\newcommand{\phantomsubfloat}[1]{
    {% apply caption setup only temporarily
        \captionsetup[subfigure]{labelformat=empty}
        \subfloat[][]{#1}
    }%
}
\hypersetup{colorlinks,
citecolor=blue,
filecolor=blue,
linkcolor=blue,
urlcolor=blue,
breaklinks=true
}
\usepackage{soul}

\newcommand{\RESP}[1]{#1}

\begin{document}

\title{Partial lifting of degeneracy in the $J_1-J_2-J_3$ Ising antiferromagnet on the kagome lattice}

\newcommand{\Ghent}{Department of Physics and Astronomy, University of Ghent, Krijgslaan 281, 9000 Gent, Belgium}
\newcommand{\Vienna}{Faculty of Physics, University of Vienna, 1090 Vienna, Austria}
\newcommand{\KIT}{Karlsruhe Institute of Technology,
76131 Karlsruhe, Germany}
\newcommand{\EPFL}{Institute of Physics, Ecole Polytechnique Fédérale de Lausanne (EPFL), CH-1015 Lausanne,~Switzerland}

\newcommand{\LPT}{Laboratoire de Physique Théorique, Université de Toulouse, CNRS, UPS,~France}

\author{Jeanne~Colbois}
\email{colbois@irsamc.ups-tlse.fr}
\affiliation{\EPFL}
\affiliation{\LPT}

\author{Bram~Vanhecke}
\email{bram.physics.vanhecke@gmail.com}
\affiliation{\Ghent}
\affiliation{\Vienna}
\author{Laurens~Vanderstraeten}
\affiliation{\Ghent}
\author{Andrew Smerald}
\affiliation{\KIT}
\author{Frank~Verstraete}
\affiliation{\Ghent}
\author{Fr\'{e}d\'{e}ric~Mila}
\affiliation{\EPFL}

\date{\today}
\begin{abstract}
Motivated by dipolar-coupled artificial spin systems, we present a theoretical study of the classical $J_1-J_2-J_3$ Ising antiferromagnet on the kagome lattice. We establish the ground-state phase diagram of this model for $J_1 > |J_2|, |J_3|$ based on exact results for the ground-state energies. When all the couplings are antiferromagnetic, the model has three macroscopically degenerate ground-state phases, and using tensor networks, we can calculate the entropies of these phases and of their boundaries very accurately. In two cases, the entropy appears to be a fraction of that of the triangular lattice Ising antiferromagnet, and we provide analytical arguments to support this observation. We also notice that, surprisingly enough, the dipolar ground state is not a ground state of the truncated model, but of the model with smaller $J_3$ interactions, an indication of a very strong competition between low-energy states in this model.
\end{abstract}

\maketitle

\section{Introduction}

In condensed matter physics, geometrical frustration characterizes those systems where a certain type of local order cannot propagate through space~\cite{Sadoc1999}. In frustrated systems, even a simple, local rule characterizing the ground states can give rise to complex low-energy configurations and often to a macroscopic ground-state degeneracy supporting emergent physics~\cite{Lacroix2011}. Two archetypal examples are the nearest-neighbor classical Ising antiferromagnets on the triangular (TIAFM) and kagome (KIAFM) lattices; both have a macroscopic ground-state degeneracy $W_{\rm{G.S.}}$ characterized by a finite residual entropy per site, \RESP{for which analytical expressions were obtained in the 1950's by diagonalizing the appropriate transfer matrices following Onsager and Kaufman's approach}~\cite{Onsager1944,Kaufman1949,Wannier1950,Wannier1973,Kano1953} (Appendix~\ref{sec:AppClosedForm}),
\begin{align}
    \label{eq:STIAFM}
    S_{\rm{TIAFM}} = \lim_{N \rightarrow \infty}\frac{1}{N} \ln(W_{G.S.}) &= 0.323066...\\
    S_{\rm{kagome}} &= 0.501833...
    \label{eq:SKIAFM}
\end{align}
While the former has a critical point at $T=0$, with algebraically decaying spin-spin correlations, the latter has exponentially decaying correlations with a very short correlation length $\xi = 1.2506\dots$~\cite{Suto1981,Apel2011}.

\begin{figure}
    \centering
    \includegraphics[width = \columnwidth]{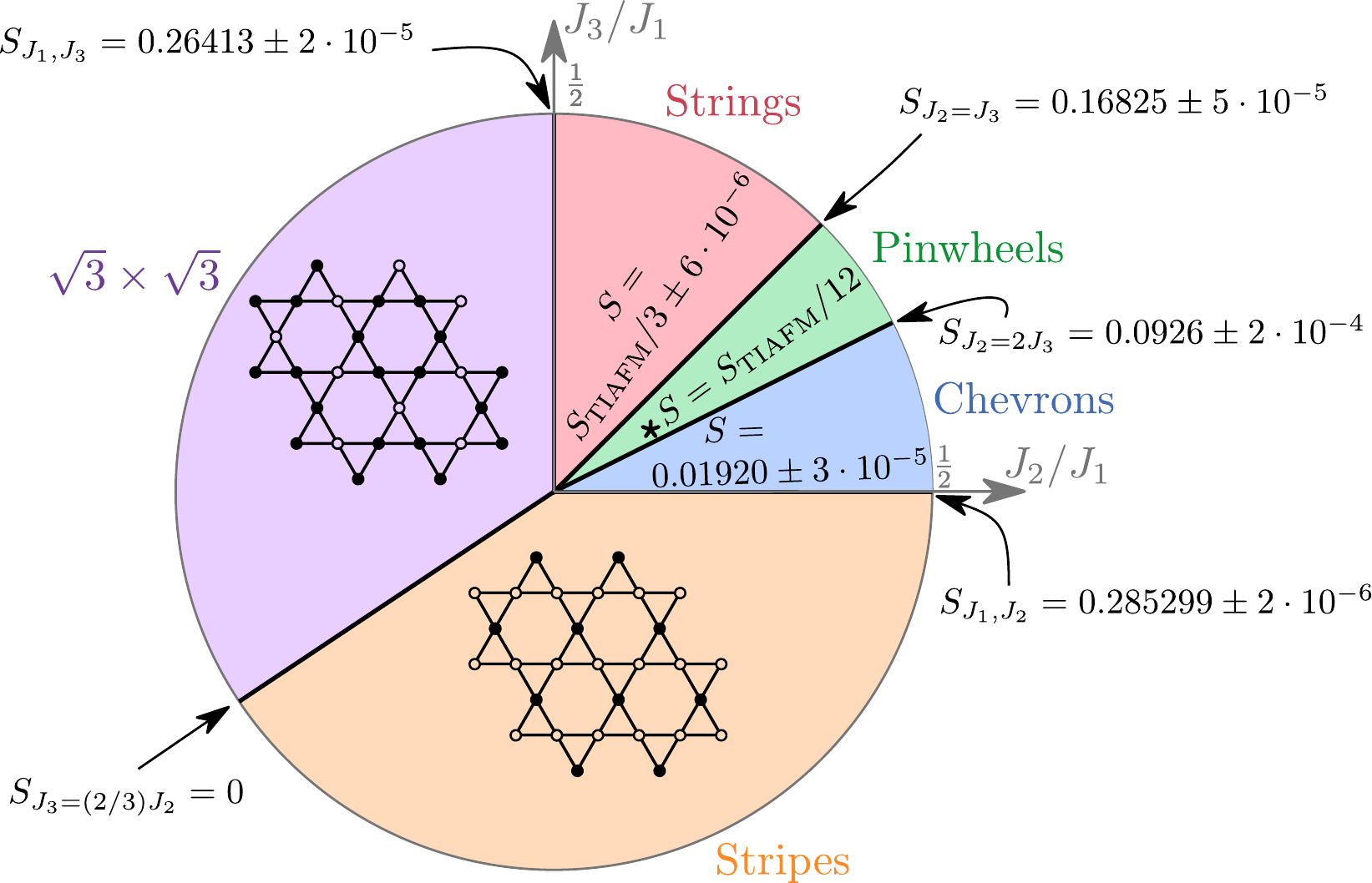}
    \caption{\label{fig:PDEntropy} Overview of the ground-state phase diagram for $J_1 \gg J_2, J_3$ (see Fig.~\ref{fig:Couplings}). The ground-state energies of the phases, given in Table~\ref{tab:ResultsOverview} and in the main text, are obtained exactly thanks to a method of inequalities~\cite{Kanamori1966}.  When $J_2$ or $J_3$ are ferromagnetic, there is no macroscopic ground-state degeneracy; when they are both antiferromagnetic, we find three phases and four boundaries with a finite residual entropy, which we evaluate using a tensor-network construction, as discussed in Sec.~\ref{sec:TNENT}. \RESP{The star indicates the value of} $J_2/J_1$ and $J_3/J_1$ \RESP{that corresponds to the dipolar model} Eq.~\ref{eq:PLGS_DKIAFM} \RESP{truncated to third-neighbor couplings} Eq.~\ref{eq:PLGS_truncatedDKIAFM}.}
\end{figure}

\par One generally assumes that the presence of a macroscopic degeneracy in the ground state of frustrated classical models is highly sensitive to perturbations. Simple examples of this sensitivity are the well-known mechanism of order-by-disorder, whereby thermal or quantum fluctuations select an ordered state over a classical macroscopically degenerate ground state, or the numerous cases where the macroscopic ground-state degeneracy in a certain model is completely lifted by farther-neighbor interactions. Although in three dimension there is a notable exception to this rule-of-thumb (the pyrochlore spin ice, which respects the ice rules down to very low temperatures \emph{because} of the dipolar interactions \cite{Gingras2011, Castelnovo2008, Isakov2004, Isakov2005}), in two-dimensional systems, the intuition that farther-neighbor interactions select a long-range ordered ground state is supported by a number of examples.

\par Such longer-range interactions arise naturally in artificial spin systems. These lithographically-patterned arrays of mesoscopic nanomagnets (fifty to a few hundred nanometers, either connected or not) allow a direct probing of the individual moments using magnetic microscopy techniques~\cite{Wang2006, Tanaka2006}, and can come in a wide variety of designs, making them particularly well suited to the study of frustration in two-dimensional systems and of celebrated models of statistical mechanics (see~\cite{Skjaervoe2020,Rougemaille2019,Schiffer2021} for recent reviews). The moments of the single-domain nanomagnets are well modeled by using Ising spins, with the axis of the Ising spins being parallel to the magnetic easy axis of the nanomagnet which is (typically) due to the shape anisotropy. In the first artificial spin systems that have been designed~\cite{Wang2006,Tanaka2006,Qi2008}, this easy axis was set parallel to the plane of the lattice (``in-plane''), but in the last ten years there has been growing interest for systems with an easy axis perpendicular to the plane of the lattice (``out-of-plane'')~\cite{Zhang2012,Chioar2014,Chioar2016,Rougemaille2019,Kempinger2021}.

\par When the magnets are not connected, the magnetostatic interactions are well approximated by a Hamiltonian with long-range, dipolar interactions~\cite{Rougemaille2011,Chioar2014,Mengotti2009}. In two dimensions, this can be considered as a fast decay (in the sense that the sum of the couplings at large distance is convergent). As a result, in general, the lifting of the ground-state degeneracy of frustrated models by the dipolar couplings, selecting a long-range ordered ground state, is already understood from the Hamiltonian truncated to short-range couplings. For instance, adding second-neighbor antiferromagnetic interactions to the TIAFM immediately reduces the residual entropy to zero and selects the stripe  state~\cite{Metcalf1974,Tanaka1975,Glosli1983,Korshunov2005,Rastelli2005,Smerald2016}. Such antiferromagnetic second-neighbor interactions arise naturally in the case of an artificial array of nanomagnets with out-of-plane magnetic anisotropy (i.e. the effective Ising spins' axis is perpendicular to the plane of the triangular lattice); and the stripe state indeed corresponds to the ground-state of the dipolar Ising model with an out-of-plane easy axis~\cite{Rossler2001,Smerald2017}. Similarly, the in-plane kagome artificial spin systems are modeled by dipolar interactions with ferromagnetic second-neighbor couplings~\cite{Tanaka2006, Moller2009, Rougemaille2011}, which  completely lift the macroscopic ground-state degeneracy of the KIAFM and select the $\sqrt{3}\times\sqrt{3}$ ordered state~\cite{Takagi1993,Chern2012,Kao2020}. Therefore, the ground state in these models is readily understood from a Hamiltonian truncated to second-neighbor interactions. In the kagome lattice case, the long-range interactions still play a significant role, by changing the nature of the finite-temperature phase transitions~\cite{Moller2009, Chern2011, Chern2012}.
\par In this paper, motivated by the physics of the dipolar out-of-plane kagome-lattice Ising antiferromagnet (DKIAFM), we study the kagome-lattice Ising antiferromagnet with up to third-neighbor interactions. In sharp contrast to the above examples, we find that large regions of the ground-state phase diagram have a macroscopic degeneracy. This is illustrated in Fig.~\ref{fig:PDEntropy}, which gives an overview of our results for the ground-state phase diagram of this model.
It is already well-known that the $J_1-J_2$ Ising antiferromagnet on the kagome lattice with antiferromagnetic $J_2$ has a macroscopic ground-state degeneracy~\cite{Takagi1993,Colbois2021}. However, the fact that the spin-spin correlations in that ground state are long-ranged~\cite{Takagi1993} makes it all the more surprising that a macroscopic degeneracy survives in a wide range of values of a third neighbor interactions, a result which is also in contrast to simple Pauling estimates that yield a negative residual entropy, suggesting an ordered ground-state~\cite{Hamp2018}. Furthermore, we show that the DKIAFM Hamiltonian truncated to third neighbors selects a ground-state manifold that \emph{does not} contain the long-range dipolar ground state. Exact results for the ground-state energies are accessible thanks to a method of inequalities developed by Kanamori in the 1960s~\cite{Kanamori1966,Ducastelle1991}, while very precise numerical results for the residual entropy are obtained using a recently introduced tensor-network approach. The latter is based on ground-state local rules~\cite{Huang2016} to construct \emph{contractible} tensor networks, allowing us to study frustrated models exhibiting macroscopic ground-state degeneracies~\cite{Vanderstraeten2018, Vanhecke2021}. 
\par The rest of the paper is organized as follows: In Sec.~\ref{sec:Model}, we introduce the model. We give an overview of its ground-state phase diagram with exact results for the ground-state energies in Sec.~\ref{sec:GSPD}; we then  give a short reminder of the tensor network approach introduced in Ref.~\onlinecite{Vanhecke2021} and an overview of our results for the entropy of the various ground-state phases. In Sec.~\ref{sec:PHASES} we give detailed discussions of the three main ground-state phases. We particularly focus on the phase corresponding to the dipolar model truncated to third-neighbor interactions, and we discuss its relation to the ground-state phase of the nearest-neighbor triangular-lattice Ising antiferromagnet (TIAFM). In Sec.~\ref{sec:DISC} we summarize our findings and put the results in perspective.

\section{DKIAFM and truncated Hamiltonian}
\label{sec:Model}

\par
\begin{figure}
    \centering
    \includegraphics[width = 0.5\columnwidth]{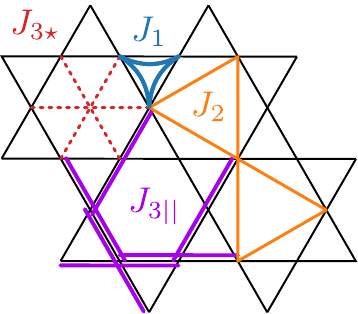}
    \caption[Couplings.]{\label{fig:Couplings} The first few couplings on the kagome lattice. In the present paper, we consider that the interaction strengths only depend on the distance, and therefore $J_{3\star}$ across the hexagon is the same as $J_{3||}$ which is parallel to the bonds of the lattice. The nearest- and second-neighbor interactions form kagome sublattices, and the third-neighbor interactions form triangular sublattices.}
\end{figure}

 The DKIAFM Hamiltonian is motivated by artificial spin systems of nanomagnets on a kagome lattice, with out-of-plane Ising anisotropy. Developed by Zhang \emph{et al}.~\cite{Zhang2012}, these systems were originally believed to show some ``universality'' between out-of-plane and in-plane artificial spin systems, since at the nearest-neighbor level both systems are described by the KIAFM Hamiltonian. However, it was soon shown that the configurations of the longer-range magnetostatic interactions must be taken into account when modeling these systems. A dipolar approximation with a correction to the nearest-neighbor interactions due to the finite-size of the nanomagnets and the effect of proximity which break the dipolar approximation seems to be sufficient~\cite{Chioar2014}. In the out-of-plane case, this yields the following Hamiltonian: 
\begin{equation}
    \label{eq:PLGS_DKIAFM}
	H_{\text{DKIAFM}} = J_0\sum_{\langle i,j \rangle} \sigma_i \sigma_j + D \sum_{(i,j)} \frac{\sigma_i \sigma_j}{|r_{i,j}|^3}, 
\end{equation}
 in which the effective nearest-neighbor interaction $J_{NN} = J_0 + D/|r_{NN}|^3$ is typically of the order of $ 1.5 (D/|r_{NN}|^3)$, with an exact value that can vary with the details of the system~\cite{Chioar2016,Colbois2021}. The first few interactions for this Hamiltonian are illustrated in Fig.~\ref{fig:Couplings}.
 
 \par The presence of dipolar couplings is actually playing an essential role, both in the ground state and at the finite temperatures that can be reached by the artificial spin systems. In the in-plane case, farther-neighbor interactions starting at the second neighbors select a long-range-ordered $\sqrt{3} \times \sqrt{3}$ ground state, which is reached when lowering the temperature by going through two successive phase transitions (either of the Kosterlitz-Thouless or Ising and Potts types, depending on whether one considers only second-neighbor interactions or the full dipolar interactions)~\cite{Moller2009, Chern2011, Chern2012}. 
 \par
\begin{figure}
    \centering
    \includegraphics[width = 0.7\columnwidth]{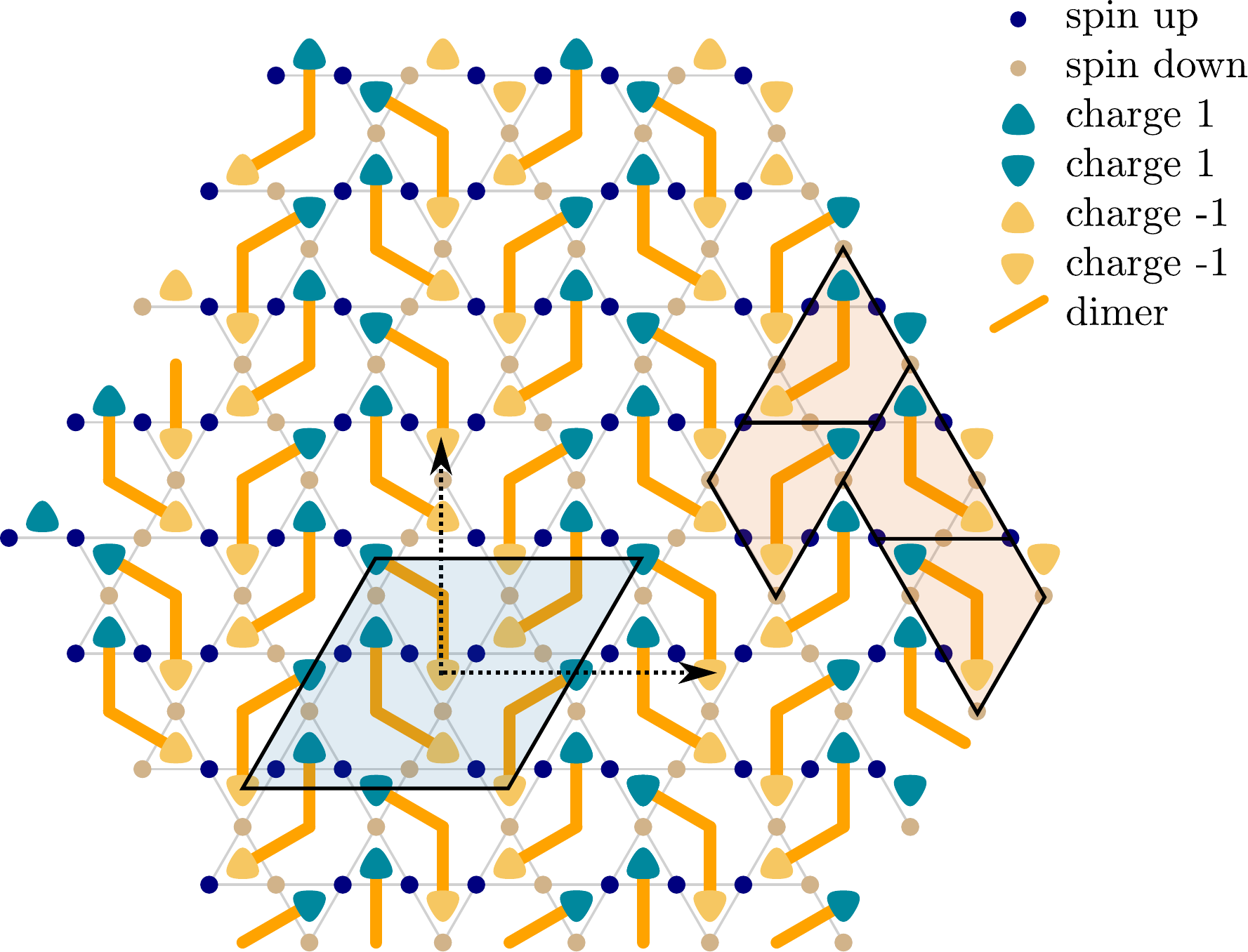}
    \caption[Ground state of the dipolar model.]{\label{fig:DipolarGS} The long-range ordered ground state of the dipolar model. The light-blue and the beige, rounded triangles describe the effective ``charges'' [Eq.~(\ref{eq:charge})], which form a pattern of zigzag-like stripes with alternating charge. The orange segments represent the dimers [Eq.~(\ref{eq:NM_dimmap})]. Two choices for the unit cell are shown : the ``7-shaped'' unit cell of Ref.~\onlinecite{Chioar2016} (light orange) and the unit cell of Ref.~\onlinecite{Hamp2018} (light blue).}
\end{figure}

 By contrast, in the out-of-plane case, Monte Carlo simulations on small systems (up to 300 spins)~\cite{Chioar2016,Hamp2018} strongly suggest that, when the long-range dipolar couplings are considered, the ground state is a long-range ordered state with a 12-site unit cell (Fig.~\ref{fig:DipolarGS}), which is reached through a direct first-order phase transition from the paramagnetic phase. The single-spin flip Monte Carlo simulations on this model have a strong tendency to fall out of equilibrium, making simulations on larger system sizes particularly challenging~\cite{Hamp2018}. The proposed ground state is characterized by an ordering of the effective ``charges''
 \begin{equation}
    \label{eq:charge}
     Q_{\bigtriangleup} = \sum_{i \in \bigtriangleup} \sigma_i\,, \quad Q_{\bigtriangledown} = -\sum_{i \in \bigtriangledown} \sigma_i\,
 \end{equation}
 following a pattern of alternating strings of positive and negative charges. Further evidence of the nature of the ground-state phase has been provided by a mean-field approach, which also gave arguments to interpret the dynamical slowing down of the model near the first order transition based on a spin-glass transition occurring at the mean-field level~\cite{Cugliandolo2020}.
\par It is well-known that the second-neighbor Ising antiferromagnet on the kagome lattice has a macroscopic ground-state degeneracy~\cite{Takagi1993, Hamp2018, Colbois2021}. This motivated an attempt by Hamp et. al.~\cite{Hamp2018} to understand the lifting of the ground-state degeneracy from the nearest-neighbor case to the DKIAFM ground state. However, a simple Pauling-like estimate, assuming that the nearest-, second-, and third-neighbor triangles all satisfy a ``two-ups one-down /two-downs one-up'' rule, yields a negative residual entropy $S_{\triangle\text{ rules}}$, perhaps suggesting that the system orders when third-neighbor interactions are considered. The authors also attempted single-spin-flip Monte Carlo simulations, which fall out of equilibrium even on small system sizes but show a non zero value of the order parameter of the DKIAFM ground state at low temperature, suggesting a possible ordering compatible with this ground state. Yet, no definitive information about the exact ground state was obtained for the dipolar Hamiltonian truncated to third-neighbor couplings.
\par At the same time, it is known that at carefully selected couplings, Ising antiferromagnets on the kagome lattice with up to third-neighbor interactions either across the hexagons or along the bonds of the lattice, but not both, can be written as a perfect sum of squares or mapped onto a charge Hamiltonian~\cite{Hamp2018, Mizoguchi2017}. At these specific points, a finite residual entropy is recovered, for instance in the $J_1-J_2-J_{3||}$ model when $J_2 = J_{3||}$~\cite{Wolf1988, Mizoguchi2017}. This line was thoroughly studied, be it for the Ising model~\cite{Wolf1988,Mizoguchi2017, Tokushuku2020}, the classical and the quantum Heisenberg models~(\cite{Mizoguchi2018,Li2021,Kiese2022,Lugan2022} and references therein), and the macroscopic degeneracy on this fine-tuned line has been showed to be particularly stable. In a different context, it was also recently noticed that the macroscopic ground-state degeneracy in the Ising model is not completely lifted when $J_2 < J_{3||}$~\cite{Colbois2021}. Furthermore, in the $J_1-J_2-J_3$ model, when $J_1$ is \emph{ferromagnetic}, there is a phase with a macroscopic ground-state degeneracy~\cite{Vanhecke2021}.
\par Together with the puzzling results for the truncated dipolar Hamiltonian, the survival of such properties related to frustration motivates us to study the following Hamiltonian for arbitrary $J_2$, $J_3$ with antiferromagnetic nearest-neighbor interactions such that $J_1 \gg |J_2|, |J_3|$\footnote{Overall, our results are valid for $\sqrt{J_2^2+J_3^2} \lesssim J_1/2$. We also note that a similar Hamiltonian was studied in Ref.~\cite{Kao2020}, with $J_{3||} \neq J_{3\star}$, but considering the case of $J_2 <0$ relevant for \emph{in-plane} artificial spin ice; the ground states in those cases are long-range ordered.}:
\begin{equation}
	\label{eq:Hamiltonian}
	\begin{split}
	H = &J_1 \sum_{\langle i,j \rangle} \sigma_i \sigma_j +J_2 \sum_{\langle i,j \rangle_2} \sigma_i \sigma_j\\ &+J_{3||} \sum_{\langle i,j \rangle_{3||}} \sigma_i \sigma_j+J_{3\star} \sum_{\langle i,j \rangle_{3\star}} \sigma_i \sigma_j,	
	\end{split}
\end{equation}
where $\langle\cdot,\cdot\rangle_k$ indicates $k$-th neighbors (Fig.~\ref{fig:Couplings}) and where we take $J_{3\star} = J_{3||} =:J_3$ corresponding to the fact that the interactions in the dipolar model only depend on the distance.

\section{Overview of the ground-state phase diagram}
\label{sec:GSPD}

\subsection{Ground-state energies}
To study the model in Eq.~\ref{eq:Hamiltonian} and establish exact results for the ground-state energies, we use Kanamori's method of inequalities (also known as the configurational polytope method), a technique introduced in Ref.~\onlinecite{Kanamori1966} and later reformulated and modified by several authors \cite{Ducastelle1991, vanDeWalle2000, Kudo1976, Wolf1988}.\footnote{Ref.~\onlinecite{Ducastelle1991} (Ch. 3, p. 109) gives a list of all papers using the Kanamori method until 1990. See also~\cite{Colbois2022}.} \RESP{The purpose of this approach is to obtain ground-state energy lower bounds by finding inequalities, imposed by the lattice structure and the nature of the local degree-of-freedom, that constrain the convex set of possible correlations of the model. } 

\begin{figure*}
\centering
\phantomsubfloat{\label{fig:Kanamori2D_a}}
\phantomsubfloat{\label{fig:Kanamori2D_b}}
\phantomsubfloat{\label{fig:Kanamori2D_c}}
\phantomsubfloat{\label{fig:Kanamori2D_d}}
\phantomsubfloat{\label{fig:Kanamori2D_e}}
\phantomsubfloat{\label{fig:Kanamori2D_f}}
\includegraphics[width = 0.95\textwidth]{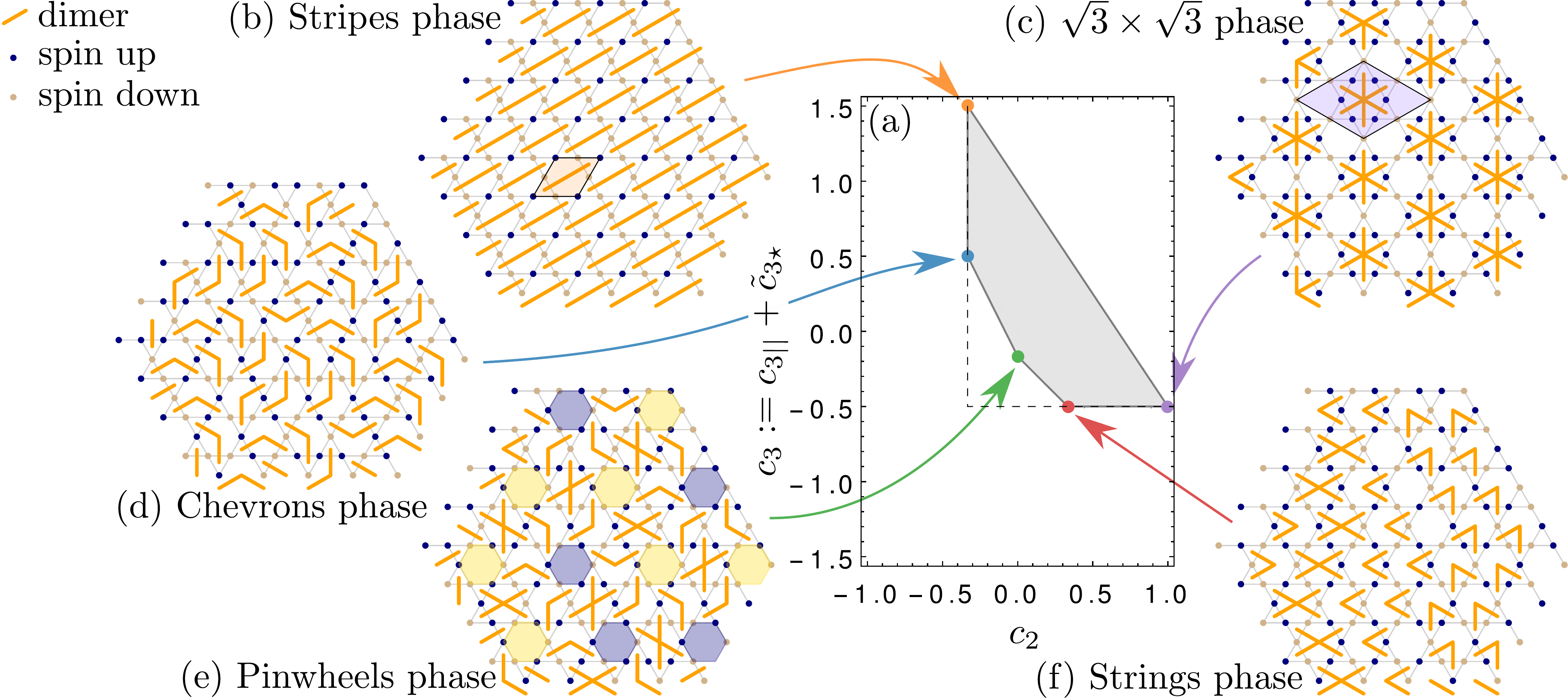}
\caption{\label{fig:Kanamori2D}(a) Polytope from Kanamori's method of inequalities reduced for $J_1 \gg J_2, J_3$, $J_1 > 0$. For reference, the dashed lines give the inequalities corresponding to minimizing the energy on all the $J_2$ or all the $J_3$ triangles respectively. For each corner we show one Monte Carlo snapshot which has these correlations. In all cases $c_1 = - \frac{1}{3}$. The correlations correspond to states as follows. (b) $(c_2, c_{3}) = (-1/3, 3/2)$: long-range ordered Stripe states; (c) $(c_2, c_3) = (1, -1/2)$: long-range ordered $\sqrt{3}\times\sqrt{3}$ states; (d) $(c_2, c_3) = (-1/3, 1/2)$: a phase where pair of dimers form chevrons; (e) $(c_2,c_3) = (0, -1/6)$: a phase where dimers rotate clockwise or counter-clockwise around empty hexagons as pinwheels (highlighted); (f) $(c_2,c_3) = (1/3, -1/2)$: a phase characterized by strings of crosses spanning the system. The corresponding ground-state phase diagram is shown in Fig.~\ref{fig:PDEntropy}.}
\end{figure*}

\par For our purposes, the method consists in using small clusters with $n$ sites ($n = 3, \dots, 8$) to construct inequalities of the form
\begin{equation}
    \label{eq:ineqloc}
    \mod(n, 2) \leq \left(\sum_{i=1}^n s_i \sigma_i\right)^2 \leq n^2
\end{equation}
(where $s_i$ are arbitrary signs). With these inequalities, one can constrain the possible values of the (disconnected) spin-spin correlations
\begin{equation}
    c_k := \frac{1}{2N} \sum_{\langle i,j \rangle_k} \sigma_i \sigma_j,
\end{equation}
where $k$ stands for type of neighbor pairs and $N$ for the total number of sites in the system. 
\par On the kagome lattice, there are two first-neighbor bonds per site, implying that $-1 \leq c_1 \leq 1$. Similarly, there are two $k$-th neighbor bonds per site for $k = 2$ and $ k = {3||}$ (Fig.~\ref{fig:Couplings}), and therefore $-1 \leq c_k \leq 1$; there is only one third-neighbor bond of the star type per site, and therefore $-1/2 \leq c_{3\star} \leq 1/2$. 
\par Associating the inequalities in Eq.~\ref{eq:ineqloc} with a cluster of sites on the lattice and translating/rotating that cluster yields stronger inequalities for the correlations. Considering all clusters which only include up to third neighbor spin pairs, one gets a set of inequalities defining a polytope in the $(c_1, c_2, c_{3||}, c_{3\star})$ space which further constrains the possible values of the correlations. Note that the inequalities might be redundant and only a few of them are the main, most constraining ones.

\par Since, for given values of the $J_k$'s, Eq.~\ref{eq:Hamiltonian} corresponds to a hyperplane in the correlation space whose offset along the vector $\vec{J} := (J_1, J_2, J_{3||}, J_{3\star})$ changes with the energy per site, the minimal energy is reached at the boundaries of the polytope. This straightforwardly provides lower bounds for the ground-state energy per site as a function of the couplings. If one can prove that a given corner of the polytope is realized (namely, there exists a configurations whose correlations correspond to the coordinates of the corner), the associated ground-state energy is not just a lower bound but an exact result. Since a range of $\vec{J}$ share the same corner as an extremal point, a validated corner corresponds to a ground-state phase; similarly, the edges of the polytope correspond to phase boundaries between two phases.

\par Applying this to the Hamiltonian of Eq.~\ref{eq:Hamiltonian} yields a list of 17 main inequalities corresponding to a polytope in the four-dimensional $(c_1, c_2, c_{3||}, c_{3\star})$ space. Being interested in the $J_{3\star} = J_{3||}$ case, we can reduce it to a three-dimensional polytope in the space $(c_1, c_2, c_{3||}+ c_{3\star})$ by taking the convex hull of the set
\begin{equation}
    \left\{(c_1^{(i)}, c_2^{(i)}, c_{3||}^{(i)}, c_{3\star}^{(i)}) \cdot (0,0,1,1)\right \}_{i\in\text{corners}}.
\end{equation} 
\par A further simplification can be obtained by considering large enough $J_1$, which corresponds to minimizing the nearest-neighbor correlations $c_1$; in practice, because $c_1 \geq -1/3$, we can focus on the section of the three-dimensional polytope that lies in the $c_1 = -1/3$ plane. This gives the polygon illustrated in Fig.~\ref{fig:Kanamori2D}.
\par \RESP{The corners of the Kanamori polytope directly correspond to ground-state energy lower bounds. Yet, one has to check that these corners can be realized; to this end, we perform small-scale Monte Carlo simulations using the dual worm algorithm of Ref.}~\onlinecite{Rakala2017}. Importantly, we use periodic boundary conditions, which means that if we find a state realizing the spin-spin correlations of the corner, these correlations can also be realized in the infinite-size limit. The Monte Carlo updates rely on a mapping from the spin model on the kagome lattice to a (classical) dimer model on the (dual) dice lattice. A dimer variable is defined by
\begin{equation}
\label{eq:NM_dimmap}
	d_{b_{i,j}} = \sigma_i \sigma_j  = \begin{cases}
	1 & \text{if } \sigma_i = \sigma_j \\
	-1 & \text{if } \sigma_i = -\sigma_j 
	\end{cases},
\end{equation}
if $b_{i,j}$ is the bond of the dual graph lying between nearest-neighbor sites $i$ and $j$ of the original lattice; we put a dimer on $b_{i,j}$ if $d_{b_{i,j}} = 1$, otherwise we do not (see e.g. Fig.~\ref{fig:DipolarGS}). We name the phases as indicated in Fig.~\ref{fig:Kanamori2D}, according to the typical Monte Carlo configuration in spin or dimer language. Note that although the Kanamori approach provides exact results for the ground-state energy, in principle it does not say anything about the degeneracy of the ground states.

\par Two of these phases, the $\sqrt{3}\times\sqrt{3}$ phase and the stripes phase, are long-range ordered and correspond to ferromagnetic second-neighbor sublattices and ferromagnetic third-neighbor sublattices, respectively. They are also present in the related $J_1-J_2-J_{3||}$ model on the kagome lattice studied by Wolf and Schotte \cite{Wolf1988}, where $J_{3\star} = 0$. Note that the presence of a non zero $J_{3\star}$ coupling in our model creates three triangular third-neighbor sublattices, instead of square sublattices in the model of Wolf and Schotte, meaning that the main differences between the ground states of the two models arise when $J_3$ is antiferromagnetic. \RESP{Because they correspond to ferromagnetic order on the $J_2$ or the $J_3$ sublattices, these phases have the same periodicity as the ground-state phases occuring in similar parameter ranges in the Heisenberg version of the $J_1-J_2-J_3$ model}~\cite{Domenge2005,Messio2011, Messio2012,Grison2020}\RESP{; the main difference between the Ising model and its Heisenberg counterpart is the way of accommodating the frustration and minimizing the energy on a nearest-neighbor triangle.}. 
\par We turn to antiferromagnetic $J_2$ and $J_3$ couplings, where the ground-state energy is obtained by minimizing the second- and third-neighbor correlations. In Fig.~\ref{fig:Kanamori2D}, we indicated by dashed lines the corner formed by the two inequalities $c_2 \geq -1/3$ and $c_3 \geq -1/2$ that correspond to satisfying the rules (2-ups 1-down, 2-downs 1-up) on each $J_2$ triangle and each $J_3$ triangle, respectively. We find more restrictive inequalities that cut this corner, which shows that there is no state satisfying simultaneously the rule on all nearest, second-nearest and third-nearest neighbor triangles at once. This explains the negative results from Hamp et. al.~\cite{Hamp2018} for the Pauling-like estimate based on this assumption. Instead, we find three corners, i.e. three ground-state phases which, to the best of our knowledge, have not been reported before (Figs.~\ref{fig:Kanamori2D_d},~\ref{fig:Kanamori2D_e} and ~\ref{fig:Kanamori2D_f}). 
\par At small values of $J_3$, we find a phase characterized by chevrons (pair of dimers at a $120^{\circ}$ angle), whose ground-state manifold contains the 7-shaped ground state of the DKIAFM. With small system sizes (144 sites), the Monte Carlo simulations stay in equilibrium and we find configurations that match the ground-state energy lower bound from the Kanamori method:
\begin{equation}
    E_{\text{Chevrons}} = - \frac{2}{3}J_1 - \frac{2}{3}J_2 + J_3.
\end{equation}
Going back to the three-dimensional Kanamori polytope, one can check that the chevrons phase is the ground state for arbitrary antiferromagnetic $J_2$, with $J_3 < \min(J_2/2, J_1/4)$.
Furthermore, the phase seems to have a sub-extensive entropy in the sense that we find only non local, torus-winding updates between the ground states. However, for larger system sizes the Monte Carlo simulations fall out-of-equilibrium despite the worm update and the parallel tempering, and fail to reach the ground state. This means that it would be extremely challenging to use the Monte Carlo simulations to determine whether there is a finite or a zero residual entropy per site in this phase.

\par
\begin{figure}
    \centering
    \phantomsubfloat{\label{fig:EntropyPinwheelsMC_a}}
    \phantomsubfloat{\label{fig:EntropyPinwheelsMC_b}}
    \phantomsubfloat{\label{fig:EntropyPinwheelsMC_c}}
    \includegraphics[width = \columnwidth]{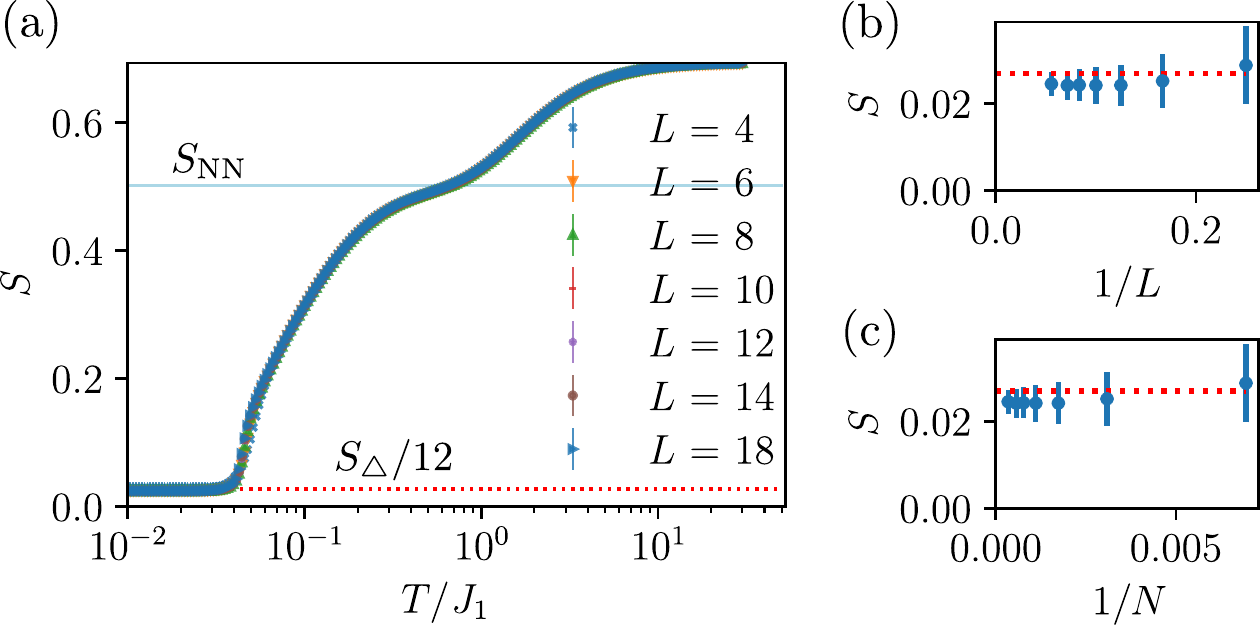}
    \caption[Residual entropy in the pinwheels phase from MC.]{\label{fig:EntropyPinwheelsMC}Residual entropy in the pinwheels phase as obtained from Monte Carlo thermodynamic integration (for $J_2/J_1 = 0.11$ and $J_3/J_1 = 0.09$). (a) Overview of the entropy as a function of the temperature. (b) and (c): Residual entropy in the lowest temperature as a function of (b) one over the linear system size $L$ and (c) one over the number of sites $N = 9L^2$.}
\end{figure}

 For $J_2/2 < J_3 < J_2$ and for $J_3 + 2/3 J_2 < J_1$, the ground states are characterized by a triangular lattice of empty hexagons (hexagons without any dimers), around which the dimers rotate either clockwise or counter-clockwise like the wings of a pinwheel. As can be seen in Fig.~\ref{fig:Kanamori2D_e}, the Monte Carlo results give hints that these pinwheels correspond to emergent Ising degrees of freedom which seem to respect a ``two-rights one-left/two-lefts one-right'' rule (three neighboring pinwheels cannot all have the same chirality), suggesting a macroscopic ground-state degeneracy that we can conjecture is $S \cong S_{\triangle}/12$. In this phase, both second- and third-neighbor ferromagnetic triangles can be found, in agreement with the ground-state energy
\begin{equation}
    \label{eq:EPinwheels}
    E_{\text{Pinwheels}} = - \frac{2}{3}J_1 - \frac{1}{3} J_3.
\end{equation}
Here, the Monte Carlo update is efficient enough and we can perform thermodynamic integration to try and verify the postulate for the value of the residual entropy. This is illustrated in Fig.~\ref{fig:EntropyPinwheelsMC} and the challenge is quite clear: Although we do find a finite residual entropy in the Monte Carlo simulations which is compatible with our prediction (Fig.~\ref{fig:EntropyPinwheelsMC_a}), the size of the error bars and the lack of control over the finite-size scaling means that we do not have a very high control on the accuracy of the residual entropy (Figs.~\ref{fig:EntropyPinwheelsMC_b} and~\ref{fig:EntropyPinwheelsMC_c}).

\par Finally, for arbitrarily large $J_3$ and for $J_2 < \min(J_3, 2 J_1/5)$, all third-neighbor triangles respect a ``two-ups one-down /two-downs one-up'' rule and the ground-state energy is given by
\begin{equation}
    \label{eq:StrGSE}
    E_{\text{Strings}} = - \frac{2}{3}J_1 + \frac{2}{3} J_2 - J_3.
\end{equation}
In the dimer picture, the ground states in this phase are characterized by directed strings of crosses spanning the system. We will see that these strings play a crucial role to understand the entropy of this phase, so we name it the strings phase. A simple sketch allows one to find a lower bound for the residual entropy per site: it is easy to construct a state with the energy of Eq.~\ref{eq:StrGSE} and such that one spin every nine sites can be set up or down at random, which gives 
\begin{equation}
    S_{\text{Strings}} \geq \frac{1}{9}\ln(2) \cong 0.077 \dots . 
\end{equation}
\par On top of proving the ground-state energy exactly, we also find that all three ground-state phases have a non trivial ground-state degeneracy. Therefore, we now focus on obtaining precise results for the residual entropy in these phases.

\subsection{Residual entropy from tensor networks}
\label{sec:TNENT}
As well evidenced by the chevrons and pinwheels phases, obtaining a few ground-state configurations with small-scale Monte Carlo simulations is far less challenging than obtaining a precise result for the residual entropy per site, which requires a good control over the whole temperature range to allow for the thermodynamic integration, as well as a good understanding of the expected finite-size scaling~\cite{Kolafa2014,Roma2004}. Furthermore, in all these phases, we expect the residual entropy to be very small, if not zero.
\par Therefore, we turn to a different technique to evaluate the residual entropy: tensor networks\footnote{See e.g. Refs.~\cite{Okunishi2021, Cirac2021} for recent reviews on tensor networks.}. However, it is now well-known that in the presence of frustration and macroscopic degeneracy, standard contraction algorithms (VUMPS~\cite{Baxter1968,Fishman2018,ZaunerStauber2018,Nietner2020}, CTMRG~\cite{Baxter1968,Baxter1978,Nishino1996}, TRG/TNR~\cite{Levin2007,Evenbly15}, ...) fail to converge when applied to the usual formulation of the partition function in terms of a tensor network of on-bond Boltzmann weight tensors and on-site delta tensors~\cite{Wang2014, zhu2019, Vanhecke2021, FriasPerez2021,Liu2021}. This problem is at least partially due to the imperfect numerical cancellation of very small and very large numbers~\cite{zhu2019,Liu2021}. Since increasing the numerical precision for the computations is, in general, not viable, one must find an alternative approach. In Ref.~\onlinecite{Liu2021}, it was shown that working with the logarithm of the Boltzmann weights (the ``tropical algebra'') solves the problem in the case of \emph{exact}, finite-size contractions. It is however not clear how to use this tropical algebra in the case of \emph{approximate} contraction algorithms such as VUMPS or CTMRG, which are very powerful to study translation invariant problems. Some of us have shown that the contraction issue can be circumvented by finding alternative formulations of the partition function based on a ground-state local rule encoded at the level of the tensor, yielding \emph{contractible} tensor networks for frustrated Ising models ~\cite{Vanderstraeten2018, Vanhecke2021}. \footnote{A method partially inspired from this approach was also successfully applied to the fully frustrated XY model~\cite{Song2022}.} 

\par Here, we follow this second approach, described in Ref.~\cite{Vanhecke2021}, where the problem is solved by making sure that the frustration is dealt with at the level of the tensor. In this method, and when the local rule is not already known, the first step is to split the Hamiltonian into a sum of local Hamiltonians on small clusters, with weights describing how each bond is shared between the local cluster Hamiltonians~\cite{Huang2016}. For any weights, the ground-state energy of the local Hamiltonian on the cluster gives a lower bound on the global ground-state energy (assuming periodic boundary conditions); this lower bound can be maximized over the set of weights (MAX-MIN approach, see~\cite{Huang2016}). \RESP{The point is that if the cluster is such that this optimal lower bound matches the exact ground-state energy, then this implies that we can minimize the energy of all the cluster Hamiltonians simultaneously, i.e. the local ground-state configurations can be arranged together in a compatible manner to construct a global ground state}~\cite{Vanhecke2021}.\RESP{ We then say that the frustration is relieved. In this case, \emph{all} the ground states are described as a tiling of the local ground-state configurations on the cluster, with these ground-state ``tiles'' corresponding to a local rule. Indeed, considering a state on the periodic lattice that is not an arrangement of these tiles, by definition, there is at least one cluster where the local configuration does not have the local ground-state energy, and this energy cannot be gained elsewhere; therefore it is not a ground-state of the full Hamiltonian}. In all the ground-state phases of interest here, we find that a star-shaped cluster is sufficient to relieve the frustration.

\par An important finding of Ref.~\onlinecite{Vanhecke2021} is that the weights have to be selected carefully so as to have as little \emph{spurious} tiles as possible. These tiles correspond to spin configurations on the selected reference cluster that have the ground-state energy but that cannot be used together with other ground-state tiles to cover the lattice. In some cases, the incompatibility of these spurious tiles with other ground-state tiles might be immediately detected, e.g. when it is not possible to find a compatible set of tiles in the immediate neighborhood of one of these tiles. In other cases, however, the impossibility to cover the lattice might only appear in a non local way. In the presence of a macroscopic ground-state degeneracy, this can make the convergence of the approximate contraction algorithm much more difficult, or even prevent it altogether.

\begin{figure}
\centering
    \phantomsubfloat{\label{fig:Matching_a}}
    \phantomsubfloat{\label{fig:Matching_b}}
    \includegraphics[width = \columnwidth]{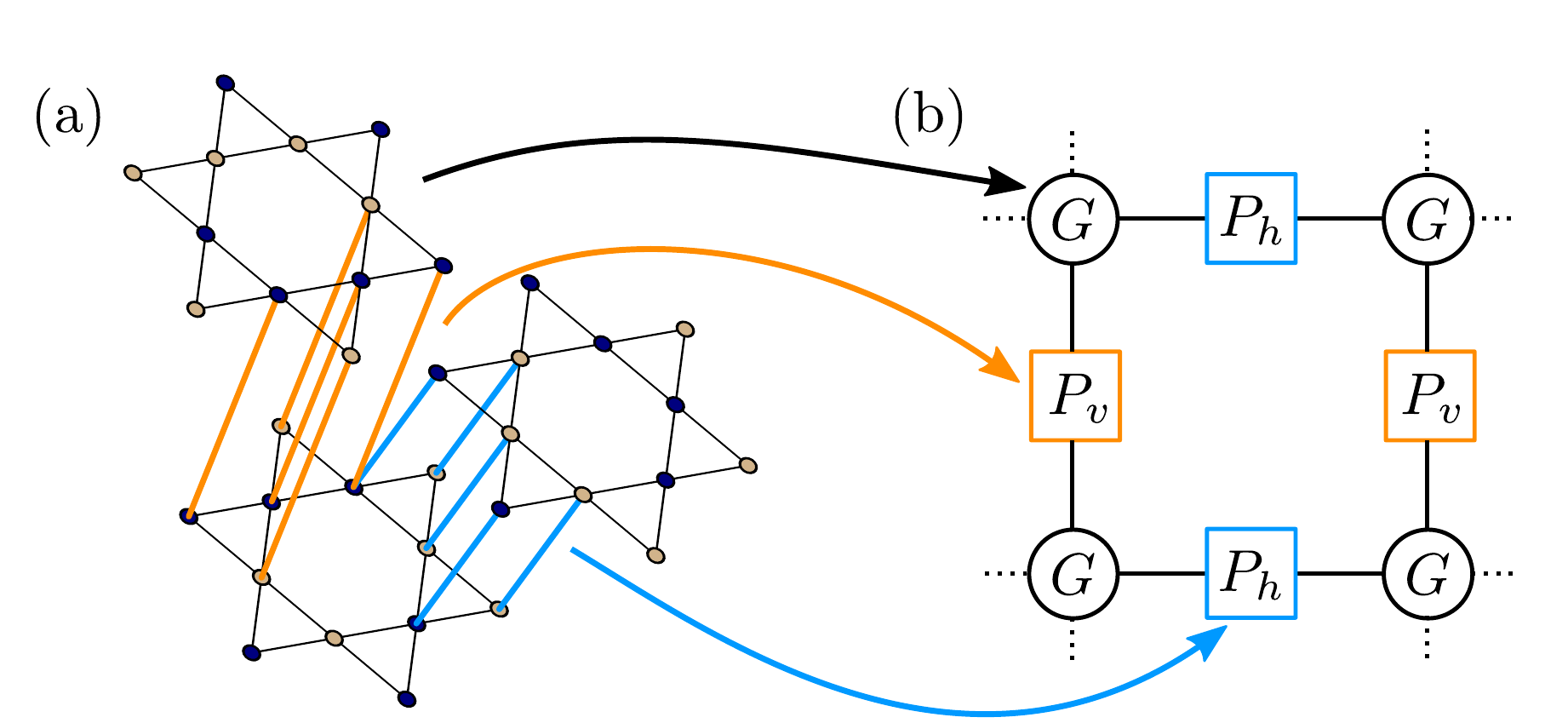}
    \caption{\label{fig:Matching}Reminder of the tensor-network construction of Ref.~\onlinecite{Vanhecke2021}. (a) Matching star-shaped tiles on the square lattice to create a kagome lattice. Horizontally and vertically, the tiles match if the five spins on the sites that they share have the same values. This automatically enforces the matching in the diagonals of the square lattice. (b) Tensor network based on this tiling. A four-legged $\Delta$-tensor $G$ associates a number to each ground-state tile. The bond tensors $P_h$ and $P_v$ enforce that neighboring tiles must match. The bond dimension can be reduced by performing an SVD on the bond matrices and grouping with the $G$ tensors.}
\end{figure}

\par Once the weights have been carefully selected according to the procedure described in Ref.~\onlinecite{Vanhecke2021}, using the ground-state tiles to build a tensor network on the square lattice is straightforward. As illustrated in Fig.~\ref{fig:Matching}, the kagome lattice can be constructed as a square-lattice tiling of overlapping clusters. To each square-lattice site, i.e. to each star-shaped cluster, we associate a $\Delta$-tensor $G$ whose indices label the allowed local ground-state configurations (or ground-state tiles). To each horizontal (vertical) square lattice bond we associate a tensor $P_h$ ($P_v$) which imposes that two neighboring ground-state tiles have the same spin configuration on the sites that they share. By performing a truncated singular-value decomposition (SVD) on these sparse bond tensors and grouping them with the on-site $G$ tensor, the bond dimension can be reduced to at most $2^{n_s}$, where $n_s$ is the number of shared sites (here, $n_s = 5$). 

\par The residual entropy per site is related to the leading eigenvalue of the row-to-row transfer matrix associated with the tensor network we just described~\cite{Baxter1968,Baxter1978,Vanderstraeten2018,Vanhecke2021}, and can be obtained efficiently using the VUMPS or the multi-site VUMPS algorithms~\cite{Baxter1968,ZaunerStauber2018,Fishman2018,Vanderstraeten2019_Tangent,Nietner2020} as appropriate. 
\par \RESP{A remark has to be made about the transfer matrix. We are considering problems on the kagome lattice with farther-neighbor interactions, and to use the VUMPS algorithm, we map them to a tensor network on the square lattice. In such cases, the standard  formulation of the partition function typically gives rise to non Hermitian 1D transfer matrices; our alternative construction also tends to yield non Hermitian transfer matrices. However, even though the VUMPS algorithm applied to such transfer matrices is not justified by a variational principle anymore, it is still valid from the point of view of the tangent space projection}~\cite{Vanderstraeten2019_Tangent, Vanhecke2021_SciPost,Vanderstraeten2022,Colbois2022}. \RESP{Furthermore, if problems are encountered, one can turn the algorithm into a sequential optimization of the normalized fidelity.}
\par Performing the approximate contraction for increasing bond dimensions of the boundary MPS, we find more and more precise values of the residual entropy. 
For each bond dimension, we get an error estimate as
\begin{equation}
    \delta S = \log\left(\sqrt{\lambda^{(2)}}\right) - \log\left(\lambda\right)
\end{equation}
where $\lambda$ is the leading eigenvalue (for one site) of the one-row transfer matrix, whereas $\lambda^{(2)}$ is evaluated as the leading eigenvalue of the \emph{two-row} transfer matrix, where the second row transfer matrix is transposed (to ensure positivity), and we re-use the same VUMPS environment. This quantity is the conceptual equivalent of the energy variance in the context of quantum spin chains, and quantifies how close the MPS is to an exact eigenstate of the transfer matrix. In order to estimate the value in the infinite bond dimension limit and to obtain a reasonable error bar in that limit, we perform a simple linear extrapolation based on a few of the largest bond dimensions, and we compare this extrapolated result to the maximal entropy obtained over all the finite bond dimensions that we have computed\footnote{One can see from the pinwheels phase in Sec.~\ref{sec:PHASES} where the numerical result can be compared to an exact result that this is a reasonable approach.}.

\begin{figure}
    \centering
    \includegraphics[width = \columnwidth]{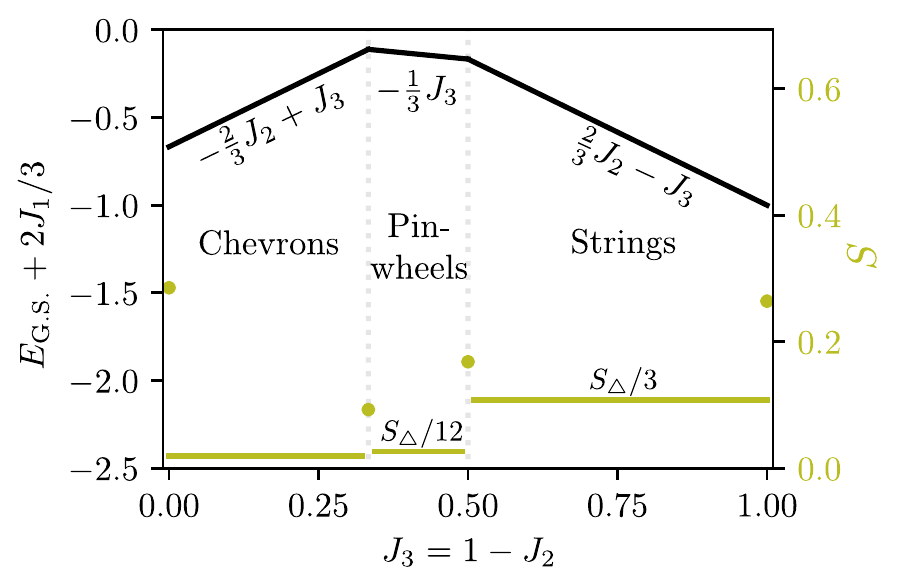}
    \caption{\label{fig:PD}Ground-state phase diagram for antiferromagnetic $J_2, J_3$. We plot the ground-state energy per site shifted by the ground-state energy of the nearest-neighbor model, as well as the residual entropy per site. The two vertical dotted lines denote the phase boundaries $J_3 = J_2/2$ and $J_3 = J_2$. At the phase boundaries, the residual entropy is larger, because in all these cases the ground-state structures of the neighboring phases can coexist (one can create domain walls which do not cost energy). In the pinwheels and in the strings phase, the numerical result for the residual entropy per site is related to the residual entropy per site of the TIAFM.}
\end{figure}

\par This yields the results summarized in Figs.~\ref{fig:PDEntropy} and~\ref{fig:PD}, and Table~\ref{tab:ResultsOverview}. Most importantly, in all the phases when $J_2$ and $J_3$ are antiferromagnetic, we find a reduced but non zero residual entropy. The chevrons phase shows a finite residual entropy that we did not detect in our (small-scale) Monte Carlo simulations. The pinwheels and strings phases have a residual entropy related to that of the triangular lattice Ising antiferromagnet (TIAFM). At the phase boundaries, several scenarios can occur, depending on whether it is possible to tessellate the plane with tiles from the two neighboring phases, and with the possibly new tiles that might belong to the ground state only at the phase boundary. At the boundary between the chevrons and the pinwheels phases and at the boundary between the stripes and the $\sqrt{3}\times\sqrt{3}$ phases, there are no such new tiles. In the former case, the tiles from the pinwheels and the chevrons phases can be matched together to cover the lattice, and the residual entropy at the boundary is larger than in those two phases. In the latter case, the tiles from the two ordered phases are not compatible, i.e. we cannot cover the lattice with a combination of tiles from \textit{both} phases, and the residual entropy is zero (in general, in such a case it would be given by the maximum of the entropies in the two neighboring phases). At all the other phase boundaries, there are new tiles which belong to the ground state only at the boundary, and these new tiles together with tiles from both neighboring phases can be used to tessellate the lattice, yielding a larger residual entropy.

\begin{table*}[]
\setlength{\extrarowheight}{12pt}
\centering
\begin{tabular}{rcccccc}
                                                   &                                                                       &                                                  & \multicolumn{4}{c}{$S$}                                                                                                                       \\ \cline{4-7} 
\multirow{-2}{*}{\makecell[r]{ Phase or\\ boundary}} & \multirow{-2}{*}{Name}                                                & \multirow{-2}{*}{$E_{\rm{G.S}}+ \frac{2}{3}J_1$} & \makecell{Pauling /\\ Lower bound}                     & Exact                                               & TN        & MPS unit cell                     \\ \hline
 $J_2 > 0, J_3 = 0$            & \multicolumn{1}{c|}{ $J_1-J_2$}                 & $- \frac{2}{3}J_2$\textsuperscript{(a)}        &  $\cong 0.31..$\textsuperscript{(b)}                   & -                            &  $0.285299 \pm 2 \times 10^{-6}$\textsuperscript{(c)} & $1\times1$  \\[8pt]
$J_2 > 2 J_3 > 0$                                    & \multicolumn{1}{c|}{Chevrons}                                         & $-\frac{2}{3}J_2 + J_{3}$                        & -                                                      & -                                                   & $0.01920 \pm 3 \times 10^{-5}$ & $1\times 2$\\[8pt]
$J_3 = J_2/2 > 0$                                    & \multicolumn{1}{c|}{-}                                                & \makecell{$-\frac{1}{6}J_2$}   & -                                                      & -                                                   & $0.0926 \pm 2 \times 10^{-4}$  &  $1\times 2$\\[8pt]
$J_2 > J_3 > J_2/2 > 0 $                             & \multicolumn{1}{c|}{Pinwheels}                                        & $ - \frac{1}{3} J_{3}$                           & -                                                      & \makecell{$S_{\rm{TIAFM}}/12$\\ $\cong 0.02692.. $} & $0.026922 \pm 3 \times 10^{-6}$ & $2\times 2$\\[8pt]
$J_3 = J_2 > 0$                                      & \multicolumn{1}{c|}{-}                                                & \makecell{$-\frac{1}{3}J_3$}     & -                                                      & -                                                   & $0.16825 \pm 5 \times 10^{-5}$ & $1\times1$  \\[8pt]
$J_3 > J_2 > 0$                                      & \multicolumn{1}{c|}{Strings}                                          & $\frac{2}{3}J_2 - J_3$                           & \makecell{$\geq \frac{1}{9}\ln{2} $\\$\cong 0.077$}    & \makecell{$S_{\rm{TIAFM}}/3$\\ $\cong 0.10769... $} & $0.107689 \pm 2 \times 10^{-6}$ & $1\times2$\\[8pt]
$J_2 = 0$                                            & \multicolumn{1}{c|}{$J_1-J_3$}                                        & $-J_3$                                           & \makecell{$\geq \frac{1}{27}\ln{559}$\\$\cong 0.2348$} & -                                                   & $0.26413 \pm 2 \times 10^{-5}$& $1\times1$ \\[8pt]
$ J_2 < 0 , J_3 > 2/3 J_2$                           & \multicolumn{1}{c|}{$\sqrt{3}\times \sqrt{3}$}  & $2J_2 - J_3$                                     & { $0$\textsuperscript{(d)}}                             & { $0$\textsuperscript{(d)}}                          & $0$      & $3\times 3$                      \\[8pt]
$J_3 = 2/3 J_2 < 0$                                  & \multicolumn{1}{c|}{-}                                                & \makecell{$\frac{4}{3}J_2$}   & -                                                      & $0$                                                 & $0$                           & $3\times 3$\\[8pt]
$J_3 < 0, J_2 > 3/2 J_3$                             & \multicolumn{1}{c|}{{ Stripes}}                  & $-\frac{2}{3}J_2 + 3 J_3$                        & { $0$\textsuperscript{(d)}}                             & { $0$\textsuperscript{(d)}}                          & $0$  & $1\times1$                         
\end{tabular}
\caption[Overview of the results]{Overview of the results. We indicate the following references for known results: (a) See Refs.~\cite{Wolf1988,Takagi1993,Wills2002, Chern2012, Hamp2018} (b) see Refs.~\cite{Wills2002,Hamp2018}, (c) see Ref.~\onlinecite{Colbois2021}, and (d) see Ref.~\onlinecite{Wolf1988} ($J_1-J_2-J_{3||}$ model) where these two ordered phases are also present. To the best of our knowledge, the rest of the results were not previously known. The last column specifies what size of unit cell was used in the VUMPS algorithm.}
\label{tab:ResultsOverview}
\end{table*}

\section{Characterizing the macroscopically degenerate phases}
\label{sec:PHASES} 
We now discuss some interesting features of the three phases for antiferromagnetic $J_2$, $J_3$. We first study the pinwheels phase, which corresponds to the DKIAFM truncated to third neighbors. Using the ground-state tiles, we demonstrate rigorously the existence of a mapping to the TIAFM ground-state manifold, allowing us to show that this phase exhibits a mixture of long-range, algebraic, and random spin-spin correlations. We then observe that the dipolar ground state does not belong to this phase and instead is a ground state in the chevrons phase. Finally, we show that the residual entropy of the strings phase, which is one third of that of the TIAFM, can be understood in terms of strings living on a honeycomb lattice, a picture qualitatively very different from that of other phases on the kagome lattice which have this residual entropy~\cite{Moessner2003, Wolf1988,Vanhecke2021}.

\subsection{Pinwheels phase}
One can easily check that the first three largest couplings in Eq.~\ref{eq:PLGS_DKIAFM} are given by
\begin{align}
\label{eq:PLGS_truncatedDKIAFM}
 J_2/J_1 =  \frac{2}{9 \sqrt{3}} \cong 0.128... \, ,& \quad J_3/J_1 = \frac{1}{12} \cong 0.083...\,
\end{align}
and that therefore, this truncated Hamiltonian corresponds to the pinwheels phase\footnote{This corresponds to $J_1 = 1.5$, $J_2 = 3^{-3/2} \cong 0.192$ and $ J_3 = 1/8 = 0.125 $, namely setting $J_0 = 0.5$ and $D = 1$ in Eq.~\ref{eq:PLGS_DKIAFM}. In Appendix A of Ref.~\onlinecite{Hamp2018}, it seems that the the couplings are $J_1 = 1.5$, $J_2' \cong 0.692$ and $J_3' = 0.625$, which corresponds to adding a $0.5$ term also for the $J_2$ and the $J_3$ couplings. Although this does not affect the ratio between $J_2$ and $J_3$, both couplings are too large compared to $J_1$ and correspond to a different ground-state phase of the $J_1$-$J_2$-$J_3$ model. However, comparing the behavior of the specific heat in the $J_1$-$J_2$ model (Fig.~9b in Ref.~\onlinecite{Hamp2018}) to known results (Fig.~8a in Ref.~\onlinecite{Takagi1993} and Fig.~26d in Ref.~\onlinecite{Colbois2021}), it seems that the couplings announced in Ref.~\onlinecite{Hamp2018} result from a typo; the results of the simulations seem more compatible with the couplings of Eq.~\ref{eq:PLGS_truncatedDKIAFM}.} (\RESP{see Fig.}~\ref{fig:PDEntropy}).

\subsubsection{Numerical results}
\par
\begin{figure}
    \centering
    \includegraphics[width = \columnwidth]{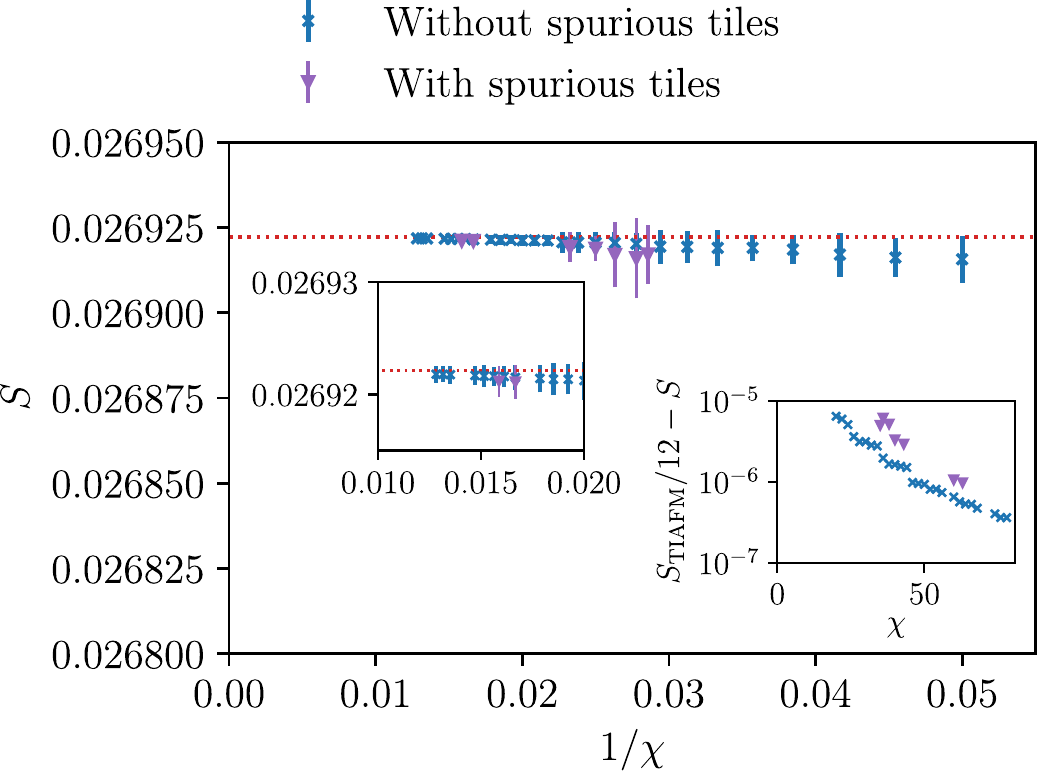}
    \caption[Residual entropy in the pinwheels phase.]{\label{fig:Pinwheels_Entropy}Residual entropy in the pinwheels phase: tensor-networks result for bond dimensions ranging from $\chi = 18$ to $\chi = 78$. As suggested in Ref.~\onlinecite{Vanhecke2021}, removing the tiles that we can identify as spurious (Type IV tiles in Fig.~\ref{fig:PinwheelsTiles}, see main text) makes it much easier to get convergence at any bond dimension; the results with these spurious tiles are however also consistent with a residual entropy of $S = \frac{1}{12}\, S_{\text{TIAFM}}$, indicated by the red dashed line. The insets show (a) a zoom on the data and (b) the difference to $S_{\rm{TIAFM}}/12$ in logarithmic scale. In the latter, the error bars are omitted for readability. }
\end{figure}

 In this phase, we have seen that the Monte Carlo results for $J_2/2 < J_3 < J_2$ suggest that the ground states could be characterized by a triangular lattice of empty hexagons surrounded by clockwise/counter-clockwise rotating dimers. These pinwheels behave like effective Ising degrees of freedom respecting a ``two-ups one-down /two-downs one-up'' rule. This intuition is further validated by the residual entropy results in this phase based on the tensor-network contraction (Fig.~\ref{fig:Pinwheels_Entropy}): 
\begin{equation}
S_{\text{Pinwheels}} = 0.026922 \pm 3 \times 10^{-6} \cong \frac{S_{\text{TIAFM}}}{12}.
\end{equation}
To obtain this result, we need to allow for two-by-two translation symmetry breaking (using a multi-site VUMPS implementation). We understand this as a consequence of the translation symmetry breaking corresponding to the triangular lattice of empty hexagons. Indeed, the snapshot in Fig.~\ref{fig:Kanamori2D_e} presents some translation symmetry breaking in terms of the location of the hexagons that do not have any dimer on them (empty hexagons). 
\begin{figure}
	\centering
	\includegraphics[width = 0.9\columnwidth]{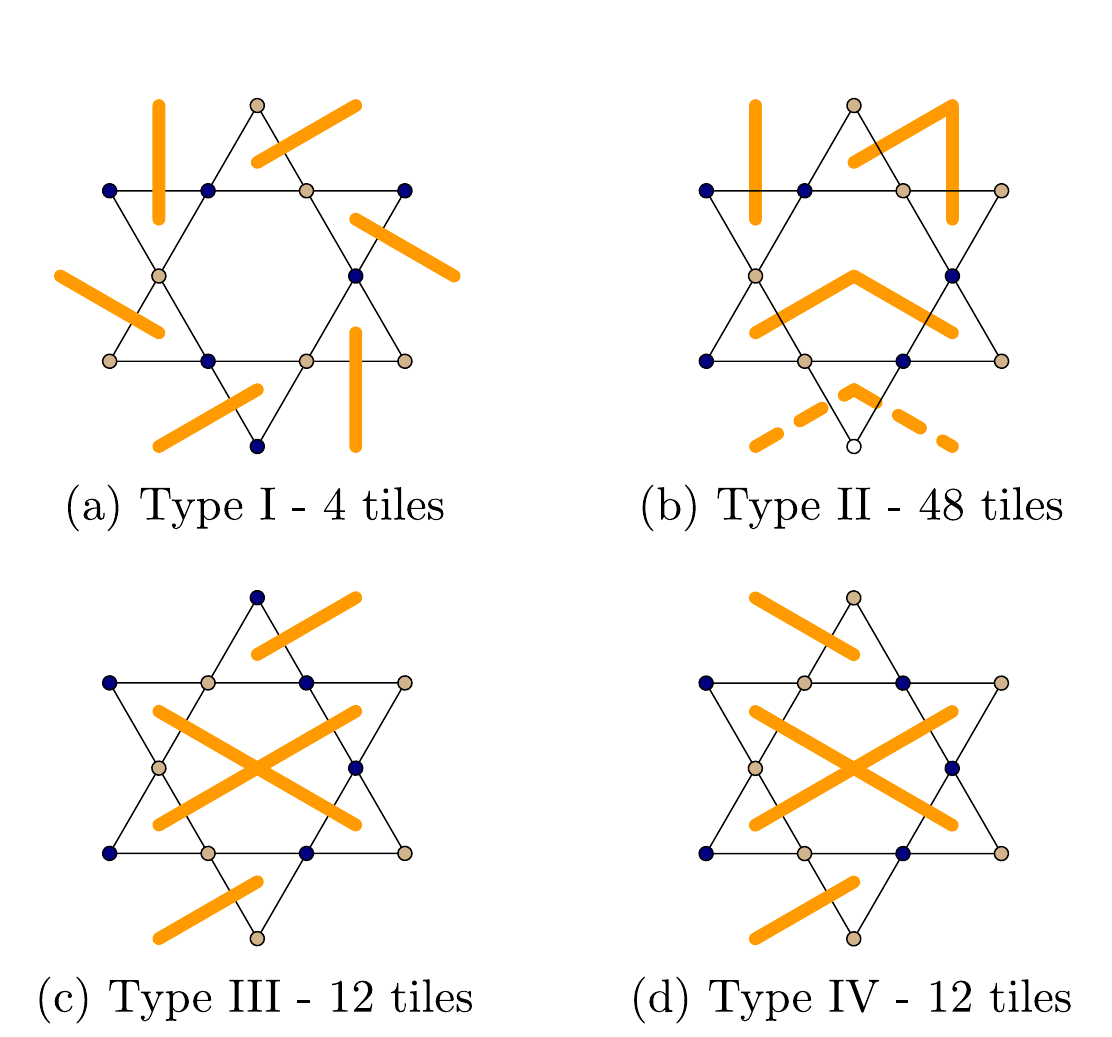}
     \caption[Ground-state tiles in the pinwheels phase.]{\label{fig:PinwheelsTiles}The 76 ground-state tiles of the pinwheels phase. See Fig.~\ref{fig:ChevronsTiles} for the convention of notation. We group the ground-state tiles in four different types. Type I tiles correspond to the pinwheels degrees of freedom. The type II and type III tiles can be matched to tessellate the plane together with the type I tiles. We show that the type IV tiles are spurious tiles. \RESP{The dashed line and white site corresponds to the dimer being either left or right, and the spin being either up or down, respectively.}}	
\end{figure}
\begin{figure*}
	\centering
    \phantomsubfloat{\label{fig:Pinwheels_ProofHelp_a}}
    \phantomsubfloat{\label{fig:Pinwheels_ProofHelp_b}}
    \phantomsubfloat{\label{fig:Pinwheels_ProofHelp_c}}
	\includegraphics[width = 0.9\textwidth]{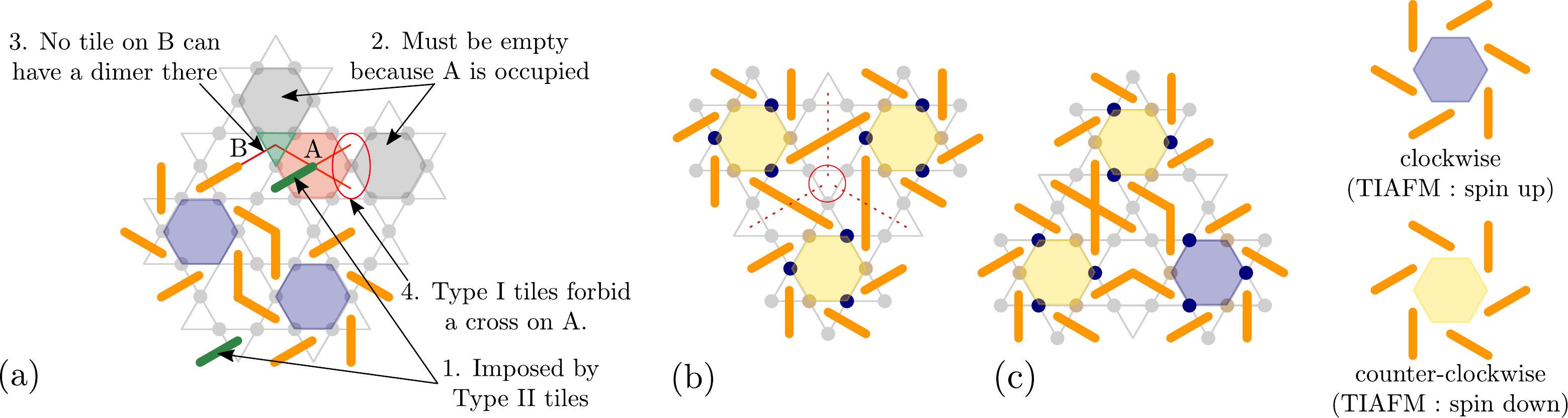}
     \caption[Help for the pinwheels phase proof]{\label{fig:Pinwheels_ProofHelp}Two illustrations to help understand the proof. (a) Graphical proof that two empty hexagons cannot be next-nearest neighbors in the pinwheels phase: we start with the configuration of the orange dimers (two second neighbor empty hexagons must have the same chirality). From the Type II tiles, the green dimers must be added. The hexagon A is thus occupied and has already three occupied nearest-neighbor hexagons. It has to have two nearest neighbors, which are empty, and these empty hexagons cannot be nearest neighbors, which fixes their positions (highlighted in gray). The green triangle must be occupied by one dimer. However, no tile that can be put in A has a dimer occupying the green triangle, because the empty hexagon forbids putting a cross on A; similarly, no tile that can be put in B has a dimer occupying this triangle. Thus the plane cannot be tessellated by tiles form the pinwheels phase starting from two second-neighbor empty hexagons. (b) Three empty hexagons which are third neighbors cannot all have the same chirality, because the corresponding tiles are not compatible. (c) In contrast, it is possible to tile three empty hexagons if one of them has a chirality opposed to the two others.}	
\end{figure*}
\subsubsection{Link with the TIAFM ground states}
\par This numerical result and the intuitive picture of a link with the TIAFM ground states based on the Monte Carlo snapshot can actually be confirmed rigorously by relying on the ground-state tiles in that phase, illustrated in Fig.~\ref{fig:PinwheelsTiles}. 
In this Figure, there are four tiles corresponding to our tentative Ising degrees of freedom; the other tiles either bear a cross or a chevron, confirming the observations from the Monte Carlo snapshots. All these tiles have in common the constraint that there must be exactly three spins up and three spins down around a hexagon. The empty hexagons (Fig.~\ref{fig:PinwheelsTiles}a) correspond to ferromagnetic second-neighbor triangles, while the chevrons (Fig.~\ref{fig:PinwheelsTiles}b) and the crosses (Fig.~\ref{fig:PinwheelsTiles}c and d) correspond to second-neighbor triangles that respect a ``two-ups one-down /two-downs one-up'' rule.

\par With this, we can prove that 
\begin{enumerate}
    \item Any ground state in the pinwheels phase corresponds to a triangular lattice of pinwheels whose chirality respects a ``two-rights one-left/two-lefts one-right'' rule on each triangle. There are eight pinwheels phase ground states mapping to the same TIAFM ground state.
    \item \emph{Any} configuration of the TIAFM ground state maps to eight ground states of the pinwheels phase.
\end{enumerate}

\par To prove the first statement, we first show that in any ground state of the pinwheels phase, the ratio of hexagons of each type is exactly
\begin{equation}
    \label{eq:ratios}
    r_{\text{pinwheel}} = \frac{1}{4},\; r_{\text{cross}} = \frac{1}{4},\; r_{\text{chevron}} = \frac{1}{2}.
\end{equation}
Indeed, the ground-state energy of the pinwheels phase [Eq.~(\ref{eq:EPinwheels})] shows that $3/4$ of the second-neighbor triangles are in their ground state and $1/4$ are ferromagnetic. With the ground-state tiles, this implies that $1/4$ of all tiles in a ground-state must be of type I, while $3/4$ must be of another type (cross or chevron). One can check with Fig.~\ref{fig:PinwheelsTiles} that the cross tiles are only compatible with chevron tiles in the cross's acute angle directions (i.e. on the left and right of type III and type IV tiles with the cross put horizontally, we can only put type II tiles, and not type I, III or IV). Furthermore, once a chevron has been matched with a cross tile, one cannot match another cross tile with this chevron. Therefore, for each cross there must be at least two chevrons. At the same time, the tiles show that one can only put a chevron tile on the lattice if there is also at least one cross as a nearest-neighbor tile. Therefore, the ratios of crosses and chevrons are fixed  to $1/4$ and $1/2$, respectively.

\par As a second step, we show that each ``occupied hexagon'' tile (type II, III, IV) must overlap with exactly two pinwheels tiles. Indeed, the fact that two pinwheel tiles cannot overlap (i.e. empty hexagons cannot be nearest neighbors) limits the number of pinwheels tiles overlapping with each occupied hexagon tile to at most two. Since a pinwheel tile can only overlap with occupied hexagon tiles, the overall ratio $r_{\text{pinwheels}} = 1/4$ can only be obtained if each occupied tile is overlapping with exactly two pinwheel tiles.

\par We still have to show that two empty hexagons cannot be second-neighbors. The impossibility of tiling the plane when starting from two pinwheel stars sharing only one site is not immediate, and we illustrate it in Fig.~\ref{fig:Pinwheels_ProofHelp}. First, one can check that the two pinwheels must have the same chirality, implying that they are separated by two chevrons. The type II tiles then enforce the existence of an occupied tile (highlighted in red, labeled $A$) which is already overlapping with three occupied tiles on one side. This fixes the position of the two pinwheel tiles that tile $A$ has to overlap with. With these constraints, it is not possible to find tiles fitting on hexagons $A$ and $B$ in Fig.~\ref{fig:Pinwheels_ProofHelp_a}. 

\par Thus, we have proven that all the pinwheels must be third neighbors; with the ratios in Eq.~\ref{eq:ratios}, the only possibility is that they \emph{have to} form a triangular lattice. 
\par At the same time, the tiling rules imply that three pinwheel stars arranged in a triangle cannot have the same chirality. This is shown in  Fig.~\ref{fig:Pinwheels_ProofHelp_b}: the only way to have three empty hexagons on a triangle with the same chirality would be to have three straight dimers; in the pinwheels phase, these have to belong to crosses, and there would be a triangle with three dimers meeting, which is forbidden. In contrast, Fig.~\ref{fig:Pinwheels_ProofHelp_c} shows that it is possible to have three empty hexagons forming a third-neighbor triangle if one of them has chirality opposing the two others. In the end, this yields a ``two-rights one-left/two-lefts one-right'' rule for these effective Ising degrees of freedom.

\par Once the triangular sublattice supporting the pinwheels is selected (four-fold translation symmetry breaking), the spin configuration is fixed by the pinwheels chirality, up to a global spin flip (two-fold $\mathbb{Z}_2$ symmetry breaking). This shows that all ground states in the pinwheels phase map to TIAFM ground states, and that eight pinwheels states map to the same TIAFM ground state. 

\begin{figure}
	\centering
	\includegraphics[width = 0.9\columnwidth]{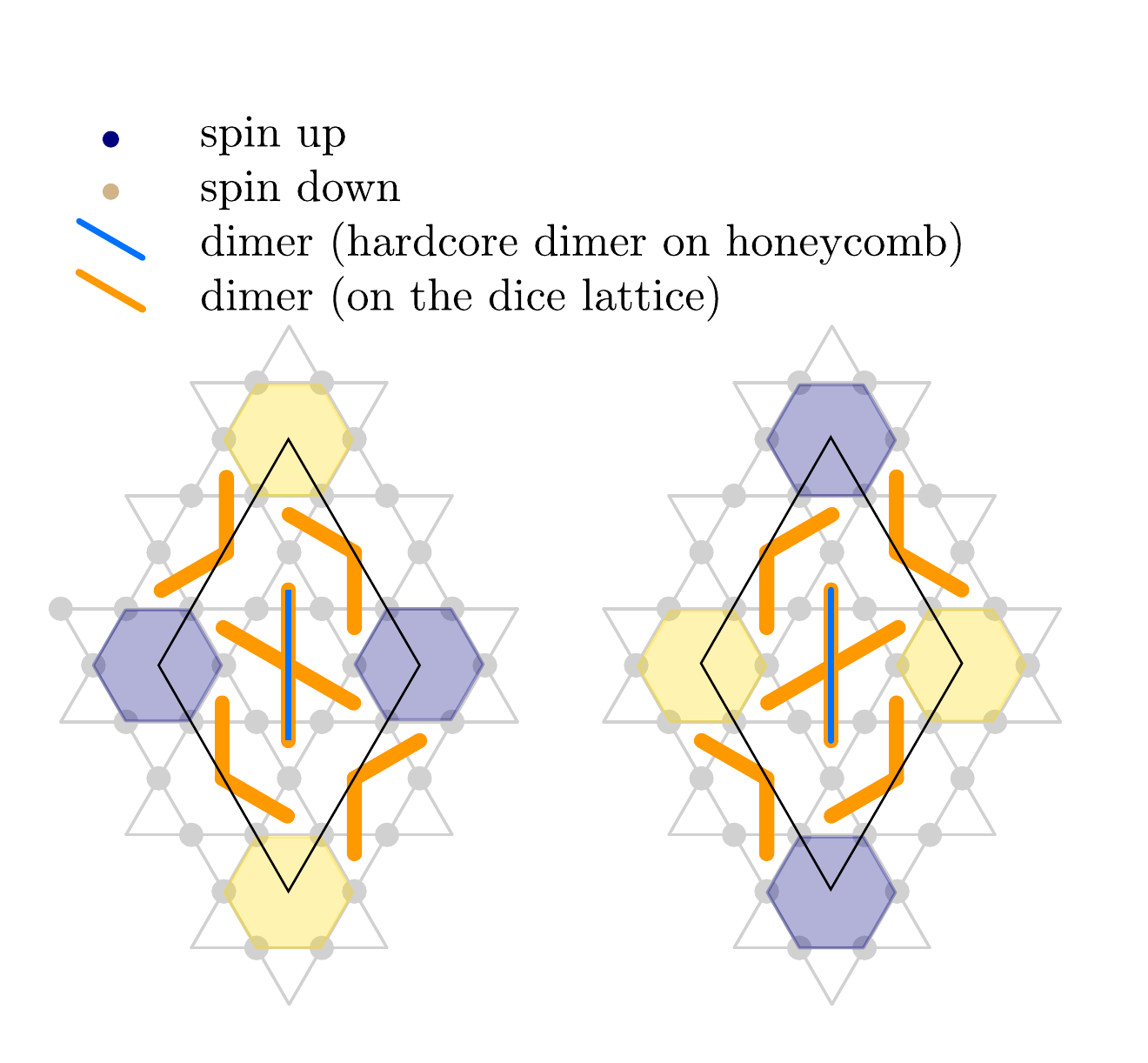}
     \caption[Rhombi and pinwheels]{\label{fig:PinwheelsRhombi}The two possible effective Ising d.o.f (TIAFM) configurations on a rhombus, associated honeycomb lattice dimer, and corresponding dimer configurations as set by the pinwheels. One can obtain all ground states of the original TIAFM model by using rotations of the rhombi shown in black. The fact that the associated dimer configurations on the dice lattice do not create additional constraints for tiling with the rhombi implies that all the TIAFM ground states have a related pinwheels ground state. }	
\end{figure}

\par We have yet to show that any TIAFM configuration is realized by the pinwheels (i.e. that there are no additional constraints on the pinwheels arrangement besides the ``two-rights one-left/two-lefts one-right'' rule). To prove this second statement, we rely on the two-to-one mapping from the TIAFM ground states to the hardcore dimer configurations on the honeycomb lattice (see e.g. Refs.~\cite{Kasteleyn1963, Fisher1966}), and the direct mapping between hardcore dimer configurations and tilings of the triangular lattice with rhombi (introduced in the context of solid-on-solid models~\cite{Blote1982,Nienhuis1984}). Indeed, to each pair of triangles sharing a bond with aligned spins in the original TIAFM ground state, one can associate a dimer on the (dual) honeycomb lattice, as well as a rhombus tile made of these two triangles. Thus, there is a two-to-one mapping between TIAFM configurations and rhombus tilings. 

\par Therefore, it is sufficient to show that (a) the two types of rhombi (up or down effective Ising on the shared bond) can be realized and (b) they can overlap without additional constraints. Fig.~\ref{fig:PinwheelsRhombi} directly shows that (a) is satisfied. The fact that the rhombi in that Figure can be matched together without additional constraints can easily be checked based on the dimer configurations. 

\par Therefore, any TIAFM configuration has an associated pinwheels state. The dimer configuration in the pinwheels phase is fixed by the associated TIAFM configuration up to translations of the pinwheels center location, and as a result the spin configuration is fixed up to an eight-fold degeneracy.

\subsubsection{Spin-spin correlations}

\begin{figure}
	\centering
    \phantomsubfloat{\label{fig:pinwheelscorr_a}}
    \phantomsubfloat{\label{fig:pinwheelscorr_b}}
    \phantomsubfloat{\label{fig:pinwheelscorr_c}}
    \phantomsubfloat{\label{fig:pinwheelscorr_d}}
	\includegraphics[width = \columnwidth]{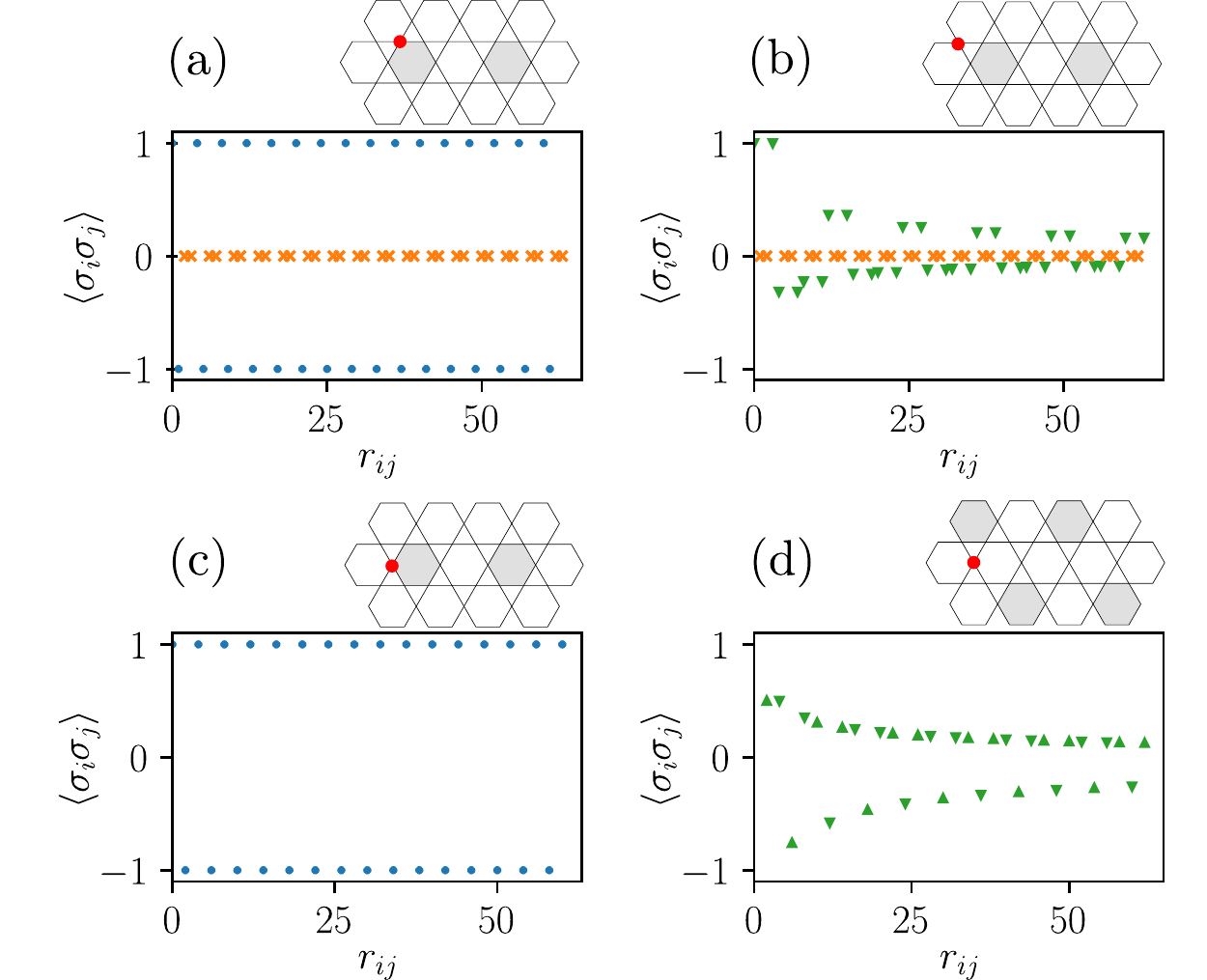}
	\caption[Spin-spin correlations in the pinwheels phase.]{\label{fig:pinwheelscorr} Horizontal spin-spin correlations in the pinwheels phase in a given translation symmetry-broken sector, as analytically predicted from the properties of the ground-state tiles and the result of Stephenson~\cite{Stephenson1970}. The correlations depend on the specific site chosen as the origin and we show the four possible set of correlations, with the red dot in the insets signaling the site selected as the origin, and the location of the pinwheels centers highlighted in gray. We show with blue dots the correlations between sites in the center of the pinwheels, with green triangles the correlations between sites in the branches of the pinwheels, and with orange crosses the mixed correlations. }
\end{figure} 

An immediate consequence of the mapping between the pinwheels states and the TIAFM ground states is that we can make an analytical prediction for the spin-spin correlations. We discuss the calculation in detail in Appendix~\ref{sec:PINCORR}, and illustrate the results in Fig.~\ref{fig:pinwheelscorr}. The essential argument relies on describing the spin configuration in translation-symmetry and $\mathbb{Z}_2$-symmetry broken sectors using the stars corresponding to the pinwheels. With this, one finds that
\begin{enumerate}
    \item The long-range order associated with the location of the pinwheels centers translates into a long-range order for the correlations between spins located on the interior hexagons of the pinwheels (blue dots in Figs.~\ref{fig:pinwheelscorr_a} and~\ref{fig:pinwheelscorr_c});
    \item There is an algebraic decay of the correlations between the pinwheels orientations, corresponding to the algebraic decay of the spin-spin correlations in the TIAFM ground-state manifold~\cite{Stephenson1970}
\begin{equation}
	\langle s_i s_j \rangle = \epsilon_0 \frac{\cos\left(\vec{q}\cdot\vec{r}\right)}{\sqrt{|\vec{r}|}} \, , \quad \vec{r} = \vec{r}_j - \vec{r}_i,
\end{equation}
where we denoted by $s_i$ the Ising spins, and where the structure factor is characterized by $\vec{q} = (\pm \frac{2 \pi}{3} , \frac{2\pi}{\sqrt{3}})$.
    \item This algebraic decay results in an algebraic decay for the correlations between spins on the branches of pinwheels with a decay exponent $\eta = 1/2$ (green triangles in Figs.~\ref{fig:pinwheelscorr_b} and~\ref{fig:pinwheelscorr_d});
    \item The correlations between a spin on the central hexagon of a pinwheel and one on the branches of a pinwheel are strictly zero on average (orange crosses in Figs.~\ref{fig:pinwheelscorr_a} and~\ref{fig:pinwheelscorr_b}). 
\end{enumerate}

\subsection{Chevrons phase}
\par  Either from the exact mapping, or more immediately from the tiling rules, it is quite clear that the ground state of the dipolar model (Fig.~\ref{fig:DipolarGS}) is not a ground state in the pinwheels phase. Instead, the Monte Carlo samples suggest that the dipolar ground state might be one of the ground states of the chevrons phase.
\par This is a somewhat surprising result since the couplings of the DKIAFM truncated to third neighbor couplings correspond to the pinwheels phase. 

\par
\begin{figure}
	\centering
	\includegraphics[width = 0.47\columnwidth]{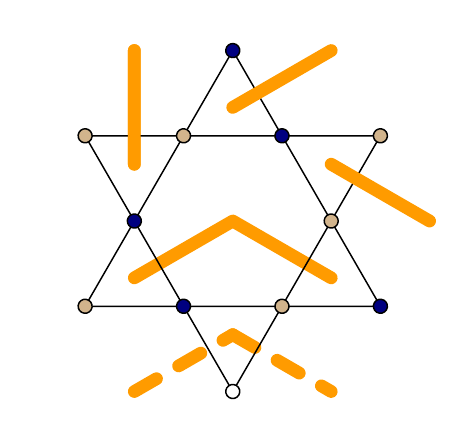}
     \caption[Ground-state tiles in the chevrons phase.]{\label{fig:ChevronsTiles}The 48 ground-state tiles of the chevrons phase. The blue and yellow dots correspond to up and down spins, respectively. The dashed lines and white site describe the fact that one can set the spin to be either up or down on the white site, and correspondingly put a dimer on one of the two dashed bonds at the exclusion of the other. Note that these tiles restrict how neighboring chevrons can be arranged. }	
\end{figure}

 The ground-state tiles confirm the intuitive insight from the Monte Carlo simulations that the phase at $J_3 < J_2/2$ is characterized by dimers at a $120^{\circ}$-angle (chevrons). Furthermore, they indicate constraints on the way the chevrons can be arranged: there are dimer configurations only made of chevrons which do not respect the constraints given by the tiles in Fig.~\ref{fig:ChevronsTiles}.

\begin{figure}
	\centering
	\includegraphics[width = 0.9\columnwidth]{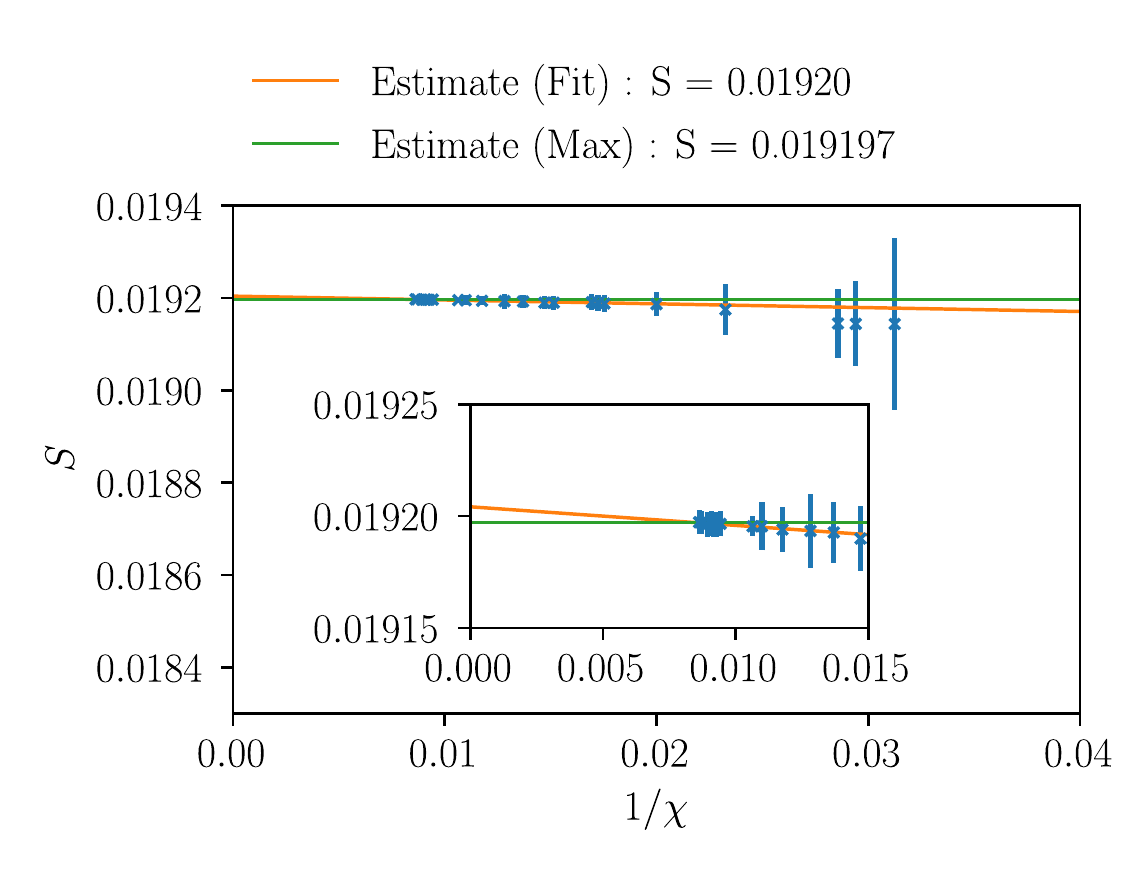}
     \caption[Residual entropy in the chevrons phase.]{\label{fig:ResidualEntropyChevrons}Extracting the residual entropy in the chevrons phase. We use a multi-site VUMPS algorithm with 1-by-2 translation symmetry breaking Ansatz. }	
\end{figure}
\par It is easy to check that the ground state of the DKIAFM corresponds to a tiling using the chevrons phase tiles, and therefore we can prove that it is one of the ground states of this phase. Yet, the ground-state degeneracy is not completely lifted. Comparing this phase with the ground-state phase of the $J_1-J_2-J_{3||}$ model (where $J_{3\star} = 0$) underlines again the determinant role of the $J_{3\star}$ interaction in this ground-state phase diagram: as compared to the latter ($S \cong 0.1439...$ \cite{Colbois2021}), the residual entropy in the chevrons phase (Fig.~\ref{fig:ResidualEntropyChevrons}, Table~\ref{tab:ResultsOverview}) is reduced by almost a factor 8 :
\begin{equation}
    S_{\text{Chevrons}} = 0.01920 \pm 3 \times 10^{-5}.
\end{equation}

\subsection{Strings phase}
\begin{figure}
	\centering
	\includegraphics[width = 0.9\columnwidth]{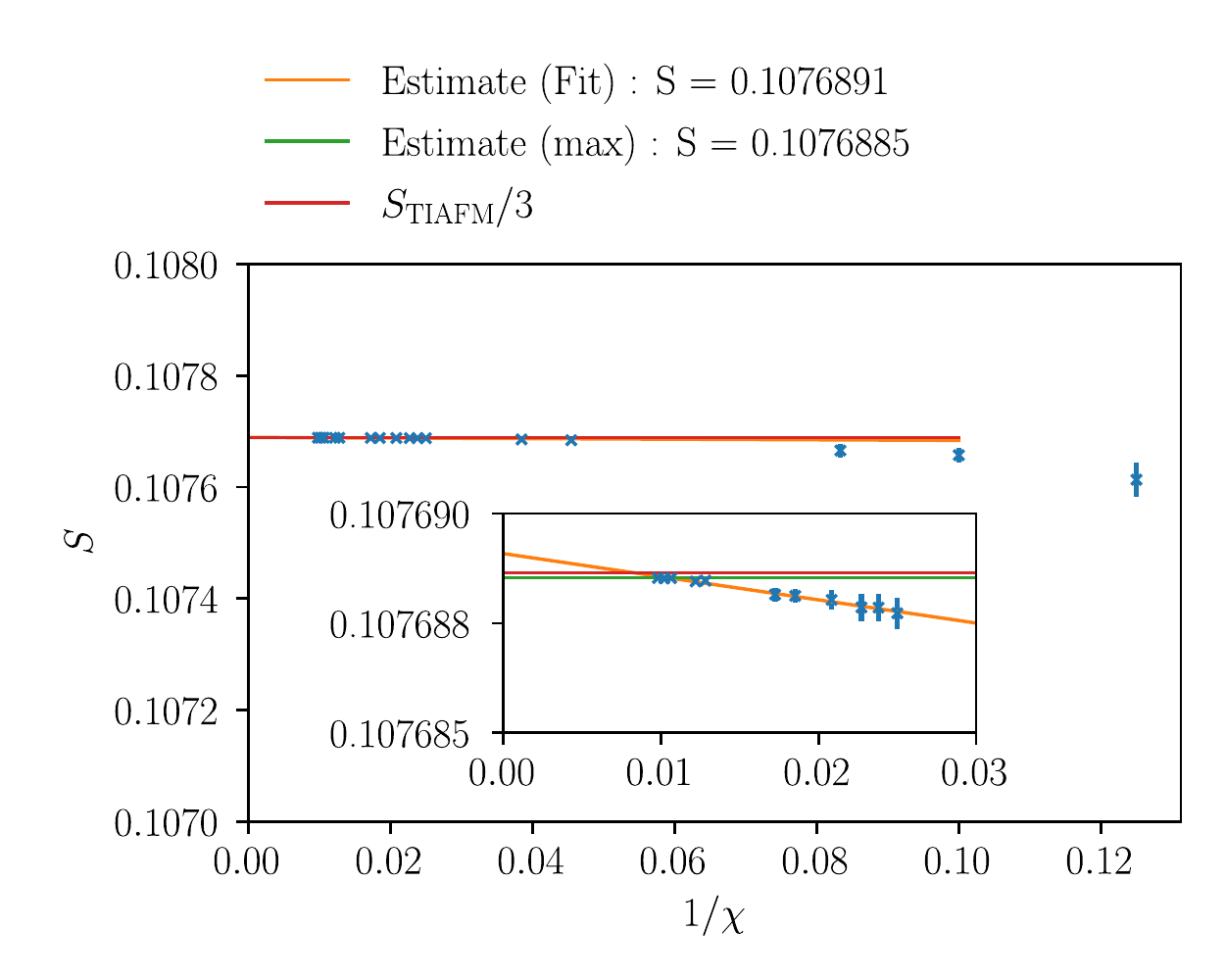}
     \caption[Residual entropy in the strings phase.]{\label{fig:ResidualEntropyStrings}Extracting the residual entropy in the strings phase. As we argue in the main text, the residual entropy corresponds to a third of that of the triangular Ising antiferromagnet, up to subextensive corrections.}	
\end{figure}

In the strings phase, we find a residual entropy corresponding to a third of the TIAFM residual entropy (Fig.~\ref{fig:ResidualEntropyStrings}). This is a result that seems to be often found in kagome Ising models. The typical example is the kagome ice phase, which occurs both in the $J_1-J_2-J_{3||}$ model when $J_2 = J_{3||} < 0 $ and in the nearest-neighbor model in a longitudinal field, and where all the nearest-neighbor triangles respect a 2-up 1-down rule which allows a mapping from the spin configurations to hardcore dimers on the honeycomb lattice~\cite{Moessner2000, Moessner2003,Wolf1988,Mizoguchi2017}. A different example is the phase at large antiferromagnetic $J_2,\, J_{3}$ and small ferromagnetic $J_1$, where the ground-state manifold is mainly understood in terms of effective Ising degrees of freedom formed by three spins on up or down nearest-neighbor triangles~\cite{Vanhecke2021}.\\

\begin{figure}
	\centering
    \phantomsubfloat{\label{fig:StringsTiles_a}}
    \phantomsubfloat{\label{fig:StringsTiles_b}}
    \phantomsubfloat{\label{fig:StringsTiles_c}}
    \phantomsubfloat{\label{fig:StringsTiles_d}}
    \phantomsubfloat{\label{fig:StringsTiles_e}}
    \phantomsubfloat{\label{fig:StringsTiles_f}}
	\includegraphics[width = 0.9\columnwidth]{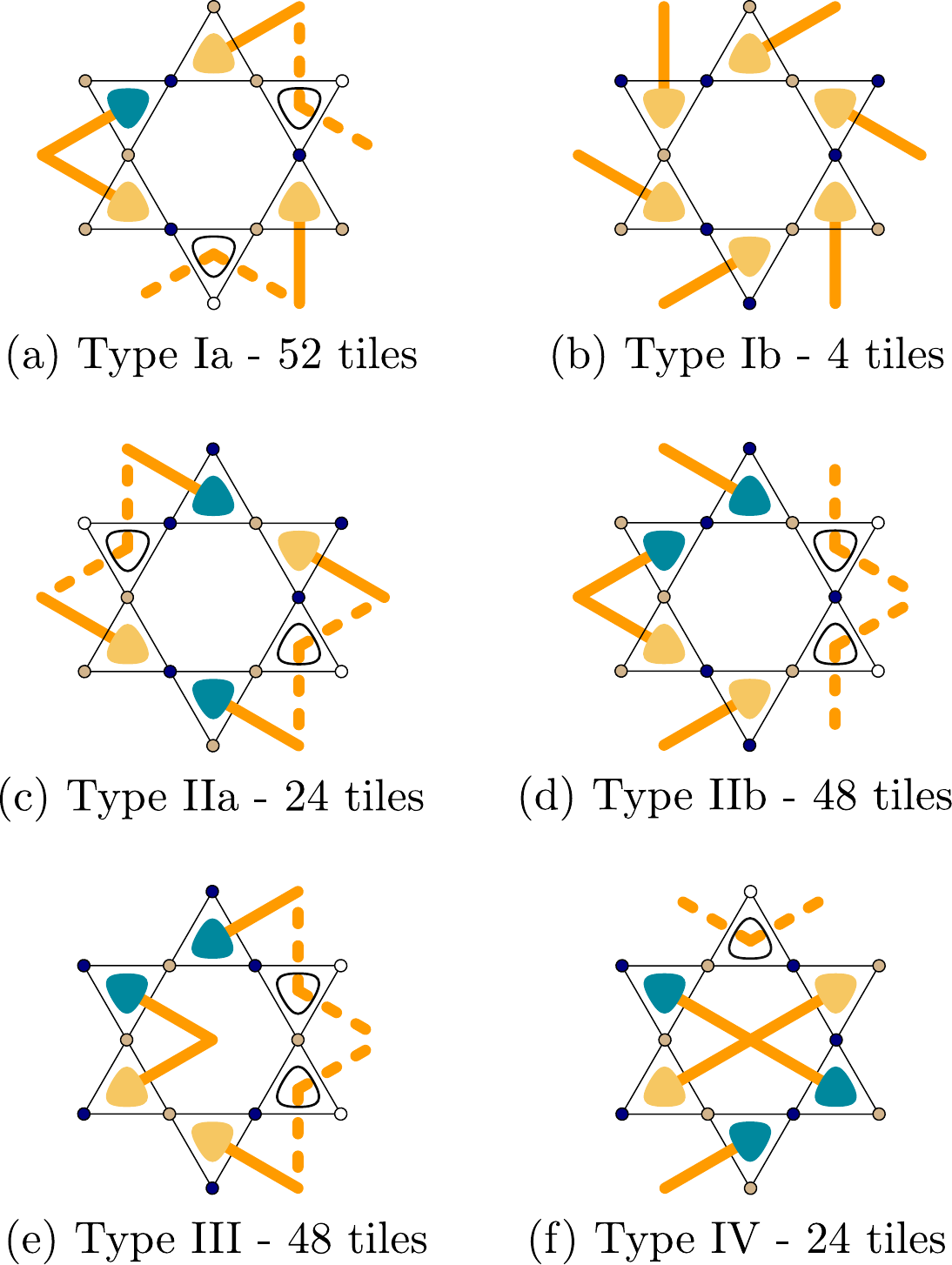}
     \caption[Ground-state tiles in the strings phase.]{\label{fig:StringsTiles}The 200 ground-state tiles of the strings phase, in the dimer picture and in the charge picture. The type I tiles (a and b) are characterized by empty hexagons and dimer configurations such that the charges on either all three up or all three down triangles (or both) all have the same sign, while in the type II tiles the charges respect a ``two-ups one-down, two-downs one-up'' rule. The type III tiles correspond to arrows while the type IV tiles correspond to crosses. We argue that the residual entropy is due to the strings formed by the type IV tiles. \RESP{See Fig.}~\ref{fig:DipolarGS}\RESP{ for the legend for the charges. Empty sites/charges and dashed lines correspond to the spin being either up or down, the charge being positive or negative, and to the dimer being placed accordingly either left or right.}}	
\end{figure}

\par In the strings phase, the simplest approach to understand the residual entropy of the phase relies on the fact that the TIAFM ground-state manifold is in two-to-one correspondence with \emph{directed} strings on the honeycomb lattice\footnote{We considered other relations to the TIAFM. An approach based on explaining the residual entropy using the charge degrees-of-freedom as effective Ising degrees-of-freedom proves challenging due to the type Ia and type Ib tiles. By building the dimer mapping starting from the strings mapping, one can see that the direct mapping between the ground states of the strings phase and dimers on honeycomb is a highly non intuitive approach.} (i.e., strings that are allowed to wiggle but do not cross, fuse or turn back)~\cite{Yokoi1986,Jiang2006, Smerald2017}.
\par A detailed discussion of the relation between the strings ground-state manifold and configurations of directed strings on the honeycomb lattice is given in Appendix~\ref{sec:AppStrings}. The key points are that
\begin{enumerate}
    \item The tiling rules (Fig.~\ref{fig:StringsTiles}) imply that tiles bearing crosses (Fig.~\ref{fig:StringsTiles_f}) have to form strings.
    \item One can take as reference configuration a state made of arrows pointing all in the same direction (Fig.~\ref{fig:StringsTiles_e}). 
    \item Periodic boundary conditions and the ground-state energy Eq.~\ref{eq:StrGSE} imply that the number of empty hexagons and the number of crosses has to be the same.
    \item To each string of arrows, one can associate a pair of directed, independent strings on the honeycomb lattice.
    \item Contracting the tensor network without the tiles bearing crosses (Fig.~\ref{fig:StringsTiles_f}) yields a zero residual entropy. 
    \item Several ground states of the strings phase are related to the same strings configuration on honeycomb.
\end{enumerate}
\par With this, we understand that the residual entropy of the strings phase is essentially explained by the crosses forming strings that are related to strings configurations on the honeycomb lattice. However, there must be corrections corresponding to the fact that the relation between the strings phase ground states and strings on honeycomb is not a strict mapping. For some particular configurations, we can show that these corrections are at least subextensive. Yet the contraction of the tensor network without the cross tiles yields a zero residual entropy, strongly suggesting that the corrections are indeed \emph{only} subextensive, and do not contribute in the thermodynamic limit.

\begin{figure*}
	\centering
	\phantomsubfloat{\label{fig:Competition_a}}
	\phantomsubfloat{\label{fig:Competition_b}}
	\phantomsubfloat{\label{fig:Competition_c}}
	\phantomsubfloat{\label{fig:Competition_d}}
	\phantomsubfloat{\label{fig:Competition_e}}
	\includegraphics[width = 0.8\textwidth]{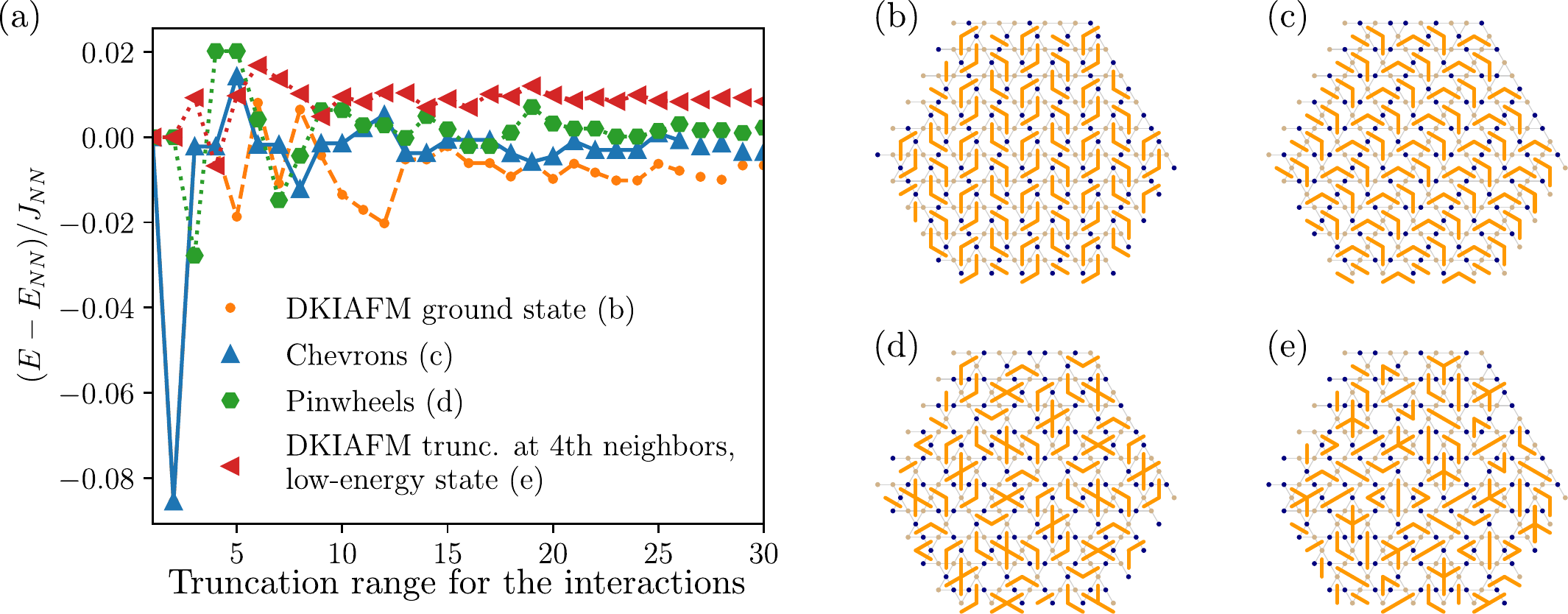}
     \caption[Competition between low energy states.]{\label{fig:Competition}(a) Competition in energy between four different states as a function of the range at which the dipolar Hamiltonian Eq.~\ref{eq:PLGS_DKIAFM} is truncated. The compared states are (b) the DKIAFM ground state, (c) another ground state in the chevrons phase, (d) a ground state of the pinwheels phase, and (e) a state obtained at low temperatures for Monte Carlo configurations of the DKIAFM truncated at fourth-neighbor couplings ($\frac{J_4}{J_1} = \frac{2}{3 \cdot 7^{3/2}}\cong 0.03599...$ ).}	
\end{figure*}

\section{Discussion and outlook}
\label{sec:DISC}

\par In this paper, we have used a method based on ground-state local rules to construct contractible tensor networks, enabling a detailed study of the macroscopic ground-state degeneracy in several  phases of the $J_1-J_2-J_3$ Ising antiferromagnet on the kagome lattice. Together with Kanamori's method of inequalities and small-scale Monte Carlo simulations, this method provides exact results for the ground-state energies of the model when $J_1 > |J_2|, |J_3|$ and very precise numerical results for the residual entropy of the model, as summarized in Figs.~\ref{fig:Kanamori2D_a},~\ref{fig:PD} and Table~\ref{tab:ResultsOverview}.

\par In studying the results, we observed that the DKIAFM ground state (Fig.~\ref{fig:DipolarGS}) is a ground state in the $J_1-J_2$ model, but does not belong to the pinwheels phase, which corresponds to the ground state of the DKIAFM Hamiltonian truncated to third-neighbor interactions. This seems also to be a consequence of the strong competition between the various frustrated couplings. It is also reminiscent of what happens in the case of the dipolar model on the triangular lattice, where the model truncated to second-neighbor couplings has the long-range stripe order corresponding to the dipolar ground state, but where the dipolar model truncated to third neighbors has a zig-zag ground state~\cite{Smerald2016}; however, in our case, both truncated models still have a macroscopic ground-state degeneracy.
\par To gain some insight into this competition, we have computed the energy of a few states of interest as a function of the range of farther-neighbor interactions included in the truncated DKIAFM Hamiltonian. In Fig.~\ref{fig:Competition_a}, we compare the truncated energies for the following states:
\begin{enumerate}
    \item the ground state of the DKIAFM (Fig.~\ref{fig:Competition_b}),
    \item another ground state in the chevrons phase (Fig.~\ref{fig:Competition_c}),
    \item a ground state of the pinwheels phase, i.e. for the couplings corresponding to Eq.~\ref{eq:PLGS_truncatedDKIAFM} (Fig.~\ref{fig:Competition_d}) 
    \item a state obtained at low temperature for the couplings in  Eq.~\ref{eq:PLGS_truncatedDKIAFM} and $\frac{J_4}{J_1} = \frac{2}{3 \cdot 7^{3/2}}$ (Fig.~\ref{fig:Competition_e}); we did not prove that it is the ground-state for the model up to fourth-neighbors, but the important point is that it has a lower energy than the other states we consider here.
\end{enumerate} 
\par Using this approach it seems that a range of at least nine neighbors is required to reach the point where the DKIAFM ground state systematically has the lowest energy. For even further ranges it is also occasionally still degenerate with the other selected ground state of the chevron phase. This strong competition suggests that a large range of neighbors would have to be taken into account to fully understand how the long-range interactions progressively lift the degeneracy and eventually select the long-range ordered 12-site DKIAFM ground state. It also suggests that a lot more states than those described by the approach in Appendix A of Ref.~\onlinecite{Hamp2018} lie in a narrow region of energy above the dipolar ground state; and it is somewhat reminiscent of the high number of low-lying excited states found in a related model~\cite{Kao2020}.

\par More broadly, our results reveal a surprising feature of the kagome lattice Ising antiferromagnet, that ought to be contrasted with other two-dimensional frustrated models: in most phases, a small but finite residual entropy survives even when $J_3 \neq 0$, and not only at fine-tuned points but in extended regions of the phase diagram. In this model, a macroscopic ground-state degeneracy seems almost to be the norm rather than the exception.

\par Such small residual entropies would be extremely challenging to evaluate with Monte Carlo simulations. In contrast, thanks to the tensor network approach, the results can be obtained down to the fifth decimal without an excessive investment of computational resources. Instead of requiring a detailed prior understanding of the model, which is typically needed to design good, ad-hoc Monte Carlo updates, the systematic approach to obtain the ground-state tiles and to evaluate the residual entropy from the tensor network construction are a crucial first step to understand important properties of the model. The description based on a ground-state local rule in the form of tiles provides a good framework to obtain a deeper understanding of the result. Sometimes, like in the pinwheels phase, we can even obtain rigorously a relation between the ground-state phase of the $J_1-J_2-J_3$ model and the triangular Ising antiferromagnet. We also find that the results of the strings phase are particularly interesting: indeed, if one would obtain such strings in an experimental setting, the first intuition might be that one is looking at excitations and that the system has not reached the ground state. Yet, here, the strings of crosses are a fundamental feature of the ground-state phase. One should therefore be careful and not immediately interpret the presence of such apparent domain walls as a failure of the system to reach its ground-state manifold.

\par Finally, let us discuss the implications of the present results for experiments. Artificial spin systems allow one to directly visualize the spin configurations in real space, and our predictions can be used to see which phase a given spin configuration might belong to. However, to check that one is dealing with a degenerate ground-state manifold, one needs to systematically compute the spin-spin correlations averaged over all ground states, and to compare it to experimental averages. Evaluating general spin-spin correlation functions can in principle be done with tensor networks (see e.g. \cite{Vanderstraeten2015, Corboz2016, Vanderstraeten2022}): the complexity is similar to computing dynamical correlation functions for one-dimensional quantum systems in imaginary time\footnote{Note that the related idea of obtaining snapshots directly from the tensor network approach has been explored in CTMRG, see Ref.~\onlinecite{Ueda2005}.}. The construction based on tiles, however, requires a specific implementation, \RESP{in particular in the case of VUMPS if the 1D transfer matrix is not Hermitian. In that case, as suggested in Ref.}~\onlinecite{Fishman2018} \RESP{the top and bottom uniform MPSs can be computed independently, just as in the iTEBD approach, but one has to be particularly careful in the cases where some symmetries are broken.} This work is left for future investigation. 

\par In addition, it can be experimentally difficult to reach low effective temperatures in artificial spin systems, although some significant recent progress has been made in the case of in-plane artificial spin systems (see e.g. Refs.~\cite{Rougemaille2019,Shcanilec2020,Hofhuis2020, Hofhuis2022}). In the case of the DKIAFM, Monte Carlo simulations have shown that the first-order transition is accompanied by a slowing down of the spin dynamics~\cite{Hamp2018} that realistically should make the long-range-ordered ground state unreachable in artificial spin-ice experiments (see for instance the effective temperatures reached in Refs.~\cite{Zhang2012,Chioar2014}). It would thus be interesting to extend the tensor-network results to finite temperature, and to compare the spin structure factor from the truncated Hamiltonian at these finite effective temperatures to the results for the full dipolar Hamiltonian~\cite{Chioar2016}. Work is in progress along these lines.

{\it Acknowledgements.} We thank Benjamin Canals, Patrick Emonts, and Mithilesh Nayak for useful discussions, and the Centro de Ciencias de Benasque Pedro Pascual for hosting conferences where significant progress was made on this paper. JC and FM are supported by the Swiss National Science Foundation (Project Number: 200020\_182179). LV is supported by the Flemish Research Foundation (FWO) via grant FWO20/PDS/115. The computations have been performed using the facilities of the Scientific IT and Application Support Center of EPFL (SCITAS).

\bibliography{main}

%apsrev4-2.bst 2019-01-14 (MD) hand-edited version of apsrev4-1.bst
%Control: key (0)
%Control: author (8) initials jnrlst
%Control: editor formatted (1) identically to author
%Control: production of article title (0) allowed
%Control: page (0) single
%Control: year (1) truncated
%Control: production of eprint (0) enabled
\begin{thebibliography}{98}%
\makeatletter
\providecommand \@ifxundefined [1]{%
 \@ifx{#1\undefined}
}%
\providecommand \@ifnum [1]{%
 \ifnum #1\expandafter \@firstoftwo
 \else \expandafter \@secondoftwo
 \fi
}%
\providecommand \@ifx [1]{%
 \ifx #1\expandafter \@firstoftwo
 \else \expandafter \@secondoftwo
 \fi
}%
\providecommand \natexlab [1]{#1}%
\providecommand \enquote  [1]{``#1''}%
\providecommand \bibnamefont  [1]{#1}%
\providecommand \bibfnamefont [1]{#1}%
\providecommand \citenamefont [1]{#1}%
\providecommand \href@noop [0]{\@secondoftwo}%
\providecommand \href [0]{\begingroup \@sanitize@url \@href}%
\providecommand \@href[1]{\@@startlink{#1}\@@href}%
\providecommand \@@href[1]{\endgroup#1\@@endlink}%
\providecommand \@sanitize@url [0]{\catcode `\\12\catcode `\$12\catcode
  `\&12\catcode `\#12\catcode `\^12\catcode `\_12\catcode `\%12\relax}%
\providecommand \@@startlink[1]{}%
\providecommand \@@endlink[0]{}%
\providecommand \url  [0]{\begingroup\@sanitize@url \@url }%
\providecommand \@url [1]{\endgroup\@href {#1}{\urlprefix }}%
\providecommand \urlprefix  [0]{URL }%
\providecommand \Eprint [0]{\href }%
\providecommand \doibase [0]{https://doi.org/}%
\providecommand \selectlanguage [0]{\@gobble}%
\providecommand \bibinfo  [0]{\@secondoftwo}%
\providecommand \bibfield  [0]{\@secondoftwo}%
\providecommand \translation [1]{[#1]}%
\providecommand \BibitemOpen [0]{}%
\providecommand \bibitemStop [0]{}%
\providecommand \bibitemNoStop [0]{.\EOS\space}%
\providecommand \EOS [0]{\spacefactor3000\relax}%
\providecommand \BibitemShut  [1]{\csname bibitem#1\endcsname}%
\let\auto@bib@innerbib\@empty
%</preamble>
\bibitem [{\citenamefont {Sadoc}\ and\ \citenamefont
  {Mosseri}(1999)}]{Sadoc1999}%
  \BibitemOpen
  \bibfield  {author} {\bibinfo {author} {\bibfnamefont {J.-F.}\ \bibnamefont
  {Sadoc}}\ and\ \bibinfo {author} {\bibfnamefont {R.}~\bibnamefont
  {Mosseri}},\ }\bibinfo {title} {Introduction to geometrical frustration},\
  in\ \href {https://doi.org/10.1017/CBO9780511599934.002} {\emph {\bibinfo
  {booktitle} {Geometrical Frustration}}},\ \bibinfo {series and number}
  {Collection Alea-Saclay: Monographs and Texts in Statistical Physics}\
  (\bibinfo  {publisher} {Cambridge University Press},\ \bibinfo {year}
  {1999})\ p.\ \bibinfo {pages} {1–13}\BibitemShut {NoStop}%
\bibitem [{\citenamefont {Lacroix}\ \emph {et~al.}(2011)\citenamefont
  {Lacroix}, \citenamefont {Mendels},\ and\ \citenamefont
  {Mila}}]{Lacroix2011}%
  \BibitemOpen
  \bibfield  {author} {\bibinfo {author} {\bibfnamefont {C.}~\bibnamefont
  {Lacroix}}, \bibinfo {author} {\bibfnamefont {P.}~\bibnamefont {Mendels}},\
  and\ \bibinfo {author} {\bibfnamefont {F.}~\bibnamefont {Mila}},\ }\href@noop
  {} {\emph {\bibinfo {title} {Introduction to {F}rustrated {M}agnetism:
  {M}aterials, {E}xperiments, {T}heory}}},\ edited by\ \bibinfo {editor}
  {\bibfnamefont {C.}~\bibnamefont {Lacroix}}, \bibinfo {editor} {\bibfnamefont
  {P.}~\bibnamefont {Mendels}},\ and\ \bibinfo {editor} {\bibfnamefont
  {F.}~\bibnamefont {Mila}}\ (\bibinfo  {publisher} {Springer-Verlag},\
  \bibinfo {address} {Berlin, Heidelberg},\ \bibinfo {year} {2011})\BibitemShut
  {NoStop}%
\bibitem [{\citenamefont {Onsager}(1944)}]{Onsager1944}%
  \BibitemOpen
  \bibfield  {author} {\bibinfo {author} {\bibfnamefont {L.}~\bibnamefont
  {Onsager}},\ }\bibfield  {title} {\bibinfo {title} {Crystal statistics. i. a
  two-dimensional model with an order-disorder transition},\ }\href
  {https://doi.org/10.1103/PhysRev.65.117} {\bibfield  {journal} {\bibinfo
  {journal} {Phys. Rev.}\ }\textbf {\bibinfo {volume} {65}},\ \bibinfo {pages}
  {117} (\bibinfo {year} {1944})}\BibitemShut {NoStop}%
\bibitem [{\citenamefont {Kaufman}(1949)}]{Kaufman1949}%
  \BibitemOpen
  \bibfield  {author} {\bibinfo {author} {\bibfnamefont {B.}~\bibnamefont
  {Kaufman}},\ }\bibfield  {title} {\bibinfo {title} {Crystal statistics. ii.
  partition function evaluated by spinor analysis},\ }\href
  {https://doi.org/10.1103/PhysRev.76.1232} {\bibfield  {journal} {\bibinfo
  {journal} {Phys. Rev.}\ }\textbf {\bibinfo {volume} {76}},\ \bibinfo {pages}
  {1232} (\bibinfo {year} {1949})}\BibitemShut {NoStop}%
\bibitem [{\citenamefont {Wannier}(1950)}]{Wannier1950}%
  \BibitemOpen
  \bibfield  {author} {\bibinfo {author} {\bibfnamefont {G.~H.}\ \bibnamefont
  {Wannier}},\ }\bibfield  {title} {\bibinfo {title} {Antiferromagnetism. the
  triangular {I}sing net},\ }\href {https://doi.org/10.1103/PhysRev.79.357}
  {\bibfield  {journal} {\bibinfo  {journal} {Phys. Rev.}\ }\textbf {\bibinfo
  {volume} {79}},\ \bibinfo {pages} {357} (\bibinfo {year} {1950})}\BibitemShut
  {NoStop}%
\bibitem [{\citenamefont {Wannier}(1973)}]{Wannier1973}%
  \BibitemOpen
  \bibfield  {author} {\bibinfo {author} {\bibfnamefont {G.~H.}\ \bibnamefont
  {Wannier}},\ }\bibfield  {title} {\bibinfo {title} {Antiferromagnetism. the
  triangular {I}sing net},\ }\href {https://doi.org/10.1103/PhysRevB.7.5017}
  {\bibfield  {journal} {\bibinfo  {journal} {Phys. Rev. B}\ }\textbf {\bibinfo
  {volume} {7}},\ \bibinfo {pages} {5017} (\bibinfo {year} {1973})}\BibitemShut
  {NoStop}%
\bibitem [{\citenamefont {Kano}\ and\ \citenamefont {Naya}(1953)}]{Kano1953}%
  \BibitemOpen
  \bibfield  {author} {\bibinfo {author} {\bibfnamefont {K.}~\bibnamefont
  {Kano}}\ and\ \bibinfo {author} {\bibfnamefont {S.}~\bibnamefont {Naya}},\
  }\bibfield  {title} {\bibinfo {title} {Antiferromagnetism. {The} {Kagome}
  {Ising} {Net}},\ }\href {https://doi.org/10.1143/ptp/10.2.158} {\bibfield
  {journal} {\bibinfo  {journal} {Progress of Theoretical Physics}\ }\textbf
  {\bibinfo {volume} {10}},\ \bibinfo {pages} {158} (\bibinfo {year}
  {1953})}\BibitemShut {NoStop}%
\bibitem [{\citenamefont {Sütö}(1981)}]{Suto1981}%
  \BibitemOpen
  \bibfield  {author} {\bibinfo {author} {\bibfnamefont {A.}~\bibnamefont
  {Sütö}},\ }\bibfield  {title} {\bibinfo {title} {Models of
  superfrustration},\ }\href {https://doi.org/10.1007/BF01292659} {\bibfield
  {journal} {\bibinfo  {journal} {Z. Phys. B}\ }\textbf {\bibinfo {volume}
  {44}},\ \bibinfo {pages} {121} (\bibinfo {year} {1981})}\BibitemShut
  {NoStop}%
\bibitem [{\citenamefont {Apel}\ and\ \citenamefont {Everts}(2011)}]{Apel2011}%
  \BibitemOpen
  \bibfield  {author} {\bibinfo {author} {\bibfnamefont {W.}~\bibnamefont
  {Apel}}\ and\ \bibinfo {author} {\bibfnamefont {H.-U.}\ \bibnamefont
  {Everts}},\ }\bibfield  {title} {\bibinfo {title} {Correlations in the
  {I}sing antiferromagnet on the anisotropic kagome lattice},\ }\href
  {https://doi.org/10.1088/1742-5468/2011/09/p09002} {\bibfield  {journal}
  {\bibinfo  {journal} {Journal of Statistical Mechanics: Theory and
  Experiment}\ }\textbf {\bibinfo {volume} {2011}},\ \bibinfo {pages} {P09002}
  (\bibinfo {year} {2011})}\BibitemShut {NoStop}%
\bibitem [{\citenamefont {Kanamori}(1966)}]{Kanamori1966}%
  \BibitemOpen
  \bibfield  {author} {\bibinfo {author} {\bibfnamefont {J.}~\bibnamefont
  {Kanamori}},\ }\bibfield  {title} {\bibinfo {title} {{Magnetization Process
  in an {I}sing Spin System}},\ }\href {https://doi.org/10.1143/PTP.35.16}
  {\bibfield  {journal} {\bibinfo  {journal} {Prog. Theor. Phys.}\ }\textbf
  {\bibinfo {volume} {35}},\ \bibinfo {pages} {16} (\bibinfo {year}
  {1966})}\BibitemShut {NoStop}%
\bibitem [{\citenamefont {Gingras}(2011)}]{Gingras2011}%
  \BibitemOpen
  \bibfield  {author} {\bibinfo {author} {\bibfnamefont {M.~J.}\ \bibnamefont
  {Gingras}},\ }\bibfield  {title} {\bibinfo {title} {Spin ice},\ }in\
  \cite{Lacroix2011},\ pp.\ \bibinfo {pages} {293--329}\BibitemShut {NoStop}%
\bibitem [{\citenamefont {Castelnovo}\ \emph {et~al.}(2008)\citenamefont
  {Castelnovo}, \citenamefont {Moessner},\ and\ \citenamefont
  {Sondhi}}]{Castelnovo2008}%
  \BibitemOpen
  \bibfield  {author} {\bibinfo {author} {\bibfnamefont {C.}~\bibnamefont
  {Castelnovo}}, \bibinfo {author} {\bibfnamefont {R.}~\bibnamefont
  {Moessner}},\ and\ \bibinfo {author} {\bibfnamefont {S.~L.}\ \bibnamefont
  {Sondhi}},\ }\bibfield  {title} {\bibinfo {title} {Magnetic monopoles in spin
  ice},\ }\href {https://doi.org/10.1038/nature06433} {\bibfield  {journal}
  {\bibinfo  {journal} {Nature}\ }\textbf {\bibinfo {volume} {451}},\ \bibinfo
  {pages} {42} (\bibinfo {year} {2008})}\BibitemShut {NoStop}%
\bibitem [{\citenamefont {Isakov}\ \emph {et~al.}(2004)\citenamefont {Isakov},
  \citenamefont {Gregor}, \citenamefont {Moessner},\ and\ \citenamefont
  {Sondhi}}]{Isakov2004}%
  \BibitemOpen
  \bibfield  {author} {\bibinfo {author} {\bibfnamefont {S.~V.}\ \bibnamefont
  {Isakov}}, \bibinfo {author} {\bibfnamefont {K.}~\bibnamefont {Gregor}},
  \bibinfo {author} {\bibfnamefont {R.}~\bibnamefont {Moessner}},\ and\
  \bibinfo {author} {\bibfnamefont {S.~L.}\ \bibnamefont {Sondhi}},\ }\bibfield
   {title} {\bibinfo {title} {Dipolar spin correlations in classical pyrochlore
  magnets},\ }\href {https://doi.org/10.1103/PhysRevLett.93.167204} {\bibfield
  {journal} {\bibinfo  {journal} {Phys. Rev. Lett.}\ }\textbf {\bibinfo
  {volume} {93}},\ \bibinfo {pages} {167204} (\bibinfo {year}
  {2004})}\BibitemShut {NoStop}%
\bibitem [{\citenamefont {Isakov}\ \emph {et~al.}(2005)\citenamefont {Isakov},
  \citenamefont {Moessner},\ and\ \citenamefont {Sondhi}}]{Isakov2005}%
  \BibitemOpen
  \bibfield  {author} {\bibinfo {author} {\bibfnamefont {S.~V.}\ \bibnamefont
  {Isakov}}, \bibinfo {author} {\bibfnamefont {R.}~\bibnamefont {Moessner}},\
  and\ \bibinfo {author} {\bibfnamefont {S.~L.}\ \bibnamefont {Sondhi}},\
  }\bibfield  {title} {\bibinfo {title} {Why spin ice obeys the ice rules},\
  }\href {https://doi.org/10.1103/PhysRevLett.95.217201} {\bibfield  {journal}
  {\bibinfo  {journal} {Phys. Rev. Lett.}\ }\textbf {\bibinfo {volume} {95}},\
  \bibinfo {pages} {217201} (\bibinfo {year} {2005})}\BibitemShut {NoStop}%
\bibitem [{\citenamefont {Wang}\ \emph {et~al.}(2006)\citenamefont {Wang},
  \citenamefont {Nisoli}, \citenamefont {Freitas}, \citenamefont {Li},
  \citenamefont {McConville}, \citenamefont {Cooley}, \citenamefont {Lund},
  \citenamefont {Samarth}, \citenamefont {Leighton}, \citenamefont {Crespi},\
  and\ \citenamefont {Schiffer}}]{Wang2006}%
  \BibitemOpen
  \bibfield  {author} {\bibinfo {author} {\bibfnamefont {R.~F.}\ \bibnamefont
  {Wang}}, \bibinfo {author} {\bibfnamefont {C.}~\bibnamefont {Nisoli}},
  \bibinfo {author} {\bibfnamefont {R.~S.}\ \bibnamefont {Freitas}}, \bibinfo
  {author} {\bibfnamefont {J.}~\bibnamefont {Li}}, \bibinfo {author}
  {\bibfnamefont {W.}~\bibnamefont {McConville}}, \bibinfo {author}
  {\bibfnamefont {B.~J.}\ \bibnamefont {Cooley}}, \bibinfo {author}
  {\bibfnamefont {M.~S.}\ \bibnamefont {Lund}}, \bibinfo {author}
  {\bibfnamefont {N.}~\bibnamefont {Samarth}}, \bibinfo {author} {\bibfnamefont
  {C.}~\bibnamefont {Leighton}}, \bibinfo {author} {\bibfnamefont {V.~H.}\
  \bibnamefont {Crespi}},\ and\ \bibinfo {author} {\bibfnamefont
  {P.}~\bibnamefont {Schiffer}},\ }\bibfield  {title} {\bibinfo {title}
  {Artificial ‘spin ice’ in a geometrically frustrated lattice of nanoscale
  ferromagnetic islands},\ }\href {https://doi.org/10.1038/nature04447}
  {\bibfield  {journal} {\bibinfo  {journal} {Nature}\ }\textbf {\bibinfo
  {volume} {439}},\ \bibinfo {pages} {303} (\bibinfo {year}
  {2006})}\BibitemShut {NoStop}%
\bibitem [{\citenamefont {Tanaka}\ \emph {et~al.}(2006)\citenamefont {Tanaka},
  \citenamefont {Saitoh}, \citenamefont {Miyajima}, \citenamefont {Yamaoka},\
  and\ \citenamefont {Iye}}]{Tanaka2006}%
  \BibitemOpen
  \bibfield  {author} {\bibinfo {author} {\bibfnamefont {M.}~\bibnamefont
  {Tanaka}}, \bibinfo {author} {\bibfnamefont {E.}~\bibnamefont {Saitoh}},
  \bibinfo {author} {\bibfnamefont {H.}~\bibnamefont {Miyajima}}, \bibinfo
  {author} {\bibfnamefont {T.}~\bibnamefont {Yamaoka}},\ and\ \bibinfo {author}
  {\bibfnamefont {Y.}~\bibnamefont {Iye}},\ }\bibfield  {title} {\bibinfo
  {title} {Magnetic interactions in a ferromagnetic honeycomb nanoscale
  network},\ }\href {https://doi.org/10.1103/PhysRevB.73.052411} {\bibfield
  {journal} {\bibinfo  {journal} {Phys. Rev. B}\ }\textbf {\bibinfo {volume}
  {73}},\ \bibinfo {pages} {052411} (\bibinfo {year} {2006})}\BibitemShut
  {NoStop}%
\bibitem [{\citenamefont {Skjærvø}\ \emph {et~al.}(2020)\citenamefont
  {Skjærvø}, \citenamefont {Marrows}, \citenamefont {Stamps},\ and\
  \citenamefont {Heyderman}}]{Skjaervoe2020}%
  \BibitemOpen
  \bibfield  {author} {\bibinfo {author} {\bibfnamefont {S.~H.}\ \bibnamefont
  {Skjærvø}}, \bibinfo {author} {\bibfnamefont {C.~H.}\ \bibnamefont
  {Marrows}}, \bibinfo {author} {\bibfnamefont {R.~L.}\ \bibnamefont
  {Stamps}},\ and\ \bibinfo {author} {\bibfnamefont {L.~J.}\ \bibnamefont
  {Heyderman}},\ }\bibfield  {title} {\bibinfo {title} {Advances in artificial
  spin ice},\ }\bibfield  {journal} {\bibinfo  {journal} {Nature Reviews
  Physics}\ }\textbf {\bibinfo {volume} {2}},\ \href
  {https://doi.org/10.1038/s42254-019-0118-3} {10.1038/s42254-019-0118-3}
  (\bibinfo {year} {2020}),\ \bibinfo {note} {number: 1 Publisher: Nature
  Publishing Group, p. 14}\BibitemShut {NoStop}%
\bibitem [{\citenamefont {Rougemaille}\ and\ \citenamefont
  {Canals}(2019)}]{Rougemaille2019}%
  \BibitemOpen
  \bibfield  {author} {\bibinfo {author} {\bibfnamefont {N.}~\bibnamefont
  {Rougemaille}}\ and\ \bibinfo {author} {\bibfnamefont {B.}~\bibnamefont
  {Canals}},\ }\bibfield  {title} {\bibinfo {title} {Cooperative magnetic
  phenomena in artificial spin systems: spin liquids, coulomb phase and
  fragmentation of magnetism – a colloquium},\ }\href
  {https://doi.org/10.1140/epjb/e2018-90346-7} {\bibfield  {journal} {\bibinfo
  {journal} {Eur. Phys. J. B}\ }\textbf {\bibinfo {volume} {92}},\ \bibinfo
  {pages} {62} (\bibinfo {year} {2019})}\BibitemShut {NoStop}%
\bibitem [{\citenamefont {Schiffer}\ and\ \citenamefont
  {Nisoli}(2021)}]{Schiffer2021}%
  \BibitemOpen
  \bibfield  {author} {\bibinfo {author} {\bibfnamefont {P.}~\bibnamefont
  {Schiffer}}\ and\ \bibinfo {author} {\bibfnamefont {C.}~\bibnamefont
  {Nisoli}},\ }\bibfield  {title} {\bibinfo {title} {Artificial spin ice: Paths
  forward},\ }\href {https://doi.org/10.1063/5.0044790} {\bibfield  {journal}
  {\bibinfo  {journal} {Applied Physics Letters}\ }\textbf {\bibinfo {volume}
  {118}},\ \bibinfo {pages} {110501} (\bibinfo {year} {2021})},\ \Eprint
  {https://arxiv.org/abs/https://doi.org/10.1063/5.0044790}
  {https://doi.org/10.1063/5.0044790} \BibitemShut {NoStop}%
\bibitem [{\citenamefont {Qi}\ \emph {et~al.}(2008)\citenamefont {Qi},
  \citenamefont {Brintlinger},\ and\ \citenamefont {Cumings}}]{Qi2008}%
  \BibitemOpen
  \bibfield  {author} {\bibinfo {author} {\bibfnamefont {Y.}~\bibnamefont
  {Qi}}, \bibinfo {author} {\bibfnamefont {T.}~\bibnamefont {Brintlinger}},\
  and\ \bibinfo {author} {\bibfnamefont {J.}~\bibnamefont {Cumings}},\
  }\bibfield  {title} {\bibinfo {title} {Direct observation of the ice rule in
  an artificial kagome spin ice},\ }\href
  {https://doi.org/10.1103/PhysRevB.77.094418} {\bibfield  {journal} {\bibinfo
  {journal} {Phys. Rev. B}\ }\textbf {\bibinfo {volume} {77}},\ \bibinfo
  {pages} {094418} (\bibinfo {year} {2008})}\BibitemShut {NoStop}%
\bibitem [{\citenamefont {Zhang}\ \emph {et~al.}(2012)\citenamefont {Zhang},
  \citenamefont {Li}, \citenamefont {Gilbert}, \citenamefont {Bartell},
  \citenamefont {Erickson}, \citenamefont {Pan}, \citenamefont {Lammert},
  \citenamefont {Nisoli}, \citenamefont {Kohli}, \citenamefont {Misra},
  \citenamefont {Crespi}, \citenamefont {Samarth}, \citenamefont {Leighton},\
  and\ \citenamefont {Schiffer}}]{Zhang2012}%
  \BibitemOpen
  \bibfield  {author} {\bibinfo {author} {\bibfnamefont {S.}~\bibnamefont
  {Zhang}}, \bibinfo {author} {\bibfnamefont {J.}~\bibnamefont {Li}}, \bibinfo
  {author} {\bibfnamefont {I.}~\bibnamefont {Gilbert}}, \bibinfo {author}
  {\bibfnamefont {J.}~\bibnamefont {Bartell}}, \bibinfo {author} {\bibfnamefont
  {M.~J.}\ \bibnamefont {Erickson}}, \bibinfo {author} {\bibfnamefont
  {Y.}~\bibnamefont {Pan}}, \bibinfo {author} {\bibfnamefont {P.~E.}\
  \bibnamefont {Lammert}}, \bibinfo {author} {\bibfnamefont {C.}~\bibnamefont
  {Nisoli}}, \bibinfo {author} {\bibfnamefont {K.~K.}\ \bibnamefont {Kohli}},
  \bibinfo {author} {\bibfnamefont {R.}~\bibnamefont {Misra}}, \bibinfo
  {author} {\bibfnamefont {V.~H.}\ \bibnamefont {Crespi}}, \bibinfo {author}
  {\bibfnamefont {N.}~\bibnamefont {Samarth}}, \bibinfo {author} {\bibfnamefont
  {C.}~\bibnamefont {Leighton}},\ and\ \bibinfo {author} {\bibfnamefont
  {P.}~\bibnamefont {Schiffer}},\ }\bibfield  {title} {\bibinfo {title}
  {Perpendicular magnetization and generic realization of the ising model in
  artificial spin ice},\ }\href
  {https://doi.org/10.1103/PhysRevLett.109.087201} {\bibfield  {journal}
  {\bibinfo  {journal} {Phys. Rev. Lett.}\ }\textbf {\bibinfo {volume} {109}},\
  \bibinfo {pages} {087201} (\bibinfo {year} {2012})}\BibitemShut {NoStop}%
\bibitem [{\citenamefont {Chioar}\ \emph {et~al.}(2014)\citenamefont {Chioar},
  \citenamefont {Rougemaille}, \citenamefont {Grimm}, \citenamefont {Fruchart},
  \citenamefont {Wagner}, \citenamefont {Hehn}, \citenamefont {Lacour},
  \citenamefont {Montaigne},\ and\ \citenamefont {Canals}}]{Chioar2014}%
  \BibitemOpen
  \bibfield  {author} {\bibinfo {author} {\bibfnamefont {I.~A.}\ \bibnamefont
  {Chioar}}, \bibinfo {author} {\bibfnamefont {N.}~\bibnamefont {Rougemaille}},
  \bibinfo {author} {\bibfnamefont {A.}~\bibnamefont {Grimm}}, \bibinfo
  {author} {\bibfnamefont {O.}~\bibnamefont {Fruchart}}, \bibinfo {author}
  {\bibfnamefont {E.}~\bibnamefont {Wagner}}, \bibinfo {author} {\bibfnamefont
  {M.}~\bibnamefont {Hehn}}, \bibinfo {author} {\bibfnamefont {D.}~\bibnamefont
  {Lacour}}, \bibinfo {author} {\bibfnamefont {F.}~\bibnamefont {Montaigne}},\
  and\ \bibinfo {author} {\bibfnamefont {B.}~\bibnamefont {Canals}},\
  }\bibfield  {title} {\bibinfo {title} {Nonuniversality of artificial
  frustrated spin systems},\ }\href
  {https://doi.org/10.1103/PhysRevB.90.064411} {\bibfield  {journal} {\bibinfo
  {journal} {Phys. Rev. B}\ }\textbf {\bibinfo {volume} {90}},\ \bibinfo
  {pages} {064411} (\bibinfo {year} {2014})}\BibitemShut {NoStop}%
\bibitem [{\citenamefont {Chioar}\ \emph {et~al.}(2016)\citenamefont {Chioar},
  \citenamefont {Rougemaille},\ and\ \citenamefont {Canals}}]{Chioar2016}%
  \BibitemOpen
  \bibfield  {author} {\bibinfo {author} {\bibfnamefont {I.~A.}\ \bibnamefont
  {Chioar}}, \bibinfo {author} {\bibfnamefont {N.}~\bibnamefont
  {Rougemaille}},\ and\ \bibinfo {author} {\bibfnamefont {B.}~\bibnamefont
  {Canals}},\ }\bibfield  {title} {\bibinfo {title} {Ground-state candidate for
  the classical dipolar kagome {{I}sing} antiferromagnet},\ }\href
  {https://doi.org/10.1103/PhysRevB.93.214410} {\bibfield  {journal} {\bibinfo
  {journal} {Phys. Rev. B}\ }\textbf {\bibinfo {volume} {93}},\ \bibinfo
  {pages} {214410} (\bibinfo {year} {2016})}\BibitemShut {NoStop}%
\bibitem [{\citenamefont {Kempinger}\ \emph {et~al.}(2021)\citenamefont
  {Kempinger}, \citenamefont {Huang}, \citenamefont {Lammert}, \citenamefont
  {Vogel}, \citenamefont {Hoffmann}, \citenamefont {Crespi}, \citenamefont
  {Schiffer},\ and\ \citenamefont {Samarth}}]{Kempinger2021}%
  \BibitemOpen
  \bibfield  {author} {\bibinfo {author} {\bibfnamefont {S.}~\bibnamefont
  {Kempinger}}, \bibinfo {author} {\bibfnamefont {Y.-S.}\ \bibnamefont
  {Huang}}, \bibinfo {author} {\bibfnamefont {P.}~\bibnamefont {Lammert}},
  \bibinfo {author} {\bibfnamefont {M.}~\bibnamefont {Vogel}}, \bibinfo
  {author} {\bibfnamefont {A.}~\bibnamefont {Hoffmann}}, \bibinfo {author}
  {\bibfnamefont {V.~H.}\ \bibnamefont {Crespi}}, \bibinfo {author}
  {\bibfnamefont {P.}~\bibnamefont {Schiffer}},\ and\ \bibinfo {author}
  {\bibfnamefont {N.}~\bibnamefont {Samarth}},\ }\bibfield  {title} {\bibinfo
  {title} {Field-tunable interactions and frustration in underlayer-mediated
  artificial spin ice},\ }\href
  {https://doi.org/10.1103/PhysRevLett.127.117203} {\bibfield  {journal}
  {\bibinfo  {journal} {Phys. Rev. Lett.}\ }\textbf {\bibinfo {volume} {127}},\
  \bibinfo {pages} {117203} (\bibinfo {year} {2021})}\BibitemShut {NoStop}%
\bibitem [{\citenamefont {Rougemaille}\ \emph {et~al.}(2011)\citenamefont
  {Rougemaille}, \citenamefont {Montaigne}, \citenamefont {Canals},
  \citenamefont {Duluard}, \citenamefont {Lacour}, \citenamefont {Hehn},
  \citenamefont {Belkhou}, \citenamefont {Fruchart}, \citenamefont
  {El~Moussaoui}, \citenamefont {Bendounan},\ and\ \citenamefont
  {Maccherozzi}}]{Rougemaille2011}%
  \BibitemOpen
  \bibfield  {author} {\bibinfo {author} {\bibfnamefont {N.}~\bibnamefont
  {Rougemaille}}, \bibinfo {author} {\bibfnamefont {F.}~\bibnamefont
  {Montaigne}}, \bibinfo {author} {\bibfnamefont {B.}~\bibnamefont {Canals}},
  \bibinfo {author} {\bibfnamefont {A.}~\bibnamefont {Duluard}}, \bibinfo
  {author} {\bibfnamefont {D.}~\bibnamefont {Lacour}}, \bibinfo {author}
  {\bibfnamefont {M.}~\bibnamefont {Hehn}}, \bibinfo {author} {\bibfnamefont
  {R.}~\bibnamefont {Belkhou}}, \bibinfo {author} {\bibfnamefont
  {O.}~\bibnamefont {Fruchart}}, \bibinfo {author} {\bibfnamefont
  {S.}~\bibnamefont {El~Moussaoui}}, \bibinfo {author} {\bibfnamefont
  {A.}~\bibnamefont {Bendounan}},\ and\ \bibinfo {author} {\bibfnamefont
  {F.}~\bibnamefont {Maccherozzi}},\ }\bibfield  {title} {\bibinfo {title}
  {Artificial kagome arrays of nanomagnets: A frozen dipolar spin ice},\ }\href
  {https://doi.org/10.1103/PhysRevLett.106.057209} {\bibfield  {journal}
  {\bibinfo  {journal} {Phys. Rev. Lett.}\ }\textbf {\bibinfo {volume} {106}},\
  \bibinfo {pages} {057209} (\bibinfo {year} {2011})}\BibitemShut {NoStop}%
\bibitem [{\citenamefont {Mengotti}\ \emph {et~al.}(2009)\citenamefont
  {Mengotti}, \citenamefont {Heyderman}, \citenamefont {Bisig}, \citenamefont
  {Fraile~Rodríguez}, \citenamefont {Le~Guyader}, \citenamefont {Nolting},\
  and\ \citenamefont {Braun}}]{Mengotti2009}%
  \BibitemOpen
  \bibfield  {author} {\bibinfo {author} {\bibfnamefont {E.}~\bibnamefont
  {Mengotti}}, \bibinfo {author} {\bibfnamefont {L.~J.}\ \bibnamefont
  {Heyderman}}, \bibinfo {author} {\bibfnamefont {A.}~\bibnamefont {Bisig}},
  \bibinfo {author} {\bibfnamefont {A.}~\bibnamefont {Fraile~Rodríguez}},
  \bibinfo {author} {\bibfnamefont {L.}~\bibnamefont {Le~Guyader}}, \bibinfo
  {author} {\bibfnamefont {F.}~\bibnamefont {Nolting}},\ and\ \bibinfo {author}
  {\bibfnamefont {H.~B.}\ \bibnamefont {Braun}},\ }\bibfield  {title} {\bibinfo
  {title} {Dipolar energy states in clusters of perpendicular magnetic
  nanoislands},\ }\href {https://doi.org/10.1063/1.3133202} {\bibfield
  {journal} {\bibinfo  {journal} {Journal of Applied Physics}\ }\textbf
  {\bibinfo {volume} {105}},\ \bibinfo {pages} {113113} (\bibinfo {year}
  {2009})},\ \Eprint {https://arxiv.org/abs/https://doi.org/10.1063/1.3133202}
  {https://doi.org/10.1063/1.3133202} \BibitemShut {NoStop}%
\bibitem [{\citenamefont {Metcalf}(1974)}]{Metcalf1974}%
  \BibitemOpen
  \bibfield  {author} {\bibinfo {author} {\bibfnamefont {B.~D.}\ \bibnamefont
  {Metcalf}},\ }\bibfield  {title} {\bibinfo {title} {Ground state spin
  orderings of the triangular {Ising} model with the nearest and next nearest
  neighbor interaction},\ }\href {https://doi.org/10.1016/0375-9601(74)90247-3}
  {\bibfield  {journal} {\bibinfo  {journal} {Physics Letters A}\ }\textbf
  {\bibinfo {volume} {46}},\ \bibinfo {pages} {325} (\bibinfo {year}
  {1974})}\BibitemShut {NoStop}%
\bibitem [{\citenamefont {Tanaka}\ and\ \citenamefont
  {Uryû}(1975)}]{Tanaka1975}%
  \BibitemOpen
  \bibfield  {author} {\bibinfo {author} {\bibfnamefont {Y.}~\bibnamefont
  {Tanaka}}\ and\ \bibinfo {author} {\bibfnamefont {N.}~\bibnamefont {Uryû}},\
  }\bibfield  {title} {\bibinfo {title} {Ground state spin configurations of
  the triangular {I}sing net with the nearest and next nearest neighbor
  interactions},\ }\href {https://doi.org/10.1143/JPSJ.39.825} {\bibfield
  {journal} {\bibinfo  {journal} {J. Phys. Soc. Jpn}\ }\textbf {\bibinfo
  {volume} {39}},\ \bibinfo {pages} {825} (\bibinfo {year} {1975})}\BibitemShut
  {NoStop}%
\bibitem [{\citenamefont {Glosli}\ and\ \citenamefont
  {Plischke}(1983)}]{Glosli1983}%
  \BibitemOpen
  \bibfield  {author} {\bibinfo {author} {\bibfnamefont {J.}~\bibnamefont
  {Glosli}}\ and\ \bibinfo {author} {\bibfnamefont {M.}~\bibnamefont
  {Plischke}},\ }\bibfield  {title} {\bibinfo {title} {A {Monte} {Carlo} and
  renormalization group study of the {Ising} model with nearest and next
  nearest neighbor interactions on the triangular lattice},\ }\href
  {https://doi.org/10.1139/p83-197} {\bibfield  {journal} {\bibinfo  {journal}
  {Canadian Journal of Physics}\ }\textbf {\bibinfo {volume} {61}},\ \bibinfo
  {pages} {1515} (\bibinfo {year} {1983})},\ \bibinfo {note} {publisher: NRC
  Research Press}\BibitemShut {NoStop}%
\bibitem [{\citenamefont {Korshunov}(2005)}]{Korshunov2005}%
  \BibitemOpen
  \bibfield  {author} {\bibinfo {author} {\bibfnamefont {S.~E.}\ \bibnamefont
  {Korshunov}},\ }\bibfield  {title} {\bibinfo {title} {Nature of phase
  transitions in the striped phase of a triangular-lattice {Ising}
  antiferromagnet},\ }\href {https://doi.org/10.1103/PhysRevB.72.144417}
  {\bibfield  {journal} {\bibinfo  {journal} {Physical Review B}\ }\textbf
  {\bibinfo {volume} {72}},\ \bibinfo {pages} {144417} (\bibinfo {year}
  {2005})},\ \bibinfo {note} {publisher: American Physical Society}\BibitemShut
  {NoStop}%
\bibitem [{\citenamefont {Rastelli}\ \emph {et~al.}(2005)\citenamefont
  {Rastelli}, \citenamefont {Regina},\ and\ \citenamefont
  {Tassi}}]{Rastelli2005}%
  \BibitemOpen
  \bibfield  {author} {\bibinfo {author} {\bibfnamefont {E.}~\bibnamefont
  {Rastelli}}, \bibinfo {author} {\bibfnamefont {S.}~\bibnamefont {Regina}},\
  and\ \bibinfo {author} {\bibfnamefont {A.}~\bibnamefont {Tassi}},\ }\bibfield
   {title} {\bibinfo {title} {Monte {Carlo} simulations on a triangular {Ising}
  antiferromagnet with nearest and next-nearest interactions},\ }\href
  {https://doi.org/10.1103/PhysRevB.71.174406} {\bibfield  {journal} {\bibinfo
  {journal} {Physical Review B}\ }\textbf {\bibinfo {volume} {71}},\ \bibinfo
  {pages} {174406} (\bibinfo {year} {2005})},\ \bibinfo {note} {publisher:
  American Physical Society}\BibitemShut {NoStop}%
\bibitem [{\citenamefont {Smerald}\ \emph {et~al.}(2016)\citenamefont
  {Smerald}, \citenamefont {Korshunov},\ and\ \citenamefont
  {Mila}}]{Smerald2016}%
  \BibitemOpen
  \bibfield  {author} {\bibinfo {author} {\bibfnamefont {A.}~\bibnamefont
  {Smerald}}, \bibinfo {author} {\bibfnamefont {S.}~\bibnamefont {Korshunov}},\
  and\ \bibinfo {author} {\bibfnamefont {F.}~\bibnamefont {Mila}},\ }\bibfield
  {title} {\bibinfo {title} {Topological aspects of symmetry breaking in
  triangular-lattice ising antiferromagnets},\ }\href
  {https://doi.org/10.1103/PhysRevLett.116.197201} {\bibfield  {journal}
  {\bibinfo  {journal} {Phys. Rev. Lett.}\ }\textbf {\bibinfo {volume} {116}},\
  \bibinfo {pages} {197201} (\bibinfo {year} {2016})}\BibitemShut {NoStop}%
\bibitem [{\citenamefont {Rößler}(2001)}]{Rossler2001}%
  \BibitemOpen
  \bibfield  {author} {\bibinfo {author} {\bibfnamefont {U.~K.}\ \bibnamefont
  {Rößler}},\ }\bibfield  {title} {\bibinfo {title} {Ising dipoles on the
  triangular lattice},\ }\href {https://doi.org/10.1063/1.1358336} {\bibfield
  {journal} {\bibinfo  {journal} {Journal of Applied Physics}\ }\textbf
  {\bibinfo {volume} {89}},\ \bibinfo {pages} {7033} (\bibinfo {year}
  {2001})},\ \bibinfo {note} {publisher: American Institute of
  Physics}\BibitemShut {NoStop}%
\bibitem [{\citenamefont {Smerald}\ and\ \citenamefont
  {Mila}(2018)}]{Smerald2017}%
  \BibitemOpen
  \bibfield  {author} {\bibinfo {author} {\bibfnamefont {A.}~\bibnamefont
  {Smerald}}\ and\ \bibinfo {author} {\bibfnamefont {F.}~\bibnamefont {Mila}},\
  }\bibfield  {title} {\bibinfo {title} {{Spin-liquid behaviour and the
  interplay between Pokrovsky-Talapov and {I}sing criticality in the distorted,
  triangular-lattice, dipolar {I}sing antiferromagnet}},\ }\href
  {https://doi.org/10.21468/SciPostPhys.5.3.030} {\bibfield  {journal}
  {\bibinfo  {journal} {SciPost Phys.}\ }\textbf {\bibinfo {volume} {5}},\
  \bibinfo {pages} {30} (\bibinfo {year} {2018})}\BibitemShut {NoStop}%
\bibitem [{\citenamefont {M\"oller}\ and\ \citenamefont
  {Moessner}(2009)}]{Moller2009}%
  \BibitemOpen
  \bibfield  {author} {\bibinfo {author} {\bibfnamefont {G.}~\bibnamefont
  {M\"oller}}\ and\ \bibinfo {author} {\bibfnamefont {R.}~\bibnamefont
  {Moessner}},\ }\bibfield  {title} {\bibinfo {title} {Magnetic multipole
  analysis of kagome and artificial spin-ice dipolar arrays},\ }\href
  {https://doi.org/10.1103/PhysRevB.80.140409} {\bibfield  {journal} {\bibinfo
  {journal} {Phys. Rev. B}\ }\textbf {\bibinfo {volume} {80}},\ \bibinfo
  {pages} {140409(R)} (\bibinfo {year} {2009})}\BibitemShut {NoStop}%
\bibitem [{\citenamefont {Takagi}\ and\ \citenamefont
  {Mekata}(1993)}]{Takagi1993}%
  \BibitemOpen
  \bibfield  {author} {\bibinfo {author} {\bibfnamefont {T.}~\bibnamefont
  {Takagi}}\ and\ \bibinfo {author} {\bibfnamefont {M.}~\bibnamefont
  {Mekata}},\ }\bibfield  {title} {\bibinfo {title} {Magnetic ordering of
  {I}sing spins on kagomé lattice with the 1st and the 2nd neighbor
  interactions},\ }\href {https://doi.org/10.1143/JPSJ.62.3943} {\bibfield
  {journal} {\bibinfo  {journal} {Journal of the Physical Society of Japan}\
  }\textbf {\bibinfo {volume} {62}},\ \bibinfo {pages} {3943} (\bibinfo {year}
  {1993})}\BibitemShut {NoStop}%
\bibitem [{\citenamefont {Chern}\ and\ \citenamefont
  {Tchernyshyov}(2012)}]{Chern2012}%
  \BibitemOpen
  \bibfield  {author} {\bibinfo {author} {\bibfnamefont {G.-W.}\ \bibnamefont
  {Chern}}\ and\ \bibinfo {author} {\bibfnamefont {O.}~\bibnamefont
  {Tchernyshyov}},\ }\bibfield  {title} {\bibinfo {title} {Magnetic charge and
  ordering in kagome spin ice},\ }\href
  {https://doi.org/10.1098/rsta.2011.0388} {\bibfield  {journal} {\bibinfo
  {journal} {Phil. Trans. the R. Soc. A}\ }\textbf {\bibinfo {volume} {370}},\
  \bibinfo {pages} {5718} (\bibinfo {year} {2012})}\BibitemShut {NoStop}%
\bibitem [{\citenamefont {Kao}\ \emph {et~al.}(2020)\citenamefont {Kao},
  \citenamefont {Chern},\ and\ \citenamefont {Kao}}]{Kao2020}%
  \BibitemOpen
  \bibfield  {author} {\bibinfo {author} {\bibfnamefont {W.-H.}\ \bibnamefont
  {Kao}}, \bibinfo {author} {\bibfnamefont {G.-W.}\ \bibnamefont {Chern}},\
  and\ \bibinfo {author} {\bibfnamefont {Y.-J.}\ \bibnamefont {Kao}},\
  }\bibfield  {title} {\bibinfo {title} {Emergent snake magnetic domains in
  canted kagome ice},\ }\href
  {https://doi.org/10.1103/PhysRevResearch.2.023046} {\bibfield  {journal}
  {\bibinfo  {journal} {Phys. Rev. Research}\ }\textbf {\bibinfo {volume}
  {2}},\ \bibinfo {pages} {023046} (\bibinfo {year} {2020})}\BibitemShut
  {NoStop}%
\bibitem [{\citenamefont {Chern}\ \emph {et~al.}(2011)\citenamefont {Chern},
  \citenamefont {Mellado},\ and\ \citenamefont {Tchernyshyov}}]{Chern2011}%
  \BibitemOpen
  \bibfield  {author} {\bibinfo {author} {\bibfnamefont {G.-W.}\ \bibnamefont
  {Chern}}, \bibinfo {author} {\bibfnamefont {P.}~\bibnamefont {Mellado}},\
  and\ \bibinfo {author} {\bibfnamefont {O.}~\bibnamefont {Tchernyshyov}},\
  }\bibfield  {title} {\bibinfo {title} {Two-stage ordering of spins in dipolar
  spin ice on the kagome lattice},\ }\href
  {https://doi.org/10.1103/PhysRevLett.106.207202} {\bibfield  {journal}
  {\bibinfo  {journal} {Phys. Rev. Lett.}\ }\textbf {\bibinfo {volume} {106}},\
  \bibinfo {pages} {207202} (\bibinfo {year} {2011})}\BibitemShut {NoStop}%
\bibitem [{\citenamefont {Colbois}\ \emph {et~al.}(2021)\citenamefont
  {Colbois}, \citenamefont {Hofhuis}, \citenamefont {Luo}, \citenamefont
  {Wang}, \citenamefont {Hrabec}, \citenamefont {Heyderman},\ and\
  \citenamefont {Mila}}]{Colbois2021}%
  \BibitemOpen
  \bibfield  {author} {\bibinfo {author} {\bibfnamefont {J.}~\bibnamefont
  {Colbois}}, \bibinfo {author} {\bibfnamefont {K.}~\bibnamefont {Hofhuis}},
  \bibinfo {author} {\bibfnamefont {Z.}~\bibnamefont {Luo}}, \bibinfo {author}
  {\bibfnamefont {X.}~\bibnamefont {Wang}}, \bibinfo {author} {\bibfnamefont
  {A.}~\bibnamefont {Hrabec}}, \bibinfo {author} {\bibfnamefont {L.~J.}\
  \bibnamefont {Heyderman}},\ and\ \bibinfo {author} {\bibfnamefont
  {F.}~\bibnamefont {Mila}},\ }\bibfield  {title} {\bibinfo {title} {Artificial
  out-of-plane {I}sing antiferromagnet on the kagome lattice with very small
  farther-neighbor couplings},\ }\href
  {https://doi.org/10.1103/PhysRevB.104.024418} {\bibfield  {journal} {\bibinfo
   {journal} {Phys. Rev. B}\ }\textbf {\bibinfo {volume} {104}},\ \bibinfo
  {pages} {024418} (\bibinfo {year} {2021})}\BibitemShut {NoStop}%
\bibitem [{\citenamefont {Hamp}\ \emph {et~al.}(2018)\citenamefont {Hamp},
  \citenamefont {Moessner},\ and\ \citenamefont {Castelnovo}}]{Hamp2018}%
  \BibitemOpen
  \bibfield  {author} {\bibinfo {author} {\bibfnamefont {J.}~\bibnamefont
  {Hamp}}, \bibinfo {author} {\bibfnamefont {R.}~\bibnamefont {Moessner}},\
  and\ \bibinfo {author} {\bibfnamefont {C.}~\bibnamefont {Castelnovo}},\
  }\bibfield  {title} {\bibinfo {title} {Supercooling and fragile glassiness in
  a dipolar kagome {I}sing magnet},\ }\href
  {https://doi.org/10.1103/PhysRevB.98.144439} {\bibfield  {journal} {\bibinfo
  {journal} {Phys. Rev. B}\ }\textbf {\bibinfo {volume} {98}},\ \bibinfo
  {pages} {144439} (\bibinfo {year} {2018})}\BibitemShut {NoStop}%
\bibitem [{\citenamefont {Ducastelle}(1991)}]{Ducastelle1991}%
  \BibitemOpen
  \bibfield  {author} {\bibinfo {author} {\bibfnamefont {F.}~\bibnamefont
  {Ducastelle}},\ }\href@noop {} {\emph {\bibinfo {title} {Order and Phase
  Stability in Alloys}}},\ edited by\ \bibinfo {editor} {\bibfnamefont
  {F.}~\bibnamefont {de~Boer}}\ and\ \bibinfo {editor} {\bibfnamefont
  {D.}~\bibnamefont {Perrifor}},\ \bibinfo {series} {Cohesion and Structure}\
  No.~\bibinfo {number} {3}\ (\bibinfo  {publisher} {North-Holland, Elsevier
  Science Publisher},\ \bibinfo {address} {Amsterdam},\ \bibinfo {year}
  {1991})\ \bibinfo {note} {(No online source found)}\BibitemShut {NoStop}%
\bibitem [{\citenamefont {Huang}\ \emph {et~al.}(2016)\citenamefont {Huang},
  \citenamefont {Kitchaev}, \citenamefont {Dacek}, \citenamefont {Rong},
  \citenamefont {Urban}, \citenamefont {Cao}, \citenamefont {Luo},\ and\
  \citenamefont {Ceder}}]{Huang2016}%
  \BibitemOpen
  \bibfield  {author} {\bibinfo {author} {\bibfnamefont {W.}~\bibnamefont
  {Huang}}, \bibinfo {author} {\bibfnamefont {D.~A.}\ \bibnamefont {Kitchaev}},
  \bibinfo {author} {\bibfnamefont {S.~T.}\ \bibnamefont {Dacek}}, \bibinfo
  {author} {\bibfnamefont {Z.}~\bibnamefont {Rong}}, \bibinfo {author}
  {\bibfnamefont {A.}~\bibnamefont {Urban}}, \bibinfo {author} {\bibfnamefont
  {S.}~\bibnamefont {Cao}}, \bibinfo {author} {\bibfnamefont {C.}~\bibnamefont
  {Luo}},\ and\ \bibinfo {author} {\bibfnamefont {G.}~\bibnamefont {Ceder}},\
  }\bibfield  {title} {\bibinfo {title} {Finding and proving the exact ground
  state of a generalized {Ising} model by convex optimization and
  {MAX}-{SAT}},\ }\href {https://doi.org/10.1103/PhysRevB.94.134424} {\bibfield
   {journal} {\bibinfo  {journal} {Physical Review B}\ }\textbf {\bibinfo
  {volume} {94}},\ \bibinfo {pages} {134424} (\bibinfo {year}
  {2016})}\BibitemShut {NoStop}%
\bibitem [{\citenamefont {Vanderstraeten}\ \emph {et~al.}(2018)\citenamefont
  {Vanderstraeten}, \citenamefont {Vanhecke},\ and\ \citenamefont
  {Verstraete}}]{Vanderstraeten2018}%
  \BibitemOpen
  \bibfield  {author} {\bibinfo {author} {\bibfnamefont {L.}~\bibnamefont
  {Vanderstraeten}}, \bibinfo {author} {\bibfnamefont {B.}~\bibnamefont
  {Vanhecke}},\ and\ \bibinfo {author} {\bibfnamefont {F.}~\bibnamefont
  {Verstraete}},\ }\bibfield  {title} {\bibinfo {title} {Residual entropies for
  three-dimensional frustrated spin systems with tensor networks},\ }\href
  {https://doi.org/10.1103/PhysRevE.98.042145} {\bibfield  {journal} {\bibinfo
  {journal} {Phys. Rev. E}\ }\textbf {\bibinfo {volume} {98}},\ \bibinfo
  {pages} {042145} (\bibinfo {year} {2018})}\BibitemShut {NoStop}%
\bibitem [{\citenamefont {Vanhecke}\ \emph
  {et~al.}(2021{\natexlab{a}})\citenamefont {Vanhecke}, \citenamefont
  {Colbois}, \citenamefont {Vanderstraeten}, \citenamefont {Verstraete},\ and\
  \citenamefont {Mila}}]{Vanhecke2021}%
  \BibitemOpen
  \bibfield  {author} {\bibinfo {author} {\bibfnamefont {B.}~\bibnamefont
  {Vanhecke}}, \bibinfo {author} {\bibfnamefont {J.}~\bibnamefont {Colbois}},
  \bibinfo {author} {\bibfnamefont {L.}~\bibnamefont {Vanderstraeten}},
  \bibinfo {author} {\bibfnamefont {F.}~\bibnamefont {Verstraete}},\ and\
  \bibinfo {author} {\bibfnamefont {F.}~\bibnamefont {Mila}},\ }\bibfield
  {title} {\bibinfo {title} {Solving frustrated {I}sing models using tensor
  networks},\ }\href {https://doi.org/10.1103/PhysRevResearch.3.013041}
  {\bibfield  {journal} {\bibinfo  {journal} {Phys. Rev. Research}\ }\textbf
  {\bibinfo {volume} {3}},\ \bibinfo {pages} {013041} (\bibinfo {year}
  {2021}{\natexlab{a}})}\BibitemShut {NoStop}%
\bibitem [{\citenamefont {Cugliandolo}\ \emph {et~al.}(2020)\citenamefont
  {Cugliandolo}, \citenamefont {Foini},\ and\ \citenamefont
  {Tarzia}}]{Cugliandolo2020}%
  \BibitemOpen
  \bibfield  {author} {\bibinfo {author} {\bibfnamefont {L.~F.}\ \bibnamefont
  {Cugliandolo}}, \bibinfo {author} {\bibfnamefont {L.}~\bibnamefont {Foini}},\
  and\ \bibinfo {author} {\bibfnamefont {M.}~\bibnamefont {Tarzia}},\
  }\bibfield  {title} {\bibinfo {title} {Mean-field phase diagram and
  spin-glass phase of the dipolar kagome {I}sing antiferromagnet},\ }\href
  {https://doi.org/10.1103/PhysRevB.101.144413} {\bibfield  {journal} {\bibinfo
   {journal} {Phys. Rev. B}\ }\textbf {\bibinfo {volume} {101}},\ \bibinfo
  {pages} {144413} (\bibinfo {year} {2020})}\BibitemShut {NoStop}%
\bibitem [{\citenamefont {Mizoguchi}\ \emph {et~al.}(2017)\citenamefont
  {Mizoguchi}, \citenamefont {Jaubert},\ and\ \citenamefont
  {Udagawa}}]{Mizoguchi2017}%
  \BibitemOpen
  \bibfield  {author} {\bibinfo {author} {\bibfnamefont {T.}~\bibnamefont
  {Mizoguchi}}, \bibinfo {author} {\bibfnamefont {L.~D.~C.}\ \bibnamefont
  {Jaubert}},\ and\ \bibinfo {author} {\bibfnamefont {M.}~\bibnamefont
  {Udagawa}},\ }\bibfield  {title} {\bibinfo {title} {Clustering of topological
  charges in a kagome classical spin liquid},\ }\href
  {https://doi.org/10.1103/PhysRevLett.119.077207} {\bibfield  {journal}
  {\bibinfo  {journal} {Phys. Rev. Lett.}\ }\textbf {\bibinfo {volume} {119}},\
  \bibinfo {pages} {077207} (\bibinfo {year} {2017})}\BibitemShut {NoStop}%
\bibitem [{\citenamefont {Wolf}\ and\ \citenamefont
  {Schotte}(1988)}]{Wolf1988}%
  \BibitemOpen
  \bibfield  {author} {\bibinfo {author} {\bibfnamefont {M.}~\bibnamefont
  {Wolf}}\ and\ \bibinfo {author} {\bibfnamefont {K.~D.}\ \bibnamefont
  {Schotte}},\ }\bibfield  {title} {\bibinfo {title} {{I}sing model with
  competing next-nearest-neighbour interactions on the kagome lattice},\ }\href
  {https://doi.org/10.1088/0305-4470/21/9/032} {\bibfield  {journal} {\bibinfo
  {journal} {J. Phys. A: Math. Gen.}\ }\textbf {\bibinfo {volume} {21}},\
  \bibinfo {pages} {2195} (\bibinfo {year} {1988})}\BibitemShut {NoStop}%
\bibitem [{\citenamefont {Tokushuku}\ \emph {et~al.}(2020)\citenamefont
  {Tokushuku}, \citenamefont {Mizoguchi},\ and\ \citenamefont
  {Udagawa}}]{Tokushuku2020}%
  \BibitemOpen
  \bibfield  {author} {\bibinfo {author} {\bibfnamefont {K.}~\bibnamefont
  {Tokushuku}}, \bibinfo {author} {\bibfnamefont {T.}~\bibnamefont
  {Mizoguchi}},\ and\ \bibinfo {author} {\bibfnamefont {M.}~\bibnamefont
  {Udagawa}},\ }\bibfield  {title} {\bibinfo {title} {Field-selective classical
  spin liquid and magnetization plateaus on kagome lattice},\ }\href
  {https://doi.org/10.7566/JPSJ.89.053708} {\bibfield  {journal} {\bibinfo
  {journal} {J. Phys. Soc. Jpn.}\ }\textbf {\bibinfo {volume} {89}},\ \bibinfo
  {pages} {053708} (\bibinfo {year} {2020})},\ \Eprint
  {https://arxiv.org/abs/https://doi.org/10.7566/JPSJ.89.053708}
  {https://doi.org/10.7566/JPSJ.89.053708} \BibitemShut {NoStop}%
\bibitem [{\citenamefont {Mizoguchi}\ \emph {et~al.}(2018)\citenamefont
  {Mizoguchi}, \citenamefont {Jaubert}, \citenamefont {Moessner},\ and\
  \citenamefont {Udagawa}}]{Mizoguchi2018}%
  \BibitemOpen
  \bibfield  {author} {\bibinfo {author} {\bibfnamefont {T.}~\bibnamefont
  {Mizoguchi}}, \bibinfo {author} {\bibfnamefont {L.~D.~C.}\ \bibnamefont
  {Jaubert}}, \bibinfo {author} {\bibfnamefont {R.}~\bibnamefont {Moessner}},\
  and\ \bibinfo {author} {\bibfnamefont {M.}~\bibnamefont {Udagawa}},\
  }\bibfield  {title} {\bibinfo {title} {Magnetic clustering, half-moons, and
  shadow pinch points as signals of a proximate coulomb phase in frustrated
  heisenberg magnets},\ }\href {https://doi.org/10.1103/PhysRevB.98.144446}
  {\bibfield  {journal} {\bibinfo  {journal} {Phys. Rev. B}\ }\textbf {\bibinfo
  {volume} {98}},\ \bibinfo {pages} {144446} (\bibinfo {year}
  {2018})}\BibitemShut {NoStop}%
\bibitem [{\citenamefont {Li}(2021)}]{Li2021}%
  \BibitemOpen
  \bibfield  {author} {\bibinfo {author} {\bibfnamefont {T.}~\bibnamefont
  {Li}},\ }\bibfield  {title} {\bibinfo {title} {A continuous family of fully
  frustrated heisenberg models on the kagome lattice},\ }\href
  {https://doi.org/10.1209/0295-5075/133/47001} {\bibfield  {journal} {\bibinfo
   {journal} {Europhysics Letters}\ }\textbf {\bibinfo {volume} {133}},\
  \bibinfo {pages} {47001} (\bibinfo {year} {2021})}\BibitemShut {NoStop}%
\bibitem [{\citenamefont {Kiese}\ \emph {et~al.}(2022)\citenamefont {Kiese},
  \citenamefont {Ferrari}, \citenamefont {Astrakhantsev}, \citenamefont
  {Niggemann}, \citenamefont {Ghosh}, \citenamefont {Müller}, \citenamefont
  {Thomale}, \citenamefont {Neupert}, \citenamefont {Reuther}, \citenamefont
  {Gingras}, \citenamefont {Trebst},\ and\ \citenamefont {Iqbal}}]{Kiese2022}%
  \BibitemOpen
  \bibfield  {author} {\bibinfo {author} {\bibfnamefont {D.}~\bibnamefont
  {Kiese}}, \bibinfo {author} {\bibfnamefont {F.}~\bibnamefont {Ferrari}},
  \bibinfo {author} {\bibfnamefont {N.}~\bibnamefont {Astrakhantsev}}, \bibinfo
  {author} {\bibfnamefont {N.}~\bibnamefont {Niggemann}}, \bibinfo {author}
  {\bibfnamefont {P.}~\bibnamefont {Ghosh}}, \bibinfo {author} {\bibfnamefont
  {T.}~\bibnamefont {Müller}}, \bibinfo {author} {\bibfnamefont
  {R.}~\bibnamefont {Thomale}}, \bibinfo {author} {\bibfnamefont
  {T.}~\bibnamefont {Neupert}}, \bibinfo {author} {\bibfnamefont
  {J.}~\bibnamefont {Reuther}}, \bibinfo {author} {\bibfnamefont {M.~J.~P.}\
  \bibnamefont {Gingras}}, \bibinfo {author} {\bibfnamefont {S.}~\bibnamefont
  {Trebst}},\ and\ \bibinfo {author} {\bibfnamefont {Y.}~\bibnamefont
  {Iqbal}},\ }\href {https://doi.org/10.48550/ARXIV.2206.00264} {\bibinfo
  {title} {Pinch-points to half-moons and up in the stars: the kagome skymap}}
  (\bibinfo {year} {2022})\BibitemShut {NoStop}%
\bibitem [{\citenamefont {Lugan}\ \emph {et~al.}(2022)\citenamefont {Lugan},
  \citenamefont {Jaubert}, \citenamefont {Udagawa},\ and\ \citenamefont
  {Ralko}}]{Lugan2022}%
  \BibitemOpen
  \bibfield  {author} {\bibinfo {author} {\bibfnamefont {T.}~\bibnamefont
  {Lugan}}, \bibinfo {author} {\bibfnamefont {L.~D.~C.}\ \bibnamefont
  {Jaubert}}, \bibinfo {author} {\bibfnamefont {M.}~\bibnamefont {Udagawa}},\
  and\ \bibinfo {author} {\bibfnamefont {A.}~\bibnamefont {Ralko}},\ }\href
  {https://doi.org/10.48550/ARXIV.2206.00547} {\bibinfo {title} {Schwinger
  boson theory of the j1,j2=j3 kagome antiferromagnet}} (\bibinfo {year}
  {2022})\BibitemShut {NoStop}%
\bibitem [{\citenamefont {van~de Walle}(2000)}]{vanDeWalle2000}%
  \BibitemOpen
  \bibfield  {author} {\bibinfo {author} {\bibfnamefont {A.}~\bibnamefont
  {van~de Walle}},\ }\emph {\bibinfo {title} {The Effect of Lattice Vibrations
  on Substitutional Alloy Thermodynamics}},\ \href
  {https://ceder.berkeley.edu/theses/2000_Axel_Van_De_Walle_Thesis.pdf} {Ph.D.
  thesis},\ \bibinfo  {school} {Massachusetts Institute of Technology}
  (\bibinfo {year} {2000})\BibitemShut {NoStop}%
\bibitem [{\citenamefont {Kudō}\ and\ \citenamefont
  {Katsura}(1976)}]{Kudo1976}%
  \BibitemOpen
  \bibfield  {author} {\bibinfo {author} {\bibfnamefont {T.}~\bibnamefont
  {Kudō}}\ and\ \bibinfo {author} {\bibfnamefont {S.}~\bibnamefont
  {Katsura}},\ }\bibfield  {title} {\bibinfo {title} {{A Method of Determining
  the Orderings of the {I}sing Model with Several Neighbor Interactions under
  the Magnetic Field and Applications to Hexagonal Lattices}},\ }\href
  {https://doi.org/10.1143/PTP.56.435} {\bibfield  {journal} {\bibinfo
  {journal} {Prog. Theor. Phys.}\ }\textbf {\bibinfo {volume} {56}},\ \bibinfo
  {pages} {435} (\bibinfo {year} {1976})}\BibitemShut {NoStop}%
\bibitem [{\citenamefont {Colbois}(2022)}]{Colbois2022}%
  \BibitemOpen
  \bibfield  {author} {\bibinfo {author} {\bibfnamefont {J.}~\bibnamefont
  {Colbois}},\ }\emph {\bibinfo {title} {Tensor network investigation of
  frustrated Ising models}},\ \href {https://doi.org/10.5075/epfl-thesis-9290}
  {Ph.D. thesis},\ \bibinfo  {school} {Ecole polytechnique fédérale de
  Lausanne (EPFL), Switzerland} (\bibinfo {year} {2022})\BibitemShut {NoStop}%
\bibitem [{\citenamefont {Rakala}\ and\ \citenamefont
  {Damle}(2017)}]{Rakala2017}%
  \BibitemOpen
  \bibfield  {author} {\bibinfo {author} {\bibfnamefont {G.}~\bibnamefont
  {Rakala}}\ and\ \bibinfo {author} {\bibfnamefont {K.}~\bibnamefont {Damle}},\
  }\bibfield  {title} {\bibinfo {title} {Cluster algorithms for frustrated
  two-dimensional {I}sing antiferromagnets via dual worm constructions},\
  }\href {https://doi.org/10.1103/PhysRevE.96.023304} {\bibfield  {journal}
  {\bibinfo  {journal} {Phys. Rev. E}\ }\textbf {\bibinfo {volume} {96}},\
  \bibinfo {pages} {023304} (\bibinfo {year} {2017})}\BibitemShut {NoStop}%
\bibitem [{\citenamefont {Domenge}\ \emph {et~al.}(2005)\citenamefont
  {Domenge}, \citenamefont {Sindzingre}, \citenamefont {Lhuillier},\ and\
  \citenamefont {Pierre}}]{Domenge2005}%
  \BibitemOpen
  \bibfield  {author} {\bibinfo {author} {\bibfnamefont {J.-C.}\ \bibnamefont
  {Domenge}}, \bibinfo {author} {\bibfnamefont {P.}~\bibnamefont {Sindzingre}},
  \bibinfo {author} {\bibfnamefont {C.}~\bibnamefont {Lhuillier}},\ and\
  \bibinfo {author} {\bibfnamefont {L.}~\bibnamefont {Pierre}},\ }\bibfield
  {title} {\bibinfo {title} {Twelve sublattice ordered phase in the
  \$\{{J}\}\_\{1\}{\textbackslash}ensuremath\{-\}\{{J}\}\_\{2\}\$ model on the
  kagom{\textbackslash}'e lattice},\ }\href
  {https://doi.org/10.1103/PhysRevB.72.024433} {\bibfield  {journal} {\bibinfo
  {journal} {Physical Review B}\ }\textbf {\bibinfo {volume} {72}},\ \bibinfo
  {pages} {024433} (\bibinfo {year} {2005})},\ \bibinfo {note} {publisher:
  American Physical Society}\BibitemShut {NoStop}%
\bibitem [{\citenamefont {Messio}\ \emph {et~al.}(2011)\citenamefont {Messio},
  \citenamefont {Lhuillier},\ and\ \citenamefont {Misguich}}]{Messio2011}%
  \BibitemOpen
  \bibfield  {author} {\bibinfo {author} {\bibfnamefont {L.}~\bibnamefont
  {Messio}}, \bibinfo {author} {\bibfnamefont {C.}~\bibnamefont {Lhuillier}},\
  and\ \bibinfo {author} {\bibfnamefont {G.}~\bibnamefont {Misguich}},\
  }\bibfield  {title} {\bibinfo {title} {Lattice symmetries and regular
  magnetic orders in classical frustrated antiferromagnets},\ }\href
  {https://doi.org/10.1103/PhysRevB.83.184401} {\bibfield  {journal} {\bibinfo
  {journal} {Physical Review B}\ }\textbf {\bibinfo {volume} {83}},\ \bibinfo
  {pages} {184401} (\bibinfo {year} {2011})},\ \bibinfo {note} {publisher:
  American Physical Society}\BibitemShut {NoStop}%
\bibitem [{\citenamefont {Messio}\ \emph {et~al.}(2012)\citenamefont {Messio},
  \citenamefont {Bernu},\ and\ \citenamefont {Lhuillier}}]{Messio2012}%
  \BibitemOpen
  \bibfield  {author} {\bibinfo {author} {\bibfnamefont {L.}~\bibnamefont
  {Messio}}, \bibinfo {author} {\bibfnamefont {B.}~\bibnamefont {Bernu}},\ and\
  \bibinfo {author} {\bibfnamefont {C.}~\bibnamefont {Lhuillier}},\ }\bibfield
  {title} {\bibinfo {title} {Kagome {Antiferromagnet}: {A} {Chiral}
  {Topological} {Spin} {Liquid}?},\ }\href
  {https://doi.org/10.1103/PhysRevLett.108.207204} {\bibfield  {journal}
  {\bibinfo  {journal} {Physical Review Letters}\ }\textbf {\bibinfo {volume}
  {108}},\ \bibinfo {pages} {207204} (\bibinfo {year} {2012})}\BibitemShut
  {NoStop}%
\bibitem [{\citenamefont {Grison}\ \emph {et~al.}(2020)\citenamefont {Grison},
  \citenamefont {Viot}, \citenamefont {Bernu},\ and\ \citenamefont
  {Messio}}]{Grison2020}%
  \BibitemOpen
  \bibfield  {author} {\bibinfo {author} {\bibfnamefont {V.}~\bibnamefont
  {Grison}}, \bibinfo {author} {\bibfnamefont {P.}~\bibnamefont {Viot}},
  \bibinfo {author} {\bibfnamefont {B.}~\bibnamefont {Bernu}},\ and\ \bibinfo
  {author} {\bibfnamefont {L.}~\bibnamefont {Messio}},\ }\bibfield  {title}
  {\bibinfo {title} {Emergent {Potts} order in the kagome
  \$\{{J}\}\_\{1\}{\textbackslash}ensuremath\{-\}\{{J}\}\_\{3\}\$ {Heisenberg}
  model},\ }\href {https://doi.org/10.1103/PhysRevB.102.214424} {\bibfield
  {journal} {\bibinfo  {journal} {Physical Review B}\ }\textbf {\bibinfo
  {volume} {102}},\ \bibinfo {pages} {214424} (\bibinfo {year} {2020})},\
  \bibinfo {note} {publisher: American Physical Society}\BibitemShut {NoStop}%
\bibitem [{\citenamefont {Kolafa}(2014)}]{Kolafa2014}%
  \BibitemOpen
  \bibfield  {author} {\bibinfo {author} {\bibfnamefont {J.}~\bibnamefont
  {Kolafa}},\ }\bibfield  {title} {\bibinfo {title} {Residual entropy of ices
  and clathrates from {M}onte {C}arlo simulation},\ }\href
  {https://doi.org/10.1063/1.4879061} {\bibfield  {journal} {\bibinfo
  {journal} {J. Chem. Phys.}\ }\textbf {\bibinfo {volume} {140}},\ \bibinfo
  {pages} {204507} (\bibinfo {year} {2014})}\BibitemShut {NoStop}%
\bibitem [{\citenamefont {Romá}\ \emph {et~al.}(2004)\citenamefont {Romá},
  \citenamefont {Nieto}, \citenamefont {Vogel},\ and\ \citenamefont
  {Ramirez-Pastor}}]{Roma2004}%
  \BibitemOpen
  \bibfield  {author} {\bibinfo {author} {\bibfnamefont {F.}~\bibnamefont
  {Romá}}, \bibinfo {author} {\bibfnamefont {F.}~\bibnamefont {Nieto}},
  \bibinfo {author} {\bibfnamefont {E.~E.}\ \bibnamefont {Vogel}},\ and\
  \bibinfo {author} {\bibfnamefont {A.~J.}\ \bibnamefont {Ramirez-Pastor}},\
  }\bibfield  {title} {\bibinfo {title} {Ground-state entropy of $\pm j$
  {I}sing lattices by monte carlo simulations},\ }\href
  {https://doi.org/10.1023/B:JOSS.0000013967.52237.6e} {\bibfield  {journal}
  {\bibinfo  {journal} {Journal of Statistical Physics}\ }\textbf {\bibinfo
  {volume} {114}},\ \bibinfo {pages} {1325} (\bibinfo {year}
  {2004})}\BibitemShut {NoStop}%
\bibitem [{\citenamefont {Okunishi}\ \emph {et~al.}(2022)\citenamefont
  {Okunishi}, \citenamefont {Nishino},\ and\ \citenamefont
  {Ueda}}]{Okunishi2021}%
  \BibitemOpen
  \bibfield  {author} {\bibinfo {author} {\bibfnamefont {K.}~\bibnamefont
  {Okunishi}}, \bibinfo {author} {\bibfnamefont {T.}~\bibnamefont {Nishino}},\
  and\ \bibinfo {author} {\bibfnamefont {H.}~\bibnamefont {Ueda}},\ }\bibfield
  {title} {\bibinfo {title} {Developments in the tensor network {\textemdash}
  from statistical mechanics to quantum entanglement},\ }\bibfield  {journal}
  {\bibinfo  {journal} {Journal of the Physical Society of Japan}\ }\textbf
  {\bibinfo {volume} {91}},\ \href {https://doi.org/10.7566/jpsj.91.062001}
  {10.7566/jpsj.91.062001} (\bibinfo {year} {2022})\BibitemShut {NoStop}%
\bibitem [{\citenamefont {Cirac}\ \emph {et~al.}(2021)\citenamefont {Cirac},
  \citenamefont {P\'erez-Garc\'{\i}a}, \citenamefont {Schuch},\ and\
  \citenamefont {Verstraete}}]{Cirac2021}%
  \BibitemOpen
  \bibfield  {author} {\bibinfo {author} {\bibfnamefont {J.~I.}\ \bibnamefont
  {Cirac}}, \bibinfo {author} {\bibfnamefont {D.}~\bibnamefont
  {P\'erez-Garc\'{\i}a}}, \bibinfo {author} {\bibfnamefont {N.}~\bibnamefont
  {Schuch}},\ and\ \bibinfo {author} {\bibfnamefont {F.}~\bibnamefont
  {Verstraete}},\ }\bibfield  {title} {\bibinfo {title} {Matrix product states
  and projected entangled pair states: Concepts, symmetries, theorems},\ }\href
  {https://doi.org/10.1103/RevModPhys.93.045003} {\bibfield  {journal}
  {\bibinfo  {journal} {Rev. Mod. Phys.}\ }\textbf {\bibinfo {volume} {93}},\
  \bibinfo {pages} {045003} (\bibinfo {year} {2021})}\BibitemShut {NoStop}%
\bibitem [{\citenamefont {Baxter}(1968)}]{Baxter1968}%
  \BibitemOpen
  \bibfield  {author} {\bibinfo {author} {\bibfnamefont {R.~J.}\ \bibnamefont
  {Baxter}},\ }\bibfield  {title} {\bibinfo {title} {Dimers on a rectangular
  lattice},\ }\href {https://doi.org/10.1063/1.1664623} {\bibfield  {journal}
  {\bibinfo  {journal} {J. Math. Phys.}\ }\textbf {\bibinfo {volume} {9}},\
  \bibinfo {pages} {650} (\bibinfo {year} {1968})}\BibitemShut {NoStop}%
\bibitem [{\citenamefont {Fishman}\ \emph {et~al.}(2018)\citenamefont
  {Fishman}, \citenamefont {Vanderstraeten}, \citenamefont {Zauner-Stauber},
  \citenamefont {Haegeman},\ and\ \citenamefont {Verstraete}}]{Fishman2018}%
  \BibitemOpen
  \bibfield  {author} {\bibinfo {author} {\bibfnamefont {M.~T.}\ \bibnamefont
  {Fishman}}, \bibinfo {author} {\bibfnamefont {L.}~\bibnamefont
  {Vanderstraeten}}, \bibinfo {author} {\bibfnamefont {V.}~\bibnamefont
  {Zauner-Stauber}}, \bibinfo {author} {\bibfnamefont {J.}~\bibnamefont
  {Haegeman}},\ and\ \bibinfo {author} {\bibfnamefont {F.}~\bibnamefont
  {Verstraete}},\ }\bibfield  {title} {\bibinfo {title} {Faster methods for
  contracting infinite two-dimensional tensor networks},\ }\href
  {https://doi.org/10.1103/PhysRevB.98.235148} {\bibfield  {journal} {\bibinfo
  {journal} {Phys. Rev. B}\ }\textbf {\bibinfo {volume} {98}},\ \bibinfo
  {pages} {235148} (\bibinfo {year} {2018})}\BibitemShut {NoStop}%
\bibitem [{\citenamefont {Zauner-Stauber}\ \emph {et~al.}(2018)\citenamefont
  {Zauner-Stauber}, \citenamefont {Vanderstraeten}, \citenamefont {Fishman},
  \citenamefont {Verstraete},\ and\ \citenamefont
  {Haegeman}}]{ZaunerStauber2018}%
  \BibitemOpen
  \bibfield  {author} {\bibinfo {author} {\bibfnamefont {V.}~\bibnamefont
  {Zauner-Stauber}}, \bibinfo {author} {\bibfnamefont {L.}~\bibnamefont
  {Vanderstraeten}}, \bibinfo {author} {\bibfnamefont {M.~T.}\ \bibnamefont
  {Fishman}}, \bibinfo {author} {\bibfnamefont {F.}~\bibnamefont
  {Verstraete}},\ and\ \bibinfo {author} {\bibfnamefont {J.}~\bibnamefont
  {Haegeman}},\ }\bibfield  {title} {\bibinfo {title} {Variational optimization
  algorithms for uniform matrix product states},\ }\href
  {https://doi.org/10.1103/PhysRevB.97.045145} {\bibfield  {journal} {\bibinfo
  {journal} {Phys. Rev. B}\ }\textbf {\bibinfo {volume} {97}},\ \bibinfo
  {pages} {045145} (\bibinfo {year} {2018})}\BibitemShut {NoStop}%
\bibitem [{\citenamefont {Nietner}\ \emph {et~al.}(2020)\citenamefont
  {Nietner}, \citenamefont {Vanhecke}, \citenamefont {Verstraete},
  \citenamefont {Eisert},\ and\ \citenamefont {Vanderstraeten}}]{Nietner2020}%
  \BibitemOpen
  \bibfield  {author} {\bibinfo {author} {\bibfnamefont {A.}~\bibnamefont
  {Nietner}}, \bibinfo {author} {\bibfnamefont {B.}~\bibnamefont {Vanhecke}},
  \bibinfo {author} {\bibfnamefont {F.}~\bibnamefont {Verstraete}}, \bibinfo
  {author} {\bibfnamefont {J.}~\bibnamefont {Eisert}},\ and\ \bibinfo {author}
  {\bibfnamefont {L.}~\bibnamefont {Vanderstraeten}},\ }\bibfield  {title}
  {\bibinfo {title} {Efficient variational contraction of two-dimensional
  tensor networks with a non-trivial unit cell},\ }\href
  {https://doi.org/10.22331/q-2020-09-21-328} {\bibfield  {journal} {\bibinfo
  {journal} {{Quantum}}\ }\textbf {\bibinfo {volume} {4}},\ \bibinfo {pages}
  {328} (\bibinfo {year} {2020})}\BibitemShut {NoStop}%
\bibitem [{\citenamefont {Baxter}(1978)}]{Baxter1978}%
  \BibitemOpen
  \bibfield  {author} {\bibinfo {author} {\bibfnamefont {R.~J.}\ \bibnamefont
  {Baxter}},\ }\bibfield  {title} {\bibinfo {title} {Variational approximations
  for square lattice models in statistical mechanics},\ }\href
  {https://doi.org/10.1007/BF01011693} {\bibfield  {journal} {\bibinfo
  {journal} {J. Stat. Phys.}\ }\textbf {\bibinfo {volume} {19}},\ \bibinfo
  {pages} {461} (\bibinfo {year} {1978})}\BibitemShut {NoStop}%
\bibitem [{\citenamefont {Nishino}\ and\ \citenamefont
  {Okunishi}(1996)}]{Nishino1996}%
  \BibitemOpen
  \bibfield  {author} {\bibinfo {author} {\bibfnamefont {T.}~\bibnamefont
  {Nishino}}\ and\ \bibinfo {author} {\bibfnamefont {K.}~\bibnamefont
  {Okunishi}},\ }\bibfield  {title} {\bibinfo {title} {Corner transfer matrix
  renormalization group method},\ }\href {https://doi.org/10.1143/JPSJ.65.891}
  {\bibfield  {journal} {\bibinfo  {journal} {Journal of the Physical Society
  of Japan}\ }\textbf {\bibinfo {volume} {65}},\ \bibinfo {pages} {891}
  (\bibinfo {year} {1996})}\BibitemShut {NoStop}%
\bibitem [{\citenamefont {Levin}\ and\ \citenamefont {Nave}(2007)}]{Levin2007}%
  \BibitemOpen
  \bibfield  {author} {\bibinfo {author} {\bibfnamefont {M.}~\bibnamefont
  {Levin}}\ and\ \bibinfo {author} {\bibfnamefont {C.~P.}\ \bibnamefont
  {Nave}},\ }\bibfield  {title} {\bibinfo {title} {Tensor renormalization group
  approach to two-dimensional classical lattice models},\ }\href
  {https://doi.org/10.1103/PhysRevLett.99.120601} {\bibfield  {journal}
  {\bibinfo  {journal} {Phys. Rev. Lett.}\ }\textbf {\bibinfo {volume} {99}},\
  \bibinfo {pages} {120601} (\bibinfo {year} {2007})}\BibitemShut {NoStop}%
\bibitem [{\citenamefont {Evenbly}\ and\ \citenamefont
  {Vidal}(2015)}]{Evenbly15}%
  \BibitemOpen
  \bibfield  {author} {\bibinfo {author} {\bibfnamefont {G.}~\bibnamefont
  {Evenbly}}\ and\ \bibinfo {author} {\bibfnamefont {G.}~\bibnamefont
  {Vidal}},\ }\bibfield  {title} {\bibinfo {title} {Tensor network
  renormalization},\ }\href {https://doi.org/10.1103/PhysRevLett.115.180405}
  {\bibfield  {journal} {\bibinfo  {journal} {Phys. Rev. Lett.}\ }\textbf
  {\bibinfo {volume} {115}},\ \bibinfo {pages} {180405} (\bibinfo {year}
  {2015})}\BibitemShut {NoStop}%
\bibitem [{\citenamefont {Wang}\ \emph {et~al.}(2014)\citenamefont {Wang},
  \citenamefont {Qin},\ and\ \citenamefont {Zhou}}]{Wang2014}%
  \BibitemOpen
  \bibfield  {author} {\bibinfo {author} {\bibfnamefont {C.}~\bibnamefont
  {Wang}}, \bibinfo {author} {\bibfnamefont {S.-M.}\ \bibnamefont {Qin}},\ and\
  \bibinfo {author} {\bibfnamefont {H.-J.}\ \bibnamefont {Zhou}},\ }\bibfield
  {title} {\bibinfo {title} {Topologically invariant tensor renormalization
  group method for the {Edwards-Anderson} spin glasses model},\ }\href
  {https://doi.org/10.1103/PhysRevB.90.174201} {\bibfield  {journal} {\bibinfo
  {journal} {Phys. Rev. B}\ }\textbf {\bibinfo {volume} {90}},\ \bibinfo
  {pages} {174201} (\bibinfo {year} {2014})}\BibitemShut {NoStop}%
\bibitem [{\citenamefont {Zhu}\ and\ \citenamefont
  {Katzgraber}(2019)}]{zhu2019}%
  \BibitemOpen
  \bibfield  {author} {\bibinfo {author} {\bibfnamefont {Z.}~\bibnamefont
  {Zhu}}\ and\ \bibinfo {author} {\bibfnamefont {H.~G.}\ \bibnamefont
  {Katzgraber}},\ }\href@noop {} {\bibinfo {title} {Do tensor renormalization
  group methods work for frustrated spin systems?}} (\bibinfo {year} {2019}),\
  \Eprint {https://arxiv.org/abs/1903.07721} {arXiv:1903.07721
  [cond-mat.dis-nn]} \BibitemShut {NoStop}%
\bibitem [{\citenamefont {Frías-Pérez}\ \emph {et~al.}(2021)\citenamefont
  {Frías-Pérez}, \citenamefont {Mariën}, \citenamefont {García},
  \citenamefont {Bañuls},\ and\ \citenamefont {Iblisdir}}]{FriasPerez2021}%
  \BibitemOpen
  \bibfield  {author} {\bibinfo {author} {\bibfnamefont {M.}~\bibnamefont
  {Frías-Pérez}}, \bibinfo {author} {\bibfnamefont {M.}~\bibnamefont
  {Mariën}}, \bibinfo {author} {\bibfnamefont {D.~P.}\ \bibnamefont
  {García}}, \bibinfo {author} {\bibfnamefont {M.~C.}\ \bibnamefont
  {Bañuls}},\ and\ \bibinfo {author} {\bibfnamefont {S.}~\bibnamefont
  {Iblisdir}},\ }\href {https://doi.org/10.48550/ARXIV.2104.13264} {\bibinfo
  {title} {Collective monte carlo updates through tensor network
  renormalization}} (\bibinfo {year} {2021})\BibitemShut {NoStop}%
\bibitem [{\citenamefont {Liu}\ \emph {et~al.}(2021)\citenamefont {Liu},
  \citenamefont {Wang},\ and\ \citenamefont {Zhang}}]{Liu2021}%
  \BibitemOpen
  \bibfield  {author} {\bibinfo {author} {\bibfnamefont {J.-G.}\ \bibnamefont
  {Liu}}, \bibinfo {author} {\bibfnamefont {L.}~\bibnamefont {Wang}},\ and\
  \bibinfo {author} {\bibfnamefont {P.}~\bibnamefont {Zhang}},\ }\bibfield
  {title} {\bibinfo {title} {Tropical tensor network for ground states of spin
  glasses},\ }\href {https://doi.org/10.1103/PhysRevLett.126.090506} {\bibfield
   {journal} {\bibinfo  {journal} {Phys. Rev. Lett.}\ }\textbf {\bibinfo
  {volume} {126}},\ \bibinfo {pages} {090506} (\bibinfo {year}
  {2021})}\BibitemShut {NoStop}%
\bibitem [{\citenamefont {Song}\ and\ \citenamefont {Zhang}(2022)}]{Song2022}%
  \BibitemOpen
  \bibfield  {author} {\bibinfo {author} {\bibfnamefont {F.-F.}\ \bibnamefont
  {Song}}\ and\ \bibinfo {author} {\bibfnamefont {G.-M.}\ \bibnamefont
  {Zhang}},\ }\bibfield  {title} {\bibinfo {title} {Tensor network approach to
  the two-dimensional fully frustrated xy model and a chiral ordered phase},\
  }\href {https://doi.org/10.1103/PhysRevB.105.134516} {\bibfield  {journal}
  {\bibinfo  {journal} {Phys. Rev. B}\ }\textbf {\bibinfo {volume} {105}},\
  \bibinfo {pages} {134516} (\bibinfo {year} {2022})}\BibitemShut {NoStop}%
\bibitem [{\citenamefont {Vanderstraeten}\ \emph {et~al.}(2019)\citenamefont
  {Vanderstraeten}, \citenamefont {Haegeman},\ and\ \citenamefont
  {Verstraete}}]{Vanderstraeten2019_Tangent}%
  \BibitemOpen
  \bibfield  {author} {\bibinfo {author} {\bibfnamefont {L.}~\bibnamefont
  {Vanderstraeten}}, \bibinfo {author} {\bibfnamefont {J.}~\bibnamefont
  {Haegeman}},\ and\ \bibinfo {author} {\bibfnamefont {F.}~\bibnamefont
  {Verstraete}},\ }\bibfield  {title} {\bibinfo {title} {{Tangent-space methods
  for uniform matrix product states}},\ }\href
  {https://doi.org/10.21468/SciPostPhysLectNotes.7} {\bibfield  {journal}
  {\bibinfo  {journal} {SciPost Phys. Lect. Notes}\ ,\ \bibinfo {pages} {7}}
  (\bibinfo {year} {2019})}\BibitemShut {NoStop}%
\bibitem [{\citenamefont {Vanhecke}\ \emph
  {et~al.}(2021{\natexlab{b}})\citenamefont {Vanhecke}, \citenamefont {Damme},
  \citenamefont {Haegeman}, \citenamefont {Vanderstraeten},\ and\ \citenamefont
  {Verstraete}}]{Vanhecke2021_SciPost}%
  \BibitemOpen
  \bibfield  {author} {\bibinfo {author} {\bibfnamefont {B.}~\bibnamefont
  {Vanhecke}}, \bibinfo {author} {\bibfnamefont {M.~V.}\ \bibnamefont {Damme}},
  \bibinfo {author} {\bibfnamefont {J.}~\bibnamefont {Haegeman}}, \bibinfo
  {author} {\bibfnamefont {L.}~\bibnamefont {Vanderstraeten}},\ and\ \bibinfo
  {author} {\bibfnamefont {F.}~\bibnamefont {Verstraete}},\ }\bibfield  {title}
  {\bibinfo {title} {{Tangent-space methods for truncating uniform MPS}},\
  }\href {https://doi.org/10.21468/SciPostPhysCore.4.1.004} {\bibfield
  {journal} {\bibinfo  {journal} {SciPost Phys. Core}\ }\textbf {\bibinfo
  {volume} {4}},\ \bibinfo {pages} {004} (\bibinfo {year}
  {2021}{\natexlab{b}})}\BibitemShut {NoStop}%
\bibitem [{\citenamefont {Vanderstraeten}\ \emph {et~al.}(2022)\citenamefont
  {Vanderstraeten}, \citenamefont {Burgelman}, \citenamefont {Ponsioen},
  \citenamefont {Van~Damme}, \citenamefont {Vanhecke}, \citenamefont {Corboz},
  \citenamefont {Haegeman},\ and\ \citenamefont
  {Verstraete}}]{Vanderstraeten2022}%
  \BibitemOpen
  \bibfield  {author} {\bibinfo {author} {\bibfnamefont {L.}~\bibnamefont
  {Vanderstraeten}}, \bibinfo {author} {\bibfnamefont {L.}~\bibnamefont
  {Burgelman}}, \bibinfo {author} {\bibfnamefont {B.}~\bibnamefont {Ponsioen}},
  \bibinfo {author} {\bibfnamefont {M.}~\bibnamefont {Van~Damme}}, \bibinfo
  {author} {\bibfnamefont {B.}~\bibnamefont {Vanhecke}}, \bibinfo {author}
  {\bibfnamefont {P.}~\bibnamefont {Corboz}}, \bibinfo {author} {\bibfnamefont
  {J.}~\bibnamefont {Haegeman}},\ and\ \bibinfo {author} {\bibfnamefont
  {F.}~\bibnamefont {Verstraete}},\ }\bibfield  {title} {\bibinfo {title}
  {Variational methods for contracting projected entangled-pair states},\
  }\href {https://doi.org/10.1103/PhysRevB.105.195140} {\bibfield  {journal}
  {\bibinfo  {journal} {Phys. Rev. B}\ }\textbf {\bibinfo {volume} {105}},\
  \bibinfo {pages} {195140} (\bibinfo {year} {2022})}\BibitemShut {NoStop}%
\bibitem [{\citenamefont {Wills}\ \emph {et~al.}(2002)\citenamefont {Wills},
  \citenamefont {Ballou},\ and\ \citenamefont {Lacroix}}]{Wills2002}%
  \BibitemOpen
  \bibfield  {author} {\bibinfo {author} {\bibfnamefont {A.~S.}\ \bibnamefont
  {Wills}}, \bibinfo {author} {\bibfnamefont {R.}~\bibnamefont {Ballou}},\ and\
  \bibinfo {author} {\bibfnamefont {C.}~\bibnamefont {Lacroix}},\ }\bibfield
  {title} {\bibinfo {title} {Model of localized highly frustrated
  ferromagnetism: The kagom\'e spin ice},\ }\href
  {https://doi.org/10.1103/PhysRevB.66.144407} {\bibfield  {journal} {\bibinfo
  {journal} {Phys. Rev. B}\ }\textbf {\bibinfo {volume} {66}},\ \bibinfo
  {pages} {144407} (\bibinfo {year} {2002})}\BibitemShut {NoStop}%
\bibitem [{\citenamefont {Moessner}\ and\ \citenamefont
  {Sondhi}(2003)}]{Moessner2003}%
  \BibitemOpen
  \bibfield  {author} {\bibinfo {author} {\bibfnamefont {R.}~\bibnamefont
  {Moessner}}\ and\ \bibinfo {author} {\bibfnamefont {S.~L.}\ \bibnamefont
  {Sondhi}},\ }\bibfield  {title} {\bibinfo {title} {Theory of the [111]
  magnetization plateau in spin ice},\ }\href
  {https://doi.org/10.1103/PhysRevB.68.064411} {\bibfield  {journal} {\bibinfo
  {journal} {Phys. Rev. B}\ }\textbf {\bibinfo {volume} {68}},\ \bibinfo
  {pages} {064411} (\bibinfo {year} {2003})}\BibitemShut {NoStop}%
\bibitem [{\citenamefont {Kasteleyn}(1963)}]{Kasteleyn1963}%
  \BibitemOpen
  \bibfield  {author} {\bibinfo {author} {\bibfnamefont {P.~W.}\ \bibnamefont
  {Kasteleyn}},\ }\bibfield  {title} {\bibinfo {title} {Dimer statistics and
  phase transitions},\ }\href {https://doi.org/10.1063/1.1703953} {\bibfield
  {journal} {\bibinfo  {journal} {J. Math. Phys.}\ }\textbf {\bibinfo {volume}
  {4}},\ \bibinfo {pages} {287} (\bibinfo {year} {1963})}\BibitemShut {NoStop}%
\bibitem [{\citenamefont {Fisher}(1966)}]{Fisher1966}%
  \BibitemOpen
  \bibfield  {author} {\bibinfo {author} {\bibfnamefont {M.~E.}\ \bibnamefont
  {Fisher}},\ }\bibfield  {title} {\bibinfo {title} {On the dimer solution of
  planar ising models},\ }\href {https://doi.org/10.1063/1.1704825} {\bibfield
  {journal} {\bibinfo  {journal} {Journal of Mathematical Physics}\ }\textbf
  {\bibinfo {volume} {7}},\ \bibinfo {pages} {1776} (\bibinfo {year} {1966})},\
  \Eprint {https://arxiv.org/abs/https://doi.org/10.1063/1.1704825}
  {https://doi.org/10.1063/1.1704825} \BibitemShut {NoStop}%
\bibitem [{\citenamefont {Blote}\ and\ \citenamefont
  {Hilborst}(1982)}]{Blote1982}%
  \BibitemOpen
  \bibfield  {author} {\bibinfo {author} {\bibfnamefont {H.~W.~J.}\
  \bibnamefont {Blote}}\ and\ \bibinfo {author} {\bibfnamefont {H.~J.}\
  \bibnamefont {Hilborst}},\ }\bibfield  {title} {\bibinfo {title} {Roughening
  transitions and the zero-temperature triangular ising antiferromagnet},\
  }\href {https://doi.org/10.1088/0305-4470/15/11/011} {\bibfield  {journal}
  {\bibinfo  {journal} {J. Phys. A: Math. Gen.}\ }\textbf {\bibinfo {volume}
  {15}},\ \bibinfo {pages} {L631} (\bibinfo {year} {1982})}\BibitemShut
  {NoStop}%
\bibitem [{\citenamefont {Nienhuis}\ \emph {et~al.}(1984)\citenamefont
  {Nienhuis}, \citenamefont {Hilhorst},\ and\ \citenamefont
  {Blote}}]{Nienhuis1984}%
  \BibitemOpen
  \bibfield  {author} {\bibinfo {author} {\bibfnamefont {B.}~\bibnamefont
  {Nienhuis}}, \bibinfo {author} {\bibfnamefont {H.~J.}\ \bibnamefont
  {Hilhorst}},\ and\ \bibinfo {author} {\bibfnamefont {H.~W.~J.}\ \bibnamefont
  {Blote}},\ }\bibfield  {title} {\bibinfo {title} {Triangular {SOS} models and
  cubic-crystal shapes},\ }\href {https://doi.org/10.1088/0305-4470/17/18/025}
  {\bibfield  {journal} {\bibinfo  {journal} {J. Phys. A: Math. Gen.}\ }\textbf
  {\bibinfo {volume} {17}},\ \bibinfo {pages} {3559} (\bibinfo {year}
  {1984})}\BibitemShut {NoStop}%
\bibitem [{\citenamefont {Stephenson}(1970)}]{Stephenson1970}%
  \BibitemOpen
  \bibfield  {author} {\bibinfo {author} {\bibfnamefont {J.}~\bibnamefont
  {Stephenson}},\ }\bibfield  {title} {\bibinfo {title} {{I}sing‐model spin
  correlations on the triangular lattice. {III}. {I}sotropic antiferromagnetic
  lattice},\ }\href {https://doi.org/10.1063/1.1665154} {\bibfield  {journal}
  {\bibinfo  {journal} {J. Math. Phys.}\ }\textbf {\bibinfo {volume} {11}},\
  \bibinfo {pages} {413} (\bibinfo {year} {1970})}\BibitemShut {NoStop}%
\bibitem [{\citenamefont {Moessner}\ \emph {et~al.}(2000)\citenamefont
  {Moessner}, \citenamefont {Sondhi},\ and\ \citenamefont
  {Chandra}}]{Moessner2000}%
  \BibitemOpen
  \bibfield  {author} {\bibinfo {author} {\bibfnamefont {R.}~\bibnamefont
  {Moessner}}, \bibinfo {author} {\bibfnamefont {S.~L.}\ \bibnamefont
  {Sondhi}},\ and\ \bibinfo {author} {\bibfnamefont {P.}~\bibnamefont
  {Chandra}},\ }\bibfield  {title} {\bibinfo {title} {Two-dimensional periodic
  frustrated {I}sing models in a transverse field},\ }\href
  {https://doi.org/10.1103/PhysRevLett.84.4457} {\bibfield  {journal} {\bibinfo
   {journal} {Phys. Rev. Lett.}\ }\textbf {\bibinfo {volume} {84}},\ \bibinfo
  {pages} {4457} (\bibinfo {year} {2000})}\BibitemShut {NoStop}%
\bibitem [{\citenamefont {Yokoi}\ \emph {et~al.}(1986)\citenamefont {Yokoi},
  \citenamefont {Nagle},\ and\ \citenamefont {Salinas}}]{Yokoi1986}%
  \BibitemOpen
  \bibfield  {author} {\bibinfo {author} {\bibfnamefont {C.~S.~O.}\
  \bibnamefont {Yokoi}}, \bibinfo {author} {\bibfnamefont {J.~F.}\ \bibnamefont
  {Nagle}},\ and\ \bibinfo {author} {\bibfnamefont {S.~R.}\ \bibnamefont
  {Salinas}},\ }\bibfield  {title} {\bibinfo {title} {Dimer pair correlations
  on the brick lattice},\ }\href {https://doi.org/10.1007/BF01011905}
  {\bibfield  {journal} {\bibinfo  {journal} {J. Stat. Phys.}\ }\textbf
  {\bibinfo {volume} {44}},\ \bibinfo {pages} {729} (\bibinfo {year}
  {1986})}\BibitemShut {NoStop}%
\bibitem [{\citenamefont {Jiang}\ and\ \citenamefont {Emig}(2006)}]{Jiang2006}%
  \BibitemOpen
  \bibfield  {author} {\bibinfo {author} {\bibfnamefont {Y.}~\bibnamefont
  {Jiang}}\ and\ \bibinfo {author} {\bibfnamefont {T.}~\bibnamefont {Emig}},\
  }\bibfield  {title} {\bibinfo {title} {Ordering of geometrically frustrated
  classical and quantum triangular {I}sing magnets},\ }\href
  {https://doi.org/10.1103/PhysRevB.73.104452} {\bibfield  {journal} {\bibinfo
  {journal} {Phys. Rev. B}\ }\textbf {\bibinfo {volume} {73}},\ \bibinfo
  {pages} {104452} (\bibinfo {year} {2006})}\BibitemShut {NoStop}%
\bibitem [{\citenamefont {Vanderstraeten}\ \emph {et~al.}(2015)\citenamefont
  {Vanderstraeten}, \citenamefont {Mari\"en}, \citenamefont {Verstraete},\ and\
  \citenamefont {Haegeman}}]{Vanderstraeten2015}%
  \BibitemOpen
  \bibfield  {author} {\bibinfo {author} {\bibfnamefont {L.}~\bibnamefont
  {Vanderstraeten}}, \bibinfo {author} {\bibfnamefont {M.}~\bibnamefont
  {Mari\"en}}, \bibinfo {author} {\bibfnamefont {F.}~\bibnamefont
  {Verstraete}},\ and\ \bibinfo {author} {\bibfnamefont {J.}~\bibnamefont
  {Haegeman}},\ }\bibfield  {title} {\bibinfo {title} {Excitations and the
  tangent space of projected entangled-pair states},\ }\href
  {https://doi.org/10.1103/PhysRevB.92.201111} {\bibfield  {journal} {\bibinfo
  {journal} {Phys. Rev. B}\ }\textbf {\bibinfo {volume} {92}},\ \bibinfo
  {pages} {201111(R)} (\bibinfo {year} {2015})}\BibitemShut {NoStop}%
\bibitem [{\citenamefont {Corboz}(2016)}]{Corboz2016}%
  \BibitemOpen
  \bibfield  {author} {\bibinfo {author} {\bibfnamefont {P.}~\bibnamefont
  {Corboz}},\ }\bibfield  {title} {\bibinfo {title} {Variational optimization
  with infinite projected entangled-pair states},\ }\href
  {https://doi.org/10.1103/PhysRevB.94.035133} {\bibfield  {journal} {\bibinfo
  {journal} {Phys. Rev. B}\ }\textbf {\bibinfo {volume} {94}},\ \bibinfo
  {pages} {035133} (\bibinfo {year} {2016})}\BibitemShut {NoStop}%
\bibitem [{\citenamefont {Ueda}\ \emph {et~al.}(2005)\citenamefont {Ueda},
  \citenamefont {Otani}, \citenamefont {Nishio}, \citenamefont {Gendiar},\ and\
  \citenamefont {Nishino}}]{Ueda2005}%
  \BibitemOpen
  \bibfield  {author} {\bibinfo {author} {\bibfnamefont {K.}~\bibnamefont
  {Ueda}}, \bibinfo {author} {\bibfnamefont {R.}~\bibnamefont {Otani}},
  \bibinfo {author} {\bibfnamefont {Y.}~\bibnamefont {Nishio}}, \bibinfo
  {author} {\bibfnamefont {A.}~\bibnamefont {Gendiar}},\ and\ \bibinfo {author}
  {\bibfnamefont {T.}~\bibnamefont {Nishino}},\ }\bibfield  {title} {\bibinfo
  {title} {Snapshot observation for 2d classical lattice models by corner
  transfer matrix renormalization group},\ }\href
  {https://doi.org/10.1143/JPSJS.74S.111} {\bibfield  {journal} {\bibinfo
  {journal} {J. Phys. Soc. Jpn}\ }\textbf {\bibinfo {volume} {74}},\ \bibinfo
  {pages} {111} (\bibinfo {year} {2005})},\ \Eprint
  {https://arxiv.org/abs/https://doi.org/10.1143/JPSJS.74S.111}
  {https://doi.org/10.1143/JPSJS.74S.111} \BibitemShut {NoStop}%
\bibitem [{\citenamefont {Sch\'anilec}\ \emph {et~al.}(2020)\citenamefont
  {Sch\'anilec}, \citenamefont {Canals}, \citenamefont
  {Uhl\'{\i}\ifmmode~\check{r}\else \v{r}\fi{}}, \citenamefont
  {Flaj\ifmmode~\check{s}\else \v{s}\fi{}man}, \citenamefont {Sad\'{\i}lek},
  \citenamefont {\ifmmode~\check{S}\else \v{S}\fi{}ikola},\ and\ \citenamefont
  {Rougemaille}}]{Shcanilec2020}%
  \BibitemOpen
  \bibfield  {author} {\bibinfo {author} {\bibfnamefont {V.}~\bibnamefont
  {Sch\'anilec}}, \bibinfo {author} {\bibfnamefont {B.}~\bibnamefont {Canals}},
  \bibinfo {author} {\bibfnamefont {V.}~\bibnamefont
  {Uhl\'{\i}\ifmmode~\check{r}\else \v{r}\fi{}}}, \bibinfo {author}
  {\bibfnamefont {L.}~\bibnamefont {Flaj\ifmmode~\check{s}\else
  \v{s}\fi{}man}}, \bibinfo {author} {\bibfnamefont {J.}~\bibnamefont
  {Sad\'{\i}lek}}, \bibinfo {author} {\bibfnamefont {T.}~\bibnamefont
  {\ifmmode~\check{S}\else \v{S}\fi{}ikola}},\ and\ \bibinfo {author}
  {\bibfnamefont {N.}~\bibnamefont {Rougemaille}},\ }\bibfield  {title}
  {\bibinfo {title} {Bypassing dynamical freezing in artificial kagome ice},\
  }\href {https://doi.org/10.1103/PhysRevLett.125.057203} {\bibfield  {journal}
  {\bibinfo  {journal} {Phys. Rev. Lett.}\ }\textbf {\bibinfo {volume} {125}},\
  \bibinfo {pages} {057203} (\bibinfo {year} {2020})}\BibitemShut {NoStop}%
\bibitem [{\citenamefont {Hofhuis}\ \emph {et~al.}(2020)\citenamefont
  {Hofhuis}, \citenamefont {Hrabec}, \citenamefont {Arava}, \citenamefont
  {Leo}, \citenamefont {Huang}, \citenamefont {Chopdekar}, \citenamefont
  {Parchenko}, \citenamefont {Kleibert}, \citenamefont {Koraltan},
  \citenamefont {Abert}, \citenamefont {Vogler}, \citenamefont {Suess},
  \citenamefont {Derlet},\ and\ \citenamefont {Heyderman}}]{Hofhuis2020}%
  \BibitemOpen
  \bibfield  {author} {\bibinfo {author} {\bibfnamefont {K.}~\bibnamefont
  {Hofhuis}}, \bibinfo {author} {\bibfnamefont {A.~c.~v.}\ \bibnamefont
  {Hrabec}}, \bibinfo {author} {\bibfnamefont {H.}~\bibnamefont {Arava}},
  \bibinfo {author} {\bibfnamefont {N.}~\bibnamefont {Leo}}, \bibinfo {author}
  {\bibfnamefont {Y.-L.}\ \bibnamefont {Huang}}, \bibinfo {author}
  {\bibfnamefont {R.~V.}\ \bibnamefont {Chopdekar}}, \bibinfo {author}
  {\bibfnamefont {S.}~\bibnamefont {Parchenko}}, \bibinfo {author}
  {\bibfnamefont {A.}~\bibnamefont {Kleibert}}, \bibinfo {author}
  {\bibfnamefont {S.}~\bibnamefont {Koraltan}}, \bibinfo {author}
  {\bibfnamefont {C.}~\bibnamefont {Abert}}, \bibinfo {author} {\bibfnamefont
  {C.}~\bibnamefont {Vogler}}, \bibinfo {author} {\bibfnamefont
  {D.}~\bibnamefont {Suess}}, \bibinfo {author} {\bibfnamefont {P.~M.}\
  \bibnamefont {Derlet}},\ and\ \bibinfo {author} {\bibfnamefont {L.~J.}\
  \bibnamefont {Heyderman}},\ }\bibfield  {title} {\bibinfo {title} {Thermally
  superactive artificial kagome spin ice structures obtained with the
  interfacial dzyaloshinskii-moriya interaction},\ }\href
  {https://doi.org/10.1103/PhysRevB.102.180405} {\bibfield  {journal} {\bibinfo
   {journal} {Phys. Rev. B}\ }\textbf {\bibinfo {volume} {102}},\ \bibinfo
  {pages} {180405(R)} (\bibinfo {year} {2020})}\BibitemShut {NoStop}%
\bibitem [{\citenamefont {Hofhuis}\ \emph {et~al.}(2022)\citenamefont
  {Hofhuis}, \citenamefont {Skjærvø}, \citenamefont {Parchenko},
  \citenamefont {Arava}, \citenamefont {Luo}, \citenamefont {Kleibert},
  \citenamefont {Derlet},\ and\ \citenamefont {Heyderman}}]{Hofhuis2022}%
  \BibitemOpen
  \bibfield  {author} {\bibinfo {author} {\bibfnamefont {K.}~\bibnamefont
  {Hofhuis}}, \bibinfo {author} {\bibfnamefont {S.~H.}\ \bibnamefont
  {Skjærvø}}, \bibinfo {author} {\bibfnamefont {S.}~\bibnamefont
  {Parchenko}}, \bibinfo {author} {\bibfnamefont {H.}~\bibnamefont {Arava}},
  \bibinfo {author} {\bibfnamefont {Z.}~\bibnamefont {Luo}}, \bibinfo {author}
  {\bibfnamefont {A.}~\bibnamefont {Kleibert}}, \bibinfo {author}
  {\bibfnamefont {P.~M.}\ \bibnamefont {Derlet}},\ and\ \bibinfo {author}
  {\bibfnamefont {L.~J.}\ \bibnamefont {Heyderman}},\ }\bibfield  {title}
  {\bibinfo {title} {Real-space imaging of phase transitions in bridged
  artificial kagome spin ice},\ }\href
  {https://doi.org/10.1038/s41567-022-01564-5} {\bibfield  {journal} {\bibinfo
  {journal} {Nature Physics}\ }\textbf {\bibinfo {volume} {18}},\ \bibinfo
  {pages} {699} (\bibinfo {year} {2022})}\BibitemShut {NoStop}%
\bibitem [{\citenamefont {Houtappel}(1950)}]{Houtappel1950}%
  \BibitemOpen
  \bibfield  {author} {\bibinfo {author} {\bibfnamefont {R.~M.~F.}\
  \bibnamefont {Houtappel}},\ }\bibfield  {title} {\bibinfo {title}
  {Order-disorder in hexagonal lattices},\ }\href
  {https://doi.org/10.1016/0031-8914(50)90130-3} {\bibfield  {journal}
  {\bibinfo  {journal} {Physica}\ }\textbf {\bibinfo {volume} {16}},\ \bibinfo
  {pages} {425} (\bibinfo {year} {1950})}\BibitemShut {NoStop}%
\end{thebibliography}%


%apsrev4-2.bst 2019-01-14 (MD) hand-edited version of apsrev4-1.bst
%Control: key (0)
%Control: author (8) initials jnrlst
%Control: editor formatted (1) identically to author
%Control: production of article title (0) allowed
%Control: page (0) single
%Control: year (1) truncated
%Control: production of eprint (0) enabled
%

\appendix
\section{Analytical expressions for the residual entropies of the nearest-neighbors models}
\label{sec:AppClosedForm}
\RESP{Following in the steps of the pioneering work by Onsager and Kaufman}~\cite{Onsager1944,Kaufman1949}, \RESP{a series of papers in the 1950s and 1960s established analytical results for planar Ising models} (see~\cite{Fisher1966}). \RESP{In particular, expressions for the residual entropy of frustrated nearest-neighbor Ising models} (Eqs.~\ref{eq:STIAFM},\ref{eq:SKIAFM})\RESP{were obtained by Wannier}~\cite{Wannier1950,Wannier1973}\RESP{and Houtappel}~\cite{Houtappel1950} \RESP{for the triangular lattice:}
\begin{equation}
    S = \frac{2}{\pi}\int_{0}^{\pi/3} \ln\left(2\cos{\omega}\right) \, \d \omega
\end{equation}
and by Kano and Naya~\cite{Kano1953} for the kagome lattice:
\begin{multline}
    S = \frac{1}{24 \pi^2} \int_0^{2\pi}\int_0^{2\pi} \ln\left[ 21 \right.\\
    \left. - 4 \left(\cos{\omega_1} + \cos{\omega_2} + \cos{\omega_1 + \omega_2} \right) \right] \, \d \omega_1 \,\d \omega_2.
\end{multline}

\section{Spin-spin correlations in the pinwheels phase}
\label{sec:PINCORR}
\begin{figure} 
	\centering
	\includegraphics[width = 0.5\columnwidth]{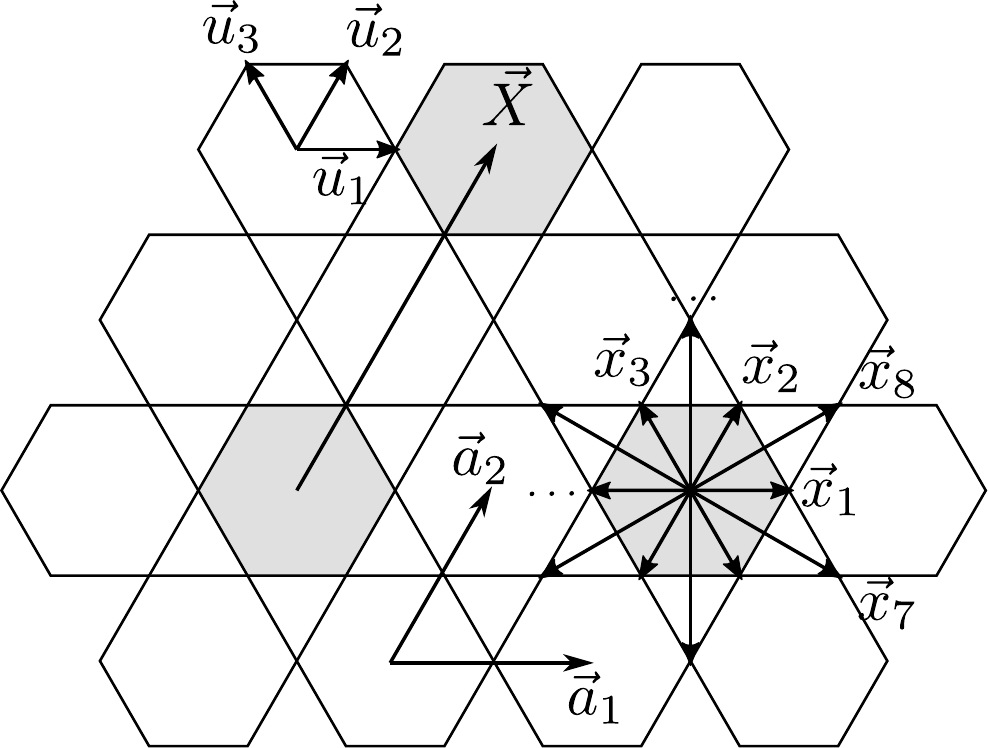}
	\caption[Convention for labeling the sites of kagome to discuss the correlations in the pinwheels phase.]{\label{fig:PLGS_Convention}Two conventions for labeling the sites of kagome.
	Either we label the hexagon centers ($\vec{R} = R_1 \vec{a}_1 + R_2 \vec{a}_2$ with $R_1$ and $R_2$ integers) and the three sites in the unit cell ($\vec{u}_1$ to $\vec{u}_3$), or we label the pinwheel centers ($\vec{X} = 2n \vec{a}_1 + 2 m\vec{a}_2$ with $n, m$ integers) and the twelve sites in the unit cell ($\vec{x}_1$ to $\vec{x}_{12}$).}
\end{figure}

By describing the spin configurations based on the David stars corresponding to pinwheels, we can show that there is a long-range order for the spins on the hexagons at the center of the pinwheels and an algebraic decay for the spin-spin correlations between spins on the branches of the pinwheels.
In the following, we label by $\vec{r}_i$ the positions of the spins, by $\vec{R}_{\alpha}$ the positions of the hexagon centers, and by $\vec{u}_k$ ($k= 1,2,3$) the positions of the spins in the unit cell, such that $\vec{r}_{k,\alpha} = \vec{R}_{\alpha} +\vec{u}_k$. We take the original kagome lattice spacing to be 1, so the basis vectors of the triangular lattice supporting the kagome lattice are $\vec{a}_1 = (2, 0)$ and $\vec{a}_2 = (1, \sqrt{3})$ (Fig.~\ref{fig:PLGS_Convention}).

\paragraph{Pinwheel centers}
The location of the pinwheels is characterized by the presence of an empty hexagon, which corresponds to two second-neighbor triangles where the ``two-ups one-down /two-downs one-up'' rule is not respected. We can introduce an operator on the hexagons
\begin{equation}
\begin{split}
    P_{\hexagon} = \frac{1}{4}\left(\sum_{\triangle_2 \in \hexagon} (\sigma_i \sigma_j + \sigma_j \sigma_k + \sigma_k \sigma_i) -2\right) \\= \begin{cases}
	1 & \text{ if the hexagon is a pinwheel center }\\
	-1 & \text{ otherwise}
	\end{cases},
\end{split}
\end{equation}
or equivalently, in terms of dimers,
\begin{equation}
	P_{\hexagon} = \frac{1}{4}\left(\sum_{i = 0}^5 (d_i d_{i+1}) -2\right),
\end{equation}
where $i$ goes through the dimers bonds touching the hexagon center and where $d_6 = d_0$.
This operator has clearly long-range order in any of the four translation-symmetry-broken sectors of the pinwheel phase, since:
\begin{equation}
	\langle P_{\alpha}P_{\beta} \rangle = \begin{cases}
	1 & \text{ if } (\vec{R}_{\alpha} - \vec{R}_{\beta}) = 2n \vec{a}_1+2m \vec{a}_2\\
	-1 & \text{ otherwise}
	\end{cases}
\end{equation}
Note that the correlations averaged over the four translation-symmetry-broken sectors give zero if $(\vec{R}_{\alpha} - \vec{R}_{\beta}) \neq 2n \vec{a}_1+2m \vec{a}_2$.

We can adjust the notation to work directly in a translation-symmetry-broken sector. Instead of labeling the sites based on $\vec{u}_i$ taking three possible values, we label the sites based on an empty star center $\vec{X}_{\alpha} = 2n \vec{a}_1 + 2m \vec{a}_2$ and a 12-sites motif describing the star $\vec{x}_i$ $i = 1, \dots, 12$. We take the convention that $i = 1, \dots, 6$ describes the hexagon sites, and $i = 7, \dots, 12$ describes the branches of the star (Fig.~\ref{fig:PLGS_Convention}).

\paragraph{Pinwheel correlations}
Only looking at pinwheels in the ground state, since around an empty hexagon the chirality is fixed once we know a single dimer position, we can easily evaluate the chirality as

\begin{equation}
	C_{\alpha} = \sigma_{\vec{X}_{\alpha} + \vec{x}_1} \sigma_{\vec{X}_{\alpha}+ \vec{x}_{8}} =  \begin{cases}
	1 & \text{ if clockwise }\\
	-1 & \text{ otherwise}
	\end{cases}.
\end{equation}

In the ground-state manifold of the TIAFM, the spin-spin correlations decay asymptotically as \cite{Stephenson1970, Smerald2017}
\begin{equation}
	\langle s_i s_j \rangle = \epsilon_0 \frac{\cos\left(\vec{q}\cdot\vec{r}\right)}{\sqrt{|\vec{r}|}} \, , \quad \vec{r} = \vec{r}_j - \vec{r}_i,
\end{equation}
where we denoted by $s_i$ the effective (emergent) Ising spins, and where the structure factor is characterized by $\vec{q} = (\pm \frac{2 \pi}{3} , \frac{2\pi}{\sqrt{3}})$.  It has been argued using numerical evidence \cite{Stephenson1970} that along a row of the triangular lattice, the proportionality factor $\epsilon_0$ related to the decay amplitude $E_0^T$ of the pair correlations at the critical point of the triangular \textit{ferromagnet} as 
\begin{equation}
	\epsilon_0 = \sqrt{2}(E_0^T)^2 \cong 0.632226080...
\end{equation}
(This can be checked directly using a tensor network contraction).

\par This implies that in the pinwheels ground-state manifold, in a given symmetry-broken sector, the pinwheel chirality correlations decay as
\begin{equation}
	\langle C_{\alpha} C_{\beta}\rangle = \epsilon_0 \delta_{P_{\alpha}, 1} \delta_{P_{\beta}, 1} \frac{\cos\left(\frac{\vec{q} \cdot \vec{X}}{4}\right)}{\sqrt{|\vec{X}|/4}} \, , \quad \vec{X} = \vec{X}_{\beta} - \vec{X}_{\alpha},
\end{equation}
where we used  $\vec{X}_{\beta} - \vec{X}_{\alpha}  = 2n \vec{a}_1+2m \vec{a}_2$.

\paragraph{Partial long-range order and critical correlations}
 The pinwheels correlations imply spin-spin correlations which we now explore, and summarize in Fig.~\ref{fig:pinwheelscorr}. An empty hexagon corresponding to a pinwheel center has two possible associated spin configurations for a given chirality, corresponding to having a spin up or a spin down on $\vec{u}_1$. We now argue that, together with the long-range order in the pinwheel center locations, there comes a long-range order in the spins living on the empty hexagons. First, fixing the spin on $\vec{u}_1$ in a given empty hexagon (center of a pinwheel) fixes all the other spins on that hexagon. Second, one can see that a spin $\sigma_1$ in $\vec{r}_1 = \vec{X}_{\alpha} + \vec{u}_i$, and another $\vec{\sigma}_2$ in $\vec{r}_2 = \vec{X}_{\alpha} + \vec{a}_i+ \vec{u}_i$, where $i = 1,2,3$ and $\vec{a}_3 = -\vec{a}_1 + \vec{a}_2$, must have opposite values $\sigma_2 = - \sigma_1$. Indeed, they are separated by a hexagon bearing either a cross or a chevron, and the path connecting them must cross either the two branches of the cross or chevron, or no branches. Thus we must have $\sigma_3 = \sigma_1$ if $\vec{r}_3 = \vec{X}_{\alpha} + 2\vec{a}_i + \vec{u}_i$. 

\par This implies that two nearest-neighbor empty hexagons have the same spin configuration. Therefore, by simple extension,
\begin{equation}
\begin{split}
	\sigma_{\vec{X}_{\beta}+\vec{x}_j}\sigma_{\vec{X}_{\alpha} + \vec{x}_i} &=\begin{cases}
	1 &\text{ if } \text{mod}(i,2) = \text{mod}(j,2)\\
	-1 &\text{ otherwise } 
	\end{cases}, \\
	&\quad i, j = 1, \dots, 6,
\end{split}
\end{equation}
where $\vec{X}_{\beta} = \vec{X}_{\alpha} + 2n \vec{a}_1 + 2m \vec{a}_2$, and the long-range order in the pinwheel centers translates into a long-range order in the spins living on the empty hexagons (Figs.~\ref{fig:pinwheelscorr}a and~\ref{fig:pinwheelscorr}c). 

For the product of spins belonging to the branches of a pinwheel we have
\begin{equation}
\begin{split}
	\sigma_{\vec{X}_{\beta}+ \vec{x}_j} \sigma_{\vec{X}_{\alpha} + \vec{x}_i} &= \begin{cases}
	C_{\alpha} C_{\beta} &\text{ if } \text{mod}(i,2) = \text{mod}(j,2)\\
	- C_{\alpha} C_{\beta} &\text{ otherwise } 
	\end{cases}, \\
	&\quad   i, j = 7, \dots, 12,
\end{split}
\end{equation}
where $\vec{X}_{\beta} = \vec{X}_{\alpha} + 2n \vec{a}_1 + 2m \vec{a}_2$. Therefore, the algebraic decay of the spin-spin correlations in the TIAFM must translate into an algebraic decay of the spin-spin correlations for spins on the branches of the pinwheels in the pinwheels phase (Figs.~\ref{fig:pinwheelscorr}b and~\ref{fig:pinwheelscorr}d).

Finally, 
\begin{equation}
\begin{split}
	\sigma_{\vec{X}_{\beta}+ \vec{x}_j} \sigma_{\vec{X}_{\alpha} + \vec{x}_i} &= \begin{cases}
	-C_{\beta} &\text{ if } \text{mod}(i,2) = \text{mod}(j,2)\\
	C_{\beta} &\text{ otherwise } 
	\end{cases},\\
	&\quad   i = 1, ..., 6 \quad j = 7, \dots, 12.
\end{split}
\end{equation}

Since $\langle C_{\beta} \rangle = 0$, these correlations are identically zero in the ground-state (Figs.~\ref{fig:pinwheelscorr}a and~\ref{fig:pinwheelscorr}b). 

\section{Strings phase}
\label{sec:AppStrings}
In this Appendix we give a detailed discussion to explain the residual entropy of the strings phase. 
\subsection{Any string configuration on the honeycomb lattice maps to at least one ground state of the strings phase}

\begin{figure}
    \includegraphics[width = 0.7\columnwidth]{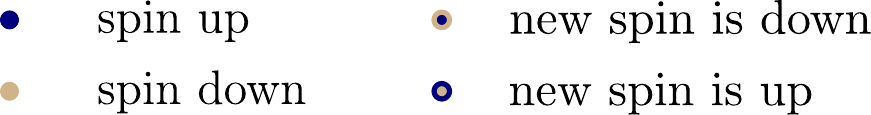}\\
    \subfloat[Reference configuration\label{fig:Strings_Ref}]{\includegraphics[width = 0.47\columnwidth]{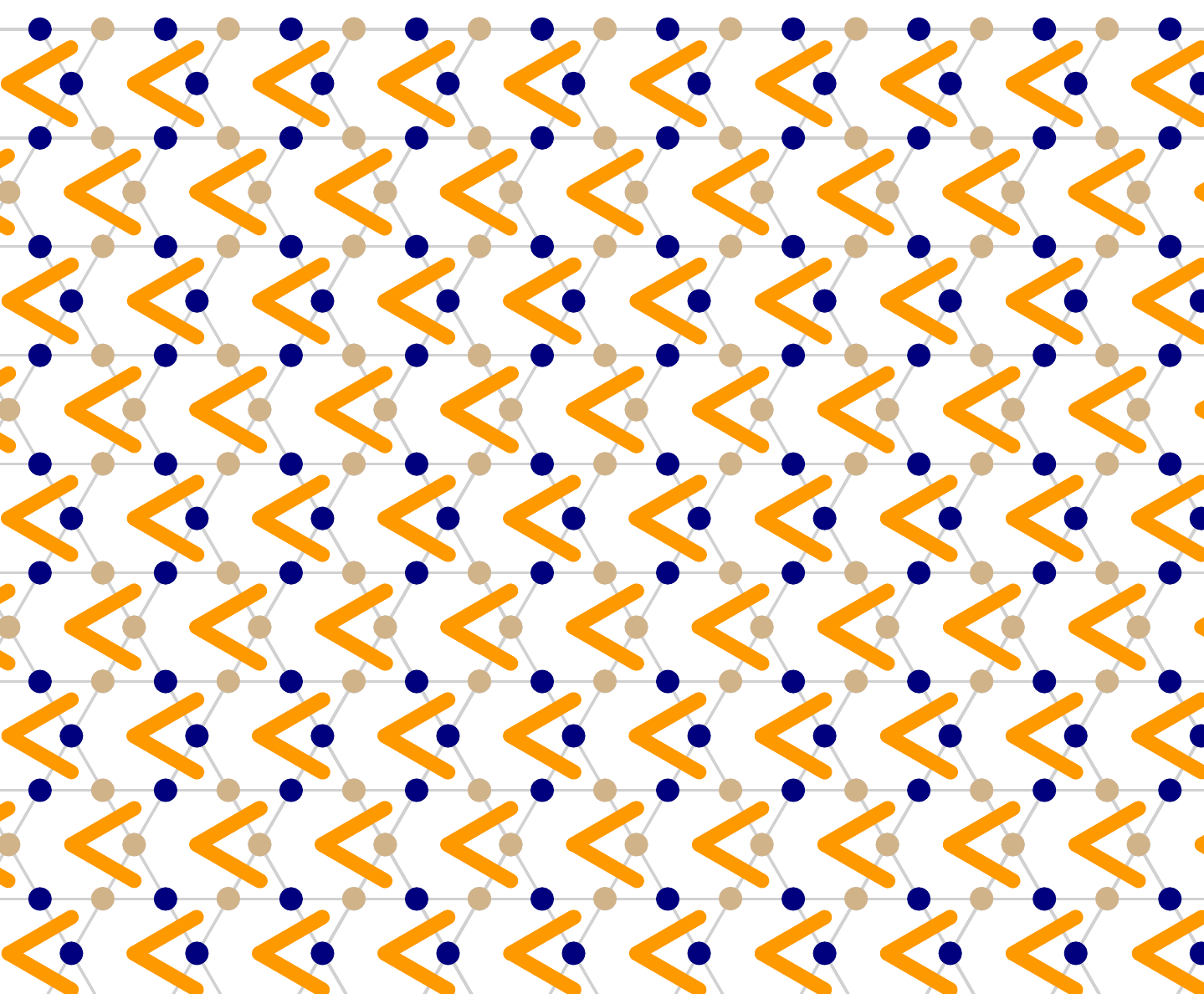}}
    \hfill
    \subfloat[Creating a pair of strings\label{fig:Strings_Pair}]{
    \includegraphics[width = 0.47\columnwidth]{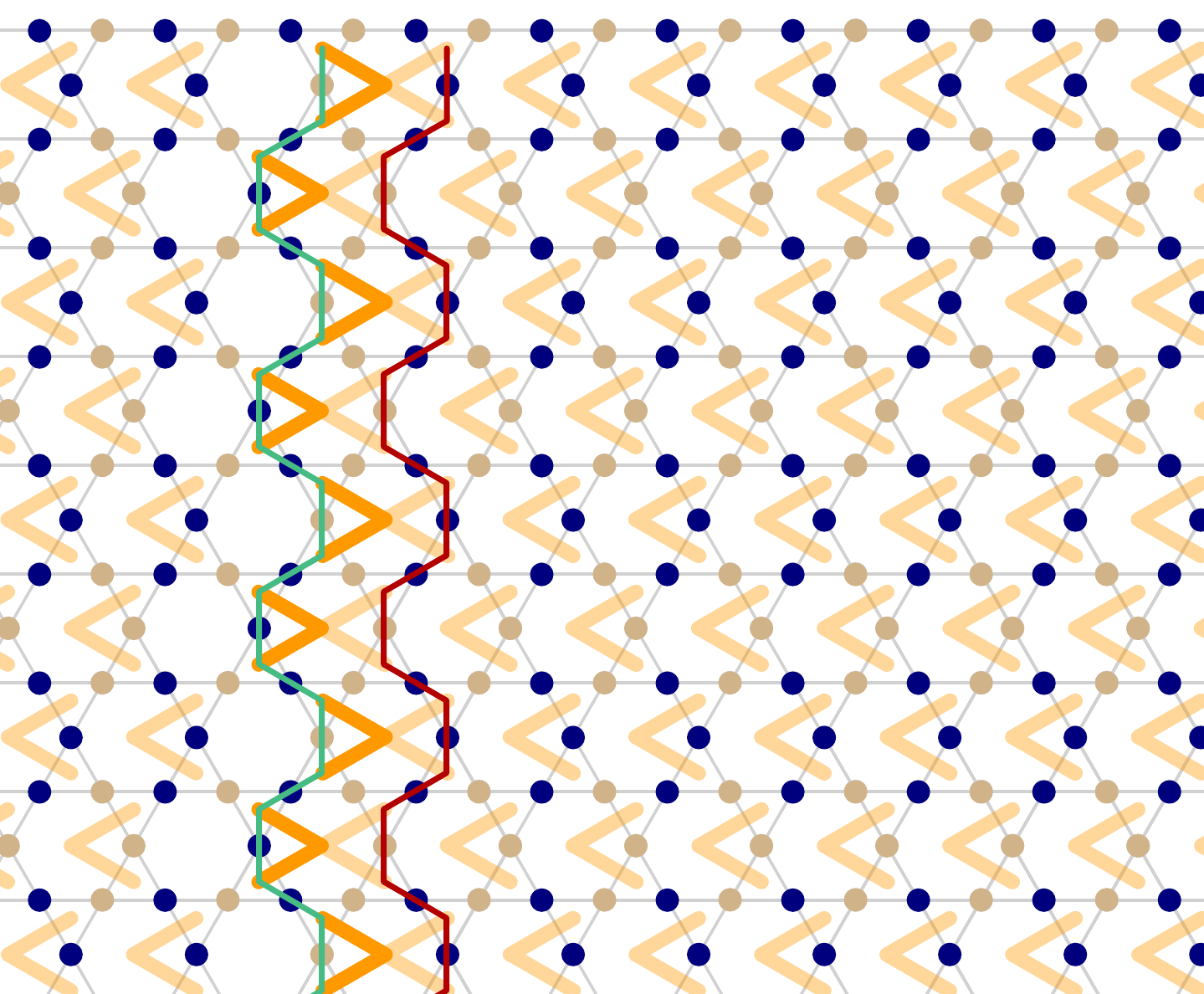}}\\
    \subfloat[Moving the left string to the left\label{fig:Strings_Pair_LeftMove}]{
    \includegraphics[width=0.47\columnwidth]{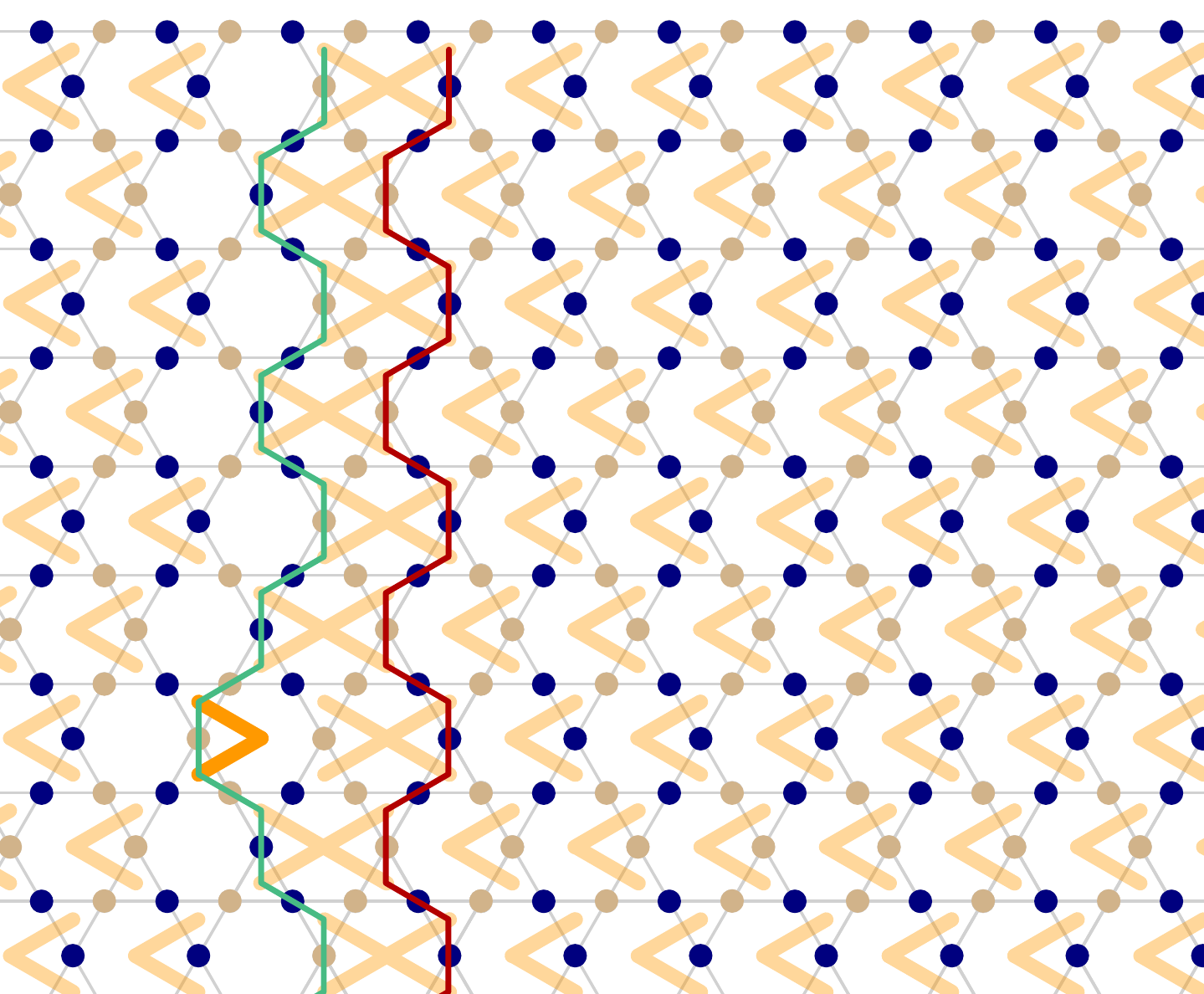}}
    \hfill
    \subfloat[Moving the right string to the right\label{fig:Strings_Pair_RightMove}]{
    \includegraphics[width=0.47\columnwidth]{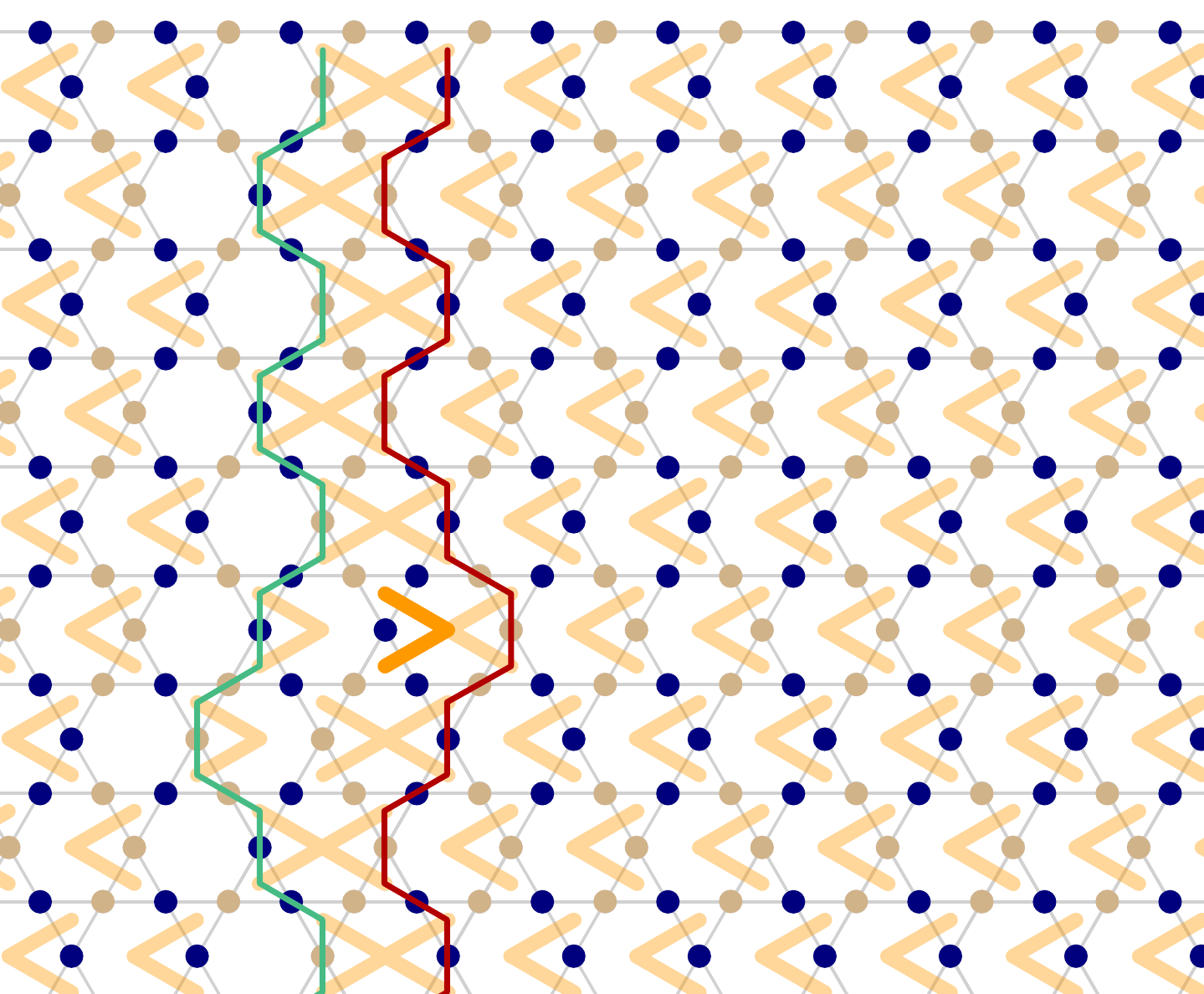}}\\
    \subfloat[A second pair of strings is easily created.\label{fig:Strings_NewPair_GreenRight}]{
    \includegraphics[width=0.47\columnwidth]{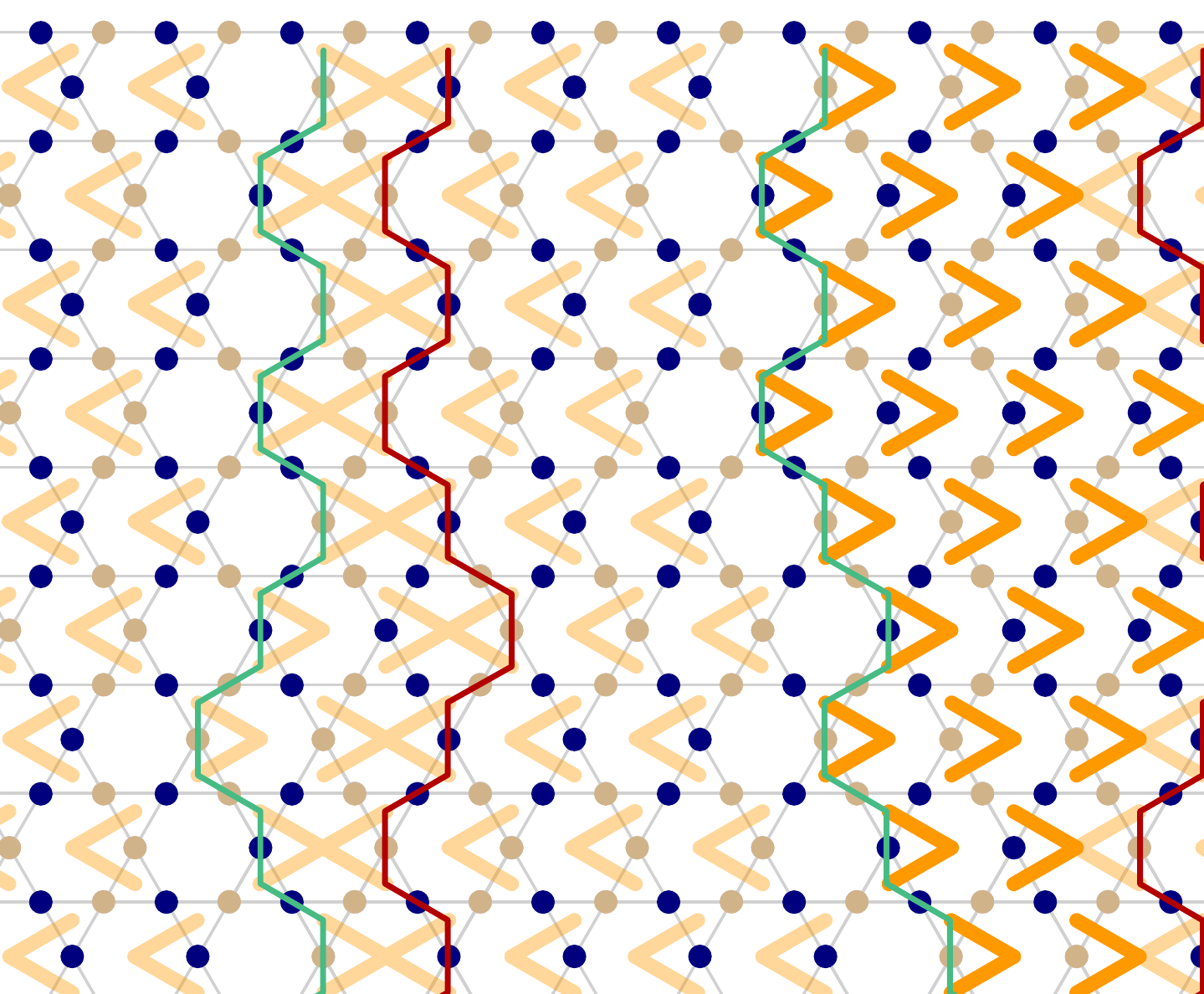}}
    \hfill
    \subfloat[Several configurations map on the same string configuration\label{fig:Strings_NewPair_Subext}]{
    \includegraphics[width=0.47\columnwidth]{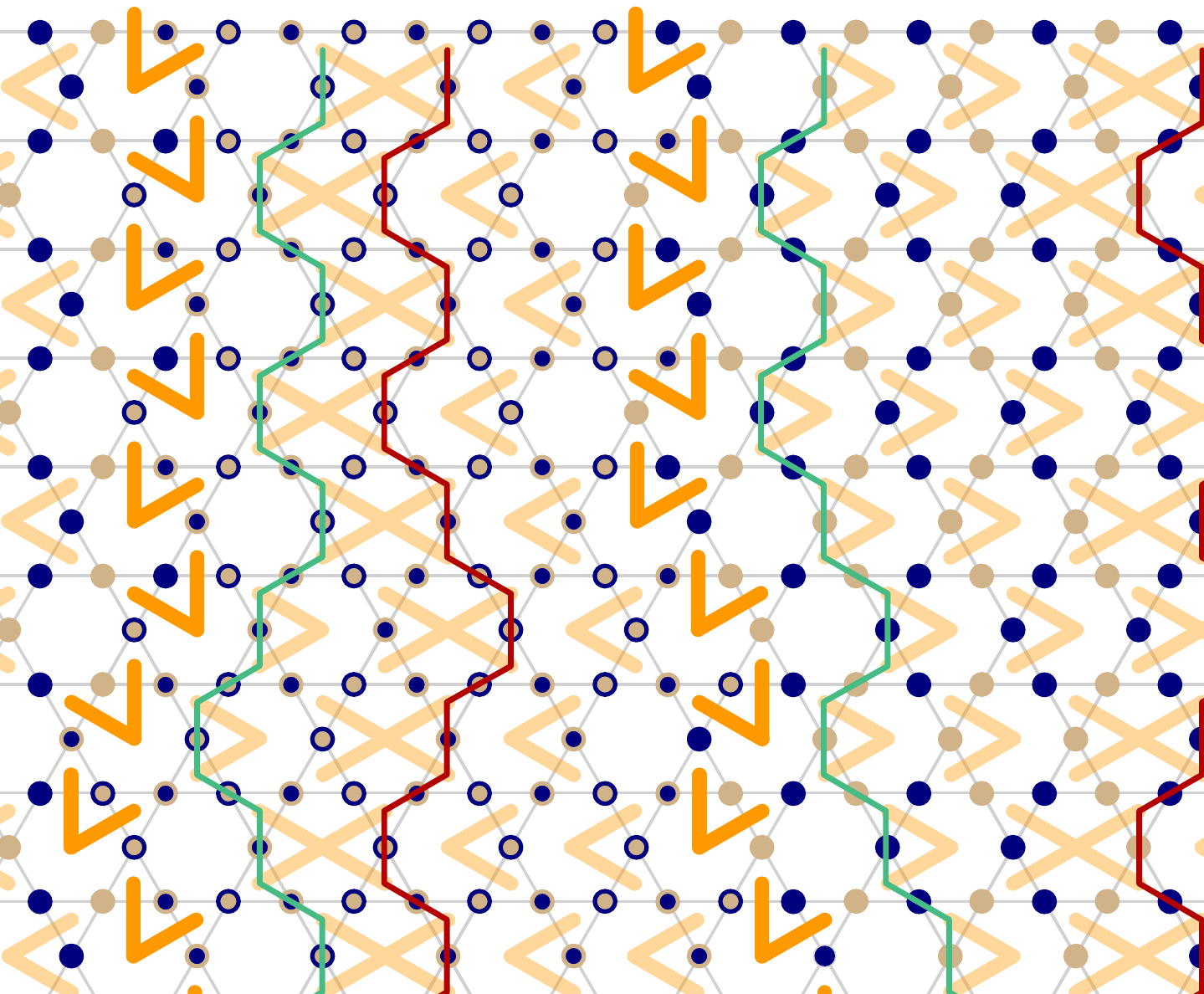}}
     \caption{\label{fig:PLGS_StringsPhaseCreatingAndMovingstrings}Creating and moving strings in the strings phase. The dark orange dimers are the ones which are modified with respect to the preceding panel, while the light-orange dimers are left untouched. In panel (h) the flipped spins are depicted differently to highlight the need of creating two strings of rotated arrows.}
\end{figure}

\subsubsection{An ordered ground state of the strings phase}
As shown in Fig.~\ref{fig:StringsTiles}, there are 200 tiles that can \textit{a priori} be used to construct ground states in the strings phase. In particular, these tiles can be differentiated based on their contribution to the second-neighbor correlations. Indeed, the tiles of Types I and II (Fig.~\ref{fig:StringsTiles_d} to~\ref{fig:StringsTiles_d}) correspond to two ferromagnetic second-neighbor triangles; the tiles of Type III to one second-neighbor triangle respecting the ``two-ups one-down /two-downs one-up'' rule and one ferromagnetic second-neighbor triangle; and the tiles of Type IV to two  second-neighbor triangles respecting the ``two-ups one-down /two-downs one-up'' rule. 

At the same time, because there are two second-neighbor triangles for three sites, the factor $\frac{2}{3}$ in front of $J_2$ in the ground-state energy of this phase [Eq.~(\ref{eq:StrGSE})] implies that half of the second-neighbor triangles must be ferromagnetic and half must be in their ground state. An easy way to satisfy this constraint together with the tiling constraints is to make a long-range ordered configuration as shown in Fig.~\ref{fig:Strings_Ref}. It is easy to check that all the nearest-neighbor triangles as well as all the third-neighbor triangles are in their ground state in this state, which proves that it is a ground state of the strings phase. This ground state will be of crucial importance in our discussion because we are going to use it as a reference configuration.

\subsubsection{Creating and moving pairs of strings}
\label{sec:PLGS_strings}
To prove that any string configuration on the honeycomb lattice maps to at least one ground state of the strings phase, we are going to explicitly show how to construct ground states associated with given string configurations.

First, we notice that flipping a column of arrows compared to the reference configuration does not change the energy, since all the tiles in the resulting configuration belong to the ground state (also, the total number of ferromagnetic $J_2$ triangles is conserved by this move). This allows one to create a pair of strings: as depicted in Fig.~\ref{fig:Strings_Pair}, we adopt the convention that a red string runs along the right ends of crosses, and that the corresponding green string runs along the left end of crosses or of the leftmost right-pointing arrows from a column of crosses. This means that the green string can be moved to the left by flipping a left-pointing arrow which stands to its left, as in Fig.~\ref{fig:Strings_Pair_LeftMove}, without changing the energy. 
Similarly, the red string can be moved to the right by flipping a left-pointing arrow which stands to its left (destroying a cross and creating a new one), as in Fig.~\ref{fig:Strings_Pair_RightMove}. This implies that any valid configuration of two pairs of strings on honeycomb maps to at least one ground state of the small~$J_2$~phase.

For completeness we now want to show that without changing the reference configuration we can have a green string to the right of a red string. The same prescription as in Fig.~\ref{fig:Strings_Pair} allows one to create a new pair of strings next to the already existing one. The red string in this new pair can be moved all the way to the right, leaving behind a trail of right-pointing arrows. The green string can also be moved to the right by flipping these arrows again (Fig.~\ref{fig:Strings_NewPair_GreenRight}).

Finally, we show in Fig.~\ref{fig:Strings_NewPair_Subext} how trying to rotate an arrow creates a string of arrows spanning the whole system and following the neighboring string of crosses. It also imposes a global spin flip to the left that can only be ``absorbed'' by another string of rotated arrows. This suggests that there is a sub-extensive number of configurations that map onto the same string configuration. To convince ourselves that this number is growing exponentially only with the linear system size, we compute the residual entropy per site associated with the set of ground-state tiles without the tiles containing crosses. In that case, VUMPS tends to struggle to converge but systematically eventually gives an eigenvalue of one, and thus a zero residual entropy (Note that this corresponds to working with open boundary conditions and thus the absence of crosses in the bulk does not remove the possibility of non local moves due to the presence of empty hexagons on the boundary). Although this is not a rigorous proof, we consider it sufficient numerical evidence. Notice also that it is because of the initial choice of reference configuration that such global updates can be made between a red string on the left and a green string on the right, but not between a green string on the left and a red string on the right. There is no fundamental asymmetry there, only a matter of convention.

\subsection{Any ground-state configuration of the strings phase maps to a string configuration on the honeycomb lattice}
We have thus shown that any valid string configuration on the honeycomb lattice maps to a ground state of the strings phase. The converse still has to be proven, however. 
First, we consider ground-state configurations which do not have any crosses or empty hexagons. With periodic boundary conditions, there are 12 such ground states, all rotations or global spin flip with respect to our reference configuration in Fig.~\ref{fig:Strings_Ref}. These ground states all map to the configuration with no strings.
We now must take care of showing that all the other ground states - which have crosses and empty hexagons - map to a valid string configuration.

\begin{figure} 
	\centering
	\includegraphics[width = 0.8\columnwidth]{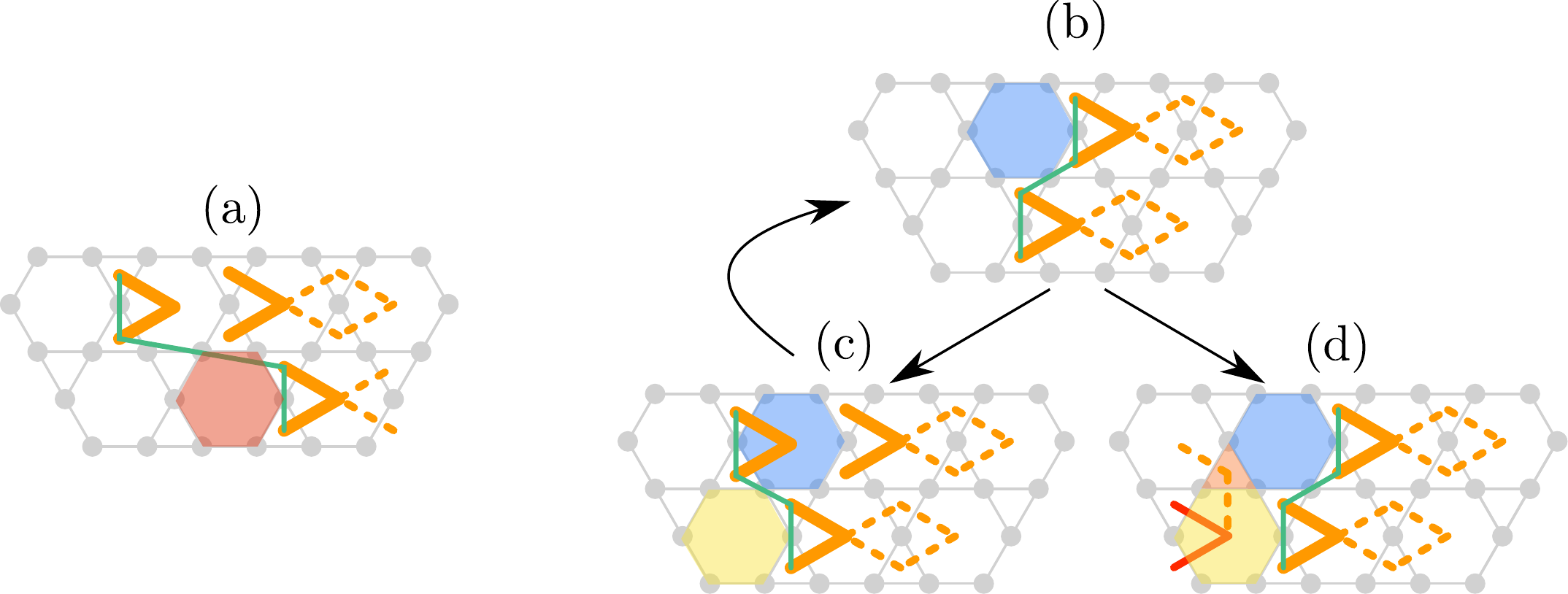}
	\caption[Support to prove that the strings are well constructed in the strings phase.]{\label{fig:PLGS_StringsPhaseProofHelp}Proving that the strings to the left of strings of crosses are well constructed. We want to show that the situation in (a), where the red hexagon does \textit{not} contain a right-pointing arrow, \textit{never} occurs. The orange segments stand for dimers, the dashed orange segments stand for two dimer possibilities. The initial position of the green string, connecting the left edges of crosses, is well-defined. Pushing the green string to the left creates situation like (b), where one has to decide whether to stop or to bring the green string across the hexagon highlighted in blue. There are two possible cases depending on whether a right-pointing arrow is in that hexagon. If there is one, as in (c), the green string can be brought across the blue hexagon, and we are back in a situation like in (b), and we can iterate. If there is none, then we are in situation (d) and cannot bring the green string further to the left. We have to show that (a) does not occur. However, since we are not in situation (c), the red triangle has to be occupied by some other dimer, which forbids putting a right-pointing arrow in the yellow triangle. Thus, the green string cannot be pushed further to the left, and the process stops.}
\end{figure}

\subsubsection{Crosses have to form strings}
We start by showing that crosses have to form strings. For this, we consider the tiles in Fig.~\ref{fig:StringsTiles}, and in particular the tiles of Type IV. These are the only tiles which bear crosses. Given such a tile, with the cross oriented horizontally, it has two lower nearest-neighbor tiles, of which at least one has to bear a dimer. From looking at the tiles of Types III and IV, it is obvious that this dimer has to belong to a cross, since no tile of Type III can fit there. Thus, one of the nearest-neighbor tiles below a cross has to be another cross. It is obvious that only one of these two tiles can be a cross, and thus crosses have to form strings on a triangular lattice of hexagons. It is also immediately clear that, given the orientation of the first cross, these strings can then only progress in one direction. According to the mapping we introduced in Appendix~\ref{sec:PLGS_strings}, a string of crosses immediately defines a string on the honeycomb lattice: one only needs to connect the right ends of the crosses. Thus, we have shown how to find the configuration of ``red strings'' associated with a ground state of the strings phase. But this is only half of the description of the associated string configuration.

\subsubsection{Finding the second string}
As we have seen above, in the case of periodic boundary conditions, strings have to come in pairs; and we still have to show how to associate a ``green strings'' configuration to a given ground state of the strings phase. For this, it is sufficient to show how to build the green string that runs to the left of a string of crosses.

In spirit, the prescription to find the location of the green string to the left of a string of crosses is simple: start from the string of crosses, and push the green string to the left until it meets a hexagon which is not a right-pointing arrow. We only have to prove that this way, the green string is always well-constructed; more precisely, that a situation like in Figure \ref{fig:PLGS_StringsPhaseProofHelp}a, where the green string has a ``jump'' and does not live on the honeycomb lattice, is prevented.

 This is easily seen from the following procedure: at any point in the process of pushing the green string to the left, one finds the situation depicted in Fig.~\ref{fig:PLGS_StringsPhaseProofHelp}b (or its vertical mirror), where one has to see if the green string can be pushed past the hexagon in blue. There are two possibilities. One is that the blue hexagon contains a right-pointing arrow, and the process has to be repeated with a new hexagon (highlighted in yellow in Fig.~\ref{fig:PLGS_StringsPhaseProofHelp}b).The other is that it corresponds to a tile which leaves empty the triangle highlighted in red. In this case the green string cannot be pushed further to the left. As illustrated in Fig.~\ref{fig:PLGS_StringsPhaseProofHelp}d, the green string is then well constructed: indeed, the red triangle can be occupied only in two ways (dotted line) which both forbid a right-pointing arrow in the triangle highlighted in yellow. Thus, iterating the procedure, one gets a well-defined green string. This shows that any ground state of the strings phase corresponds to a well-defined configuration of directed strings on the honeycomb lattice, and explain the residual entropy that we obtain.

\end{document}